\documentclass[10pt,nocut,fleqn,twoside,openright]{article}

\usepackage{packages}
\usepackage{notations}

\title{Mean-field inference methods for neural networks}
\author[1,2]{Marylou Gabrié}
\affil[1]{Center for Data Science, New York University}
\affil[2]{Center for Computational Mathematics, Flatiron Institute}
\date{}

\let\origdoublepage\cleardoublepage
\newcommand{\clearemptydoublepage}{%
  \clearpage
  {\pagestyle{empty}\origdoublepage}%
}

\let\cleardoublepage\clearemptydoublepage

\begin{document}

\maketitle{}

\abstract{
Machine learning algorithms relying on deep neural networks recently allowed a great leap forward in artificial intelligence. Despite the popularity of their applications, the efficiency of these algorithms remains largely unexplained from a theoretical point of view. The mathematical description of learning problems involves very large collections of interacting random variables, difficult to handle analytically as well as numerically. This complexity is precisely the object of study of statistical physics. Its mission, originally pointed towards natural systems, is to understand how macroscopic behaviors arise from microscopic laws. Mean-field methods are one type of approximation strategy developed in this view. We review a selection of classical mean-field methods and recent progress relevant for inference in neural networks. In particular, we remind the principles of derivations of high-temperature expansions, the replica method and message passing algorithms, highlighting their equivalences and complementarities. We also provide references for past and current directions of research on neural networks relying on mean-field methods. 
}

% \pagenumbering{roman}
{%
	%\singlespacing
	\hypersetup{linkcolor=black}
	\setcounter{tocdepth}{2}
	\tableofcontents
}

% \newpage
\section{Introduction}

With the continuous improvement of storage techniques, the amount of available data is currently growing exponentially. While it is not humanly feasible to  treat all the data created, \emph{machine learning}, as a class of algorithms that allows to automatically infer structure in large data sets, is one possible response.
In particular, \emph{deep learning} methods, based on neural networks, have drastically improved performances in key fields of artificial intelligence such as image processing, speech recognition or text mining. A good review of the first successes of this technology published in 2015 is \cite{LeCun2015a}. A few years later, the current state-of-the-art of this very active line of research is difficult to envision globally.
However, the complexity of deep neural networks remains an obstacle to the understanding of their great efficiency. Made of many layers, each of which constituted of many neurons, themselves accompanied by a collection of parameters, the set of variables describing completely a typical neural network is impossible to only visualize. Instead, aggregated quantities must be considered to characterize these models and hopefully help and explain the learning process. The first open challenge is therefore to identify the relevant observables to focus on. Often enough, what seems interesting is also what is hard to calculate. In the high-dimensional regime we need to consider, exact analytical forms are unknown most of the time and numerical computations are ruled out.  
, ways of approximation that are simultaneously simple enough to be tractable and fine enough to retain interesting features are highly needed.

In the context where dimensionality is an issue, physicists have experimented that macroscopic behaviors are typically well described by the theoretical limit of infinitely large systems. Under this \emph{thermodynamic} limit, the statistical physics of disordered systems offers powerful frameworks of approximation called \emph{mean-field theories}.
Interactions between physics and neural network theory already have a long history as we will discuss in \citesec~\ref{sec:chap1-nn-and-mf}. Yet, interconnections have been re-heightened by the recent progress in deep learning, which also brought new theoretical challenges.

Here, we wish to provide a concise methodological review of fundamental mean-field inference methods with their application to neural networks in mind. Our aim is also to provide a unified presentation of the different approximations allowing to understand how they relate and differ. 
Readers may also be interested in related review papers. Another methodological review is \cite{Advani2013}, particularly interested in applications to neurobiology. Methods presented in the latter reference have a significant overlap with what will be covered in the following. Some elements of random matrix theory are there additionally introduced. 
The approximations and algorithms which will be discussed here are also largely reviewed in \cite{Zdeborova2016}. 
% yet with emphasis on different applications in signal processing. 
This previous paper includes more details on spin glass theory, which originally motivated the development of the classical mean-field methods, and particularly focuses on community detection and linear estimation. 
Despite the significant overlap and beyond their differing motivational applications, the two previous references are also anterior to some recent exciting developments in mean-field inference covered in the present review, in particular extensions towards multi-layer networks. An older, yet very interesting, reference is the workshop proceedings \cite{opper2001advanced}, which collected both insightful introductory papers and research developments for the applications of mean-field methods in machine learning. Finally, the recent \cite{Carleo} covers more generally the connections between physical sciences and machine learning yet without detailing the methodologies. This review provides a very good list of references where statistical physics methods were used for learning theory, but also where machine learning helped in turn physics research.

Given the literature presented below is at the cross-roads of deep learning and disordered systems physics, we include short introductions to the fundamental concepts of both domains. These \citesecs~\ref{sec:chap1} and  \ref{sec:chap2} will help readers with one or the other background, but can be skipped by experts. In \citesec~\ref{sec:chap3}, classical mean-field inference approximations are derived on neural network examples. \citesec~\ref{sec:chap3further} covers some recent extensions of the classical methods that are of particular interest for applications to neural networks. We review in \citesec~\ref{sec:chapex-all} a selection of important historical and current directions of research in neural networks leveraging mean-field methods. As a conclusion, strengths, limitations and perspectives of mean-field methods for neural networks are discussed in \citesec~\ref{sec:conclu}.

\section[Machine learning with neural networks
% \\ {\small (Context and motivation)}
]{Machine learning with neural networks 
% \\ {\small (Context and motivation)}
}
\label{sec:chap1}

% \addcontentsline{toc}{section}{Machine learning with Neural networks and Mean-field approximations -Context, Motivation and Scope}%c

Machine learning is traditionally divided into three classes of problems: supervised, unsupervised and reinforcement learning. For all of them, the advent of deep learning techniques, relying on deep neural networks, has brought great leaps forward in terms of performance and opened the way to new applications. 
Nevertheless, the utterly efficient machinery of these algorithms remains full of theoretical puzzles.
This \citechap~provides fundamental concepts in machine learning for the unfamiliar reader willing to approach the literature at the crossroads of statistical physics and deep learning.
We also take this \citechap~as an opportunity to introduce the current challenges in building a strong theoretical understanding of deep learning.  
A comprehensive reference is \cite{Goodfellow2016}, while \cite{Mehta2018} offers a broad introduction to machine learning specifically addressed to physicists. 

% In this Section we quickly cover key concepts in supervised and unsupervised machine learning that will be useful to understand the context of this thesis and motivate its contributions.

\subsection{Supervised learning}
\paragraph{Learning under supervision}
Supervised learning aims at discovering systematic input to output mappings from examples. Classification is a typical supervised learning problem: for instance, from a set of pictures of cats and dogs labelled accordingly, the goal is to find a function able to predict in any new picture the species of the displayed pet. 

In practice, the \emph{training set} is a collection of $P$ example pairs $\cD = \{\x\kk, \y\kk\}_{k=1}^P$ from an input data space $\mathcal{X} \subseteq \R^N$ and an output data space $\mathcal{Y} \subseteq \R^M$. Formally, they are assumed to be i.i.d. samples from a joint distribution $p(\x,\y)$.  The predictor $h$ is chosen by a training algorithm from a \emph{hypothesis class}, a set of functions from $\mathcal{X}$ to  $\mathcal{Y}$, so as to minimize the error on the training set. This error is formalized as the \emph{empirical risk}
\begin{gather}
    \hat\cR(h, \ell, \cD) = \frac{1}{P}\sum_{k=1}^P \ell(\y\kk, h(\x\kk)) , 
\end{gather}
where the definition involves a loss function $\ell: \mathcal{Y} \times \mathcal{Y} \rightarrow \R$ measuring differences in the output space. 
This learning objective nevertheless does not guarantee \emph{generalization}, i.e. the ability of the predictor $h$ to be accurate on inputs $\x$ that are not in the training set. It is a surrogate for the ideal, but unavailable, \emph{population risk} 
\begin{gather}
    \cR(h, \ell) = \E_{\x, \y} \left[ \ell(\y, h(\x))\right] = \int_{\mathcal{X}, \mathcal{Y}} \dd \x \dd \y p(\x, \y) \ell(\y, h(\x)),
\end{gather}
expressed as an expectation over the joint distribution $p(\x,\y)$.
The different choices of hypothesis classes and training algorithms yield the now crowded zoo of supervised learning algorithms. 

\begin{figure}
    \centering
    \includegraphics[width=0.9\textwidth]{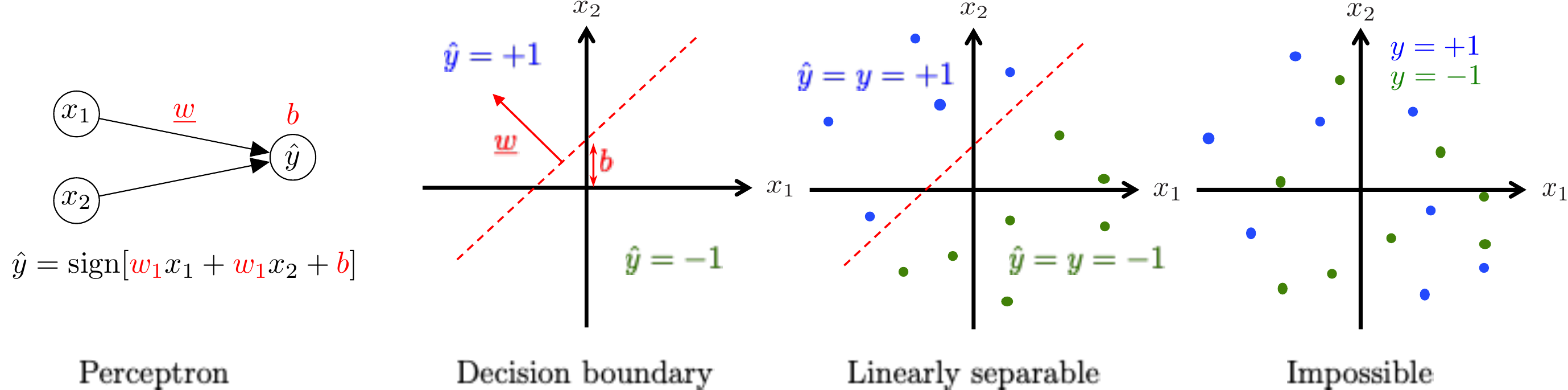}
    \caption{Let's assume we wish to classify data points $\x \in \R^2$ with labels $y=\pm 1$. We choose as an hypothesis class the perceptron sketched on the left with sign activation. For given weight vector $\vect{w}$ and bias $b$ the plane is divided by a decision boundary assigning labels. If the training data are linearly separable, then it is possible to perfectly predict all the labels with the perceptron, otherwise it is impossible.  \label{fig:chap1-perceptron}}
\end{figure}

\paragraph{Representation ability of deep neural networks}
In the context of supervised learning, deep neural networks enter the picture in the quality of a parametrized hypothesis class. Let us first quickly recall the simplest network, the \emph{perceptron}, including only a single neuron. It is formalized as a function from $\R^{N}$ to $\mathcal{Y} \subset \R$ applying an activation function $f$ to a weighted sum of its inputs shifted by a bias $b \in \R$, 
\begin{gather}
    \label{eq:chap1-perceptron}
    \hat{y} = h_{\vect{w}, b}(\x) = f(\vect{w}\T \x + b)
\end{gather}
where the weights are collected in the vector $\vect{w} \in \R^N$. From a practical standpoint, this very simple model can only solve the classification of linearly separable groups (see \citefig~\ref{fig:chap1-perceptron}). Yet from the point of view of learning theory, it has been the starting point of a rich statistical physics literature that will be discussed in \citesec~\ref{sec:chap1-nn-and-mf}.

Combining several neurons into networks defines more complex functions. The universal approximation theorem \cite{Cybenko1989, Hornik1991} proves that the following two-layer network architecture can approximate any well-behaved function with a finite number of neurons,
\begin{gather}
    \hat{y} = h_{\theta}(\x) = {\vect{w}^{(2)}}\T f(\W^{(1)} \x + \vect{b}) = \sum_{\alpha = 1}^M w^{(2)}_\alpha f({\vect{w}_\alpha^{(1)}}\T\x + b_\alpha), \qquad \theta = \{\vect{w}^{(2)}, \W^{(1)}, \vect{b}\}
\end{gather}
for $f$ a bounded, non-constant, continuous scalar function, acting component-wise. In the language of deep learning this network has one hidden layer of $M$ units. Input weight vectors $w^{(1)}_\alpha \in \R^N$ are collected in a weight matrix $\W^{(1)} \in \R^{M \times N}$. Here, and in the following, the notation $\theta$ is used as short for the collection of adjustable parameters. The universal approximation theorem is a strong result in terms of representative power of neural networks but it is not constructive. 
It does not quantify the size of the network, i.e. the number $M$ of hidden units, to approximate a given function, nor does it prescribe how to obtain the values of the parameters $\vect{w}^{(2)}, \W^{(1)}$ and $\vect{b}$ for the optimal approximation. While building an approximation theory is still ongoing (see e.g. \cite{Grohs2019}). Practice, led by empirical considerations, has nevertheless demonstrated the efficiency of neural networks.

In applications, neural networks with multiple hidden layers, deep neural networks, are preferred. A generic neural network of \emph{depth} $L$ is the function
\begin{gather}
    \label{eq:chap1-dnn}
    \hat{\y} = h_{\theta}(\x) = f(\W^{(L)}f(\W^{(L-1)} \cdots f(\W^{(1)} \x + \vect{b}^{(1)}) \cdots +\vect{b}^{(L-1)})+ \vect{b}^{(L)}), \\
    \label{eq:chap1-dnn-theta}
    \theta = \{\W^{(l)} \in \R^{N_{l} \times N_{l-1}} , \, \vect{b}^{(l)} \in \R^{N_l} \;; \; l=1 \cdots L\},
\end{gather}
where $N_0 = N$ is the dimension of the input and $N_L = M$ is the dimension of the output. The architecture is fixed by specifying the number of neurons, or \emph{width}, of the hidden layers $\{N_l\}_{l=1}^{L-1}$. The latter can be denoted $\hid^{(l)} \in \R^{N_l}$ and follow the recursion
\begin{gather}
	\label{eq:chap1-dnn-rec1}
	\hid^{(1)} = f(\W^{(1)} \x + \vect{b}^{(1)}) \, , \\
	\label{eq:chap1-dnn-rec2}
    \hid^{(l)}  = f(\W^{(l)} \hid^{(l-1)} + \vect{b}^{(l)}) \, , \quad l = 2 \cdots L-1 \, ,\\
    \label{eq:chap1-dnn-rec3}
	\hat{\y}  = f(\W^{(L)} \hid^{(L-1)} + \vect{b}^{(L)}) \, .
\end{gather}
Fixing the activation functions and the architecture of a neural network defines an hypothesis class. It is crucial that activations introduce non-linearities; the most common are the hyperbolic tangent tanh and the rectified linear unit defined as $\mathrm{relu}(x)= \max(0,x)$. Note that it is also possible to define stochastic neural networks by using noisy activation functions, uncommon in supervised learning applications except  at training time so as to encourage generalization \cite{Poole2014, Srivastava2014}.

An originally proposed intuition for the advantage of depth is that it enables to treat the information in a hierarchical manner; either looking at different scales in different layers, or learning more and more abstract representations \cite{Bengio2013}. Nevertheless, getting a clear theoretical understanding why in practice `the deeper the better' is still an ongoing direction of research (see e.g. \cite{Telgarsky2016, Daniely2017, Safran2019}).

\paragraph{Neural network training}
Given an architecture defining $h_\theta$, the supervised learning objective is to minimize the empirical risk $\hat \cR$ with respect to the parameters $\theta$. This optimization problem lives in the dimension of the number of parameters which can range from tens to millions. The idea underlying the majority of training algorithms is to perform a gradient descent (GD) starting at parameters drawn randomly from an initialization distribution:
\begin{gather}
    \theta_0 \sim p_{\theta_0}(\theta_0) \\
    \theta_{t+1} \leftarrow \theta_t - \eta \nabla_\theta \hat \cR = \theta_t - \eta  \frac{1}{P}\sum_{k=1}^P \nabla_\theta \ell\left(\y\kk, h_{\theta_t}\left(\x\kk\right)\right) \,.
\end{gather}
The parameter $\eta$ is the learning rate, controlling the size of the step in the direction of decreasing gradient per iteration. The computation of the gradients can be performed in time scaling linearly with depth by applying the derivative chain-rule leading to the \emph{back-propagation} algorithm \cite{Goodfellow2016}. A popular alternative to gradient descent is stochastic gradient descent (SGD) where the sum over the gradients for the entire training set is replaced by the sum over a small number of samples, randomly selected at each step \cite{RobbinsHMonro1951,Bottou2010}.

During the training iterations, one typically monitors the \emph{training error} (another name for the empirical risk given a training data set) and the \emph{validation error}. The latter corresponds to the empirical risk computed on a set of points held-out from the training set, the validation set, to assess the generalization ability of the model either along the training or in order to select hyperparameters of training such as the value of the learning rate. A posteriori, the performance of the model is judged from the \emph{generalization error}, which is evaluated on the never seen \emph{test set}.
While two different training algorithms (e.g. GD vs SGD) may achieve zero training error, they may differ in the level of generalization they typically reach.

% \missing{cross entropy loss?}

\paragraph{Open questions and challenges}
Building on the fundamental concepts presented in the previous paragraphs, practitioners managed to bring deep learning to unanticipated performances in the automatic processing of images, speech and text (see \cite{LeCun2015a} for a few years old review). 
Still, many of the greatest successes in the field of neural network were obtained using ingenious tricks while many fundamental theoretical questions remain unresolved. 

Regarding the optimization first, (S)GD training generally discovers parameters close to zero risk. Yet, gradient descent is guaranteed to converge to the neighborhood of a global minimum only for a convex function and is otherwise expected to get stuck in a local minimum. Therefore, the efficiency of gradient-based optimization is a priori a paradox given the empirical risk $\hat R$ is non-convex in the parameters $\theta$. 
% Second, the representation capacity of deep learning models is also poorly understood. Results in the literature that relate the size and architecture of a network to a measure of its ability to learn are too far from realistic settings to guide choices of practitioners. 
Second, the generalization ability of deep neural networks trained by (S)GD is still poorly understood.
The size of training data sets is limited by the cost of labelling by humans,  experts or heavy computations.
Thus training a deep and wide network amounts in practice to fitting a model of millions of degrees of freedom against a somehow relatively small amount of data points. Nevertheless it does not systematically lead to \emph{overfitting}. The resulting neural networks can have surprisingly good predictions both on inputs seen during training and on new inputs \cite{Zhang2017}.   
Results in the literature that relate the size and architecture of a network to a measure of its ability to generalize are too far from realistic settings to guide choices of practitioners. 
On the one hand, traditional bounds in statistics, considering worst cases, appear overly pessimistic \cite{Vapnik2000,Bartlett2002,Shalev-Shwartz2014,Abbara2019}. On the other hand, historical statistical physics analyses of learning, briefly reviewed in \citesec~\ref{sec:chap1-nn-and-mf}, only concern simple architectures and synthetic data. This lack of theory results in potentially important waste: in terms of time lost by engineers in trial and error to optimize their solution, and in terms of electrical resources used to train and re-train possibly oversized networks while storing potentially unnecessarily large training data sets.

The success of deep learning, beyond these apparent theoretical puzzles, certainly lies in the interplay of advantageous properties of training algorithms, the neural network hypothesis class and structures in typical data (e.g. real images, conversations). Disentangling the role of the different ingredients is a very active line of research (see \cite{Giryes2016} for a review).
% among which this thesis is a modest contribution to. In particular, \citechap~ \ref{sec:chap5} touches the question of the generalization capability of neural networks.

\subsection{Unsupervised learning}
\label{sec:chap1-unsupervised}
\paragraph{Density estimation and generative modelling}
The goal of unsupervised learning is to directly extract structure from data. 
Compared to the supervised learning setting, the training data set is made of a set of example inputs $\cD = \{\vect{x}\kk\}_{k=1}^P$ without corresponding outputs. A simple example of unsupervised learning is clustering, consisting in the discovery of unlabelled subgroups in the training data. 
Most unsupervised learning algorithms either implicitly or explicitly adopt a probabilistic viewpoint and implement \emph{density estimation}. The idea is to approximate the true density $p(\x)$ from which the training data was sampled by the closest (in various senses) element among a family of parametrized distributions over the input space $\{ p_\theta(.),\;  \theta \in \R^{N_\theta} \}$. The selected $p_{\theta}$ is then a model of the data.
% Structural properties of the data can sometimes be inferred from it, such as dependencies according to its factorization. 
If the model $p_{\theta}$ is easy to sample, it can be used to generate new inputs comparable to the training data points - which leads to the terminology of \emph{generative models}. In this context, \emph{unsupervised deep learning} exploits the representational power of deep neural networks to create sophisticated candidate $p_\theta$.
% \missing{Could also do clustering etc.. But really ad hoc modelling of complex data with nn is more directly what will interest us.}

A common formalization of the learning objective is to maximize the \emph{likelihood}, defined as the probability of i.i.d. draws from the model $p_\theta$ to have generated the training data $\cD = \{\vect{x}\kk\}_{k=1}^P$, or equivalently its logarithm,
\begin{gather}
    \maxx{\theta} \prod_{k=1}^P p_\theta(\x\kk) \quad \iff \quad \maxx{\theta} \sum_{k=1}^P \log p_\theta(\x\kk).
\end{gather} 
The second logarithmic additive formulation is generally preferred. 
It can be interpreted as the minimization of the Kullback-Leibler divergence between the empirical distribution $p_\cD(\x) = \sum_{k=1}^P \delta(\x - \x\kk) / P$ and the model $p_\theta$:
\begin{gather}
    \minn{\theta} \KL(p_\cD || p_\theta) = \minn{\theta} \int \dd{\x}  p_\cD(\x) \log \frac{p_\cD(\x) }{p_\theta(\x)} \quad \iff  \quad \maxx{\theta} \sum_{k=1}^P \log p_\theta(\x\kk) \,, 
\end{gather}
although considering the divergence with the discrete empirical measure is slightly abusive.
The detail of the optimization algorithm here depends on the specification of $p_\theta$. As we will see, the likelihood in itself is often intractable and learning consists in a gradient ascent on at best a lower bound, otherwise an approximation, of the likelihood. 

A few years ago, an alternative strategy called adversarial training was introduced by \cite{Goodfellow2014}. Here an additional trainable model called the discriminator, for instance parametrized by $\phi$ and denoted $d_\phi(\cdot)$, computes the probability for points in the input space $\mathcal{X}$ of belonging to the training set $\cD$ rather than being generated by the model $p_\theta(\cdot)$. The parameters $\theta$ and $\phi$ are trained simultaneously such that, the generator learns to fool the discriminator and the discriminator learns not to be fooled by the generator. The optimization problem usually considered is
\begin{gather}
    \minn{\theta}\maxx{\phi} \EE{\cD}{\log(d_\phi(\x))}  + \EE{p_\theta}{\log(1-d_\phi(\x))} \, ,
\end{gather}
where the sum of the expected log-probabilities according to the discriminator for examples in $\cD$ to be drawn from $\cD$ and examples generated by the model not to be drawn from $\cD$ is maximized with respect to $\phi$ and minimized with respect to $\theta$.

In the following, we present two classes of generative models based on neural networks.

\paragraph{Deep Generative Models}
\label{sec:chap1-vae}
A deep generative models defines a density $p_\theta$  obtained by propagating a simple distribution through a deep neural network. It can be formalized by introducing a latent variable $\z \in \R^N$ and a deep neural network $h_\theta$ similar to \eqref{eq:chap1-dnn} of input dimension $N$. The generative process is then 
\begin{gather}
    \label{eq:chap1-dgm-1}
    \z \sim p_z(\z) \\
    \label{eq:chap1-dgm-2}
    \x \sim p_\theta(\x |\z) = p_{\rm out}(\x| h_\theta(\z)),
\end{gather}
where $p_z$ is typically a factorized distribution on $\R^N$ easy to sample (e.g. a standard normal distribution), and $p_{\rm out}(.|h_\theta(\z))$ is for instance a multivariate Gaussian distribution with mean and covariance that are functions of $h_\theta(\z)$.
The motivation to consider this class of models for joint distributions is three-fold. First the class is highly expressive. Second, it follows from the intuition that data sets leave on low dimensional manifolds, which here can be spaned by varying the latent representation $\z$ usually much smaller than the input space dimension (for further intuition see also the reconstruction objective of the first autoencoders, see e.g. Chapter 14 of \cite{Goodfellow2016}). Third, yet perhaps more importantly, the class can be optimized over easily using back-propagation, unlike the Restricted Boltzmann Machines presented in the next paragraph largely replaced by deep generative models.
There are two main types of deep generative models. Generative Adversarial Networks (GAN) \cite{Goodfellow2014} trained following the adversarial objective mentioned above, and Variational AutoEncoders (VAE) \cite{Kingma2014, Rezende2014} trained to maximize a likelihood lower-bound.

\subparagraph{Variational AutoEncoders}
The computation of the likelihood of one training sample $\x\kk$ for a deep generative model \eqref{eq:chap1-dgm-1}-\eqref{eq:chap1-dgm-2} requires then the marginalization over the latent variable $\z$, 
\begin{gather}
    p_\theta(\x) = \int \dd{\z} p_{\rm out}(\x | h_\theta(\z)) p_z(\z).
\end{gather}
This multidimensional integral cannot be performed analytically in the general case. It is also hard to evaluate numerically as it does not factorize over the dimensions of $\z$ which are mixed by the neural network $h_\theta$. 
Yet a lower bound on the log-likelihood can be defined by introducing a tractable conditional distribution $q(\z |\x)$ that will play the role of an approximation of the intractable \emph{posterior} distribution $p_{\theta}(\z | \x)$ implicitly defined by the model:
\begin{align}
    \log p_{\theta}(\x) %& = \int \dd{\z} q(\z | \x) \log p_{\theta}(\x) \\
                        % & = \int \dd{\z} q(\z | \x) \log \frac{p_{\theta}(\x, \z)}{p_{\theta}(\z | \x)} \notag \\
                        % & = \int \dd{\z} q(\z | \x) \log \frac{p_{\theta}(\x, \z)}{p_{\theta}(\z | \x)} \times \frac{q(\z | \x)}{q(\z | \x)} \notag \\ 
                        % & = \int \dd{\z} q(\z | \x) \left[ - \log q(\z | \x) + \log p_{\theta}(\x, \z) \right] + \KL(q(\z|\x) ||  p_\theta(\z|\x)) \notag \\
                        & \geq \int \dd{\z} q(\z | \x) \left[ - \log q(\z | \x) + \log p_{\theta}(\x, \z) \right] = \mathrm{LB}(q, \theta, \x) \label{eq:chap1-vae-lb}.
\end{align}
% where the last inequality comes from the fact that a $\KL$-divergence is always positive; it is replaced by an equality if and only if $q(\z|\x) = p_\theta(\z|\x)$. 
% Provided $q$ has a tractable expression and is easy to sample, the remaining lower bound $\mathrm{LB}(q, \theta, \x)$ can be approximated by a Monte Carlo method.
% Maximum likelihood learning is then approached by the simultaneous maximization of the lower bound $\mathrm{LB}(q, \theta, \x)$ and tightening of the inequality \eqref{eq:chap1-vae-lb} by a minimization of $\KL(q(\z|\x) ||  p_\theta(\z|\x))$. This auxiliary optimization requires in practice to parametrize as well the tractable posterior $q = q_{\phi}$, typically with a neural network. 
Maximum likelihood learning is then approached by the maximization of the lower bound $\mathrm{LB}(q, \theta, \x)$, which requires in practice to parametrize the tractable posterior $q = q_{\phi}$, typically with a neural network. 
% In order to adjust the parameters $\theta$ and $\phi$ through gradient descent
% a last reformulation is still necessary. The expectation over $q_\phi$ estimated via Monte Carlo indeed involves the $\phi$ parameters. To disentangle this dependence $\z \sim q_\phi(\z|\x)$ must be rewritten as $\z = g(\epsilon, \phi, \x)$ with $\epsilon \sim p_\epsilon(\epsilon)$ independent of $\phi$. 
Using the so-called re-parametrization trick \cite{Kingma2014,Rezende2014}, the gradients of $\mathrm{LB}(q_\phi, \theta, \x)$ with respect to $\theta$ and $\phi$ can be approximated by a Monte Carlo, so that the likelihood lower bound can be optimized by gradient ascent.

\subparagraph{Generative Adversarial Networks}
The principle of adversarial training was designed directly for a deep generative model \cite{Goodfellow2014}. Using a deep neural network to parametrize the discriminator $d_\phi(\cdot)$ as well as the generator $p_\theta(\cdot)$, it leads to a remarkable quality of produced samples and is now one of the most studied generative model.

% \missing{An historical note, to point that VAEs are inspired by AE which are not probabilistic, and that somehow it was an improvment - or maybe later in open questions?}

% In this thesis we will mention variational autoencoders in the perspectives of \citechap~\ref{sec:chap4} and \citechap~\ref{sec:chap6}. We will discuss a different proposition to approximate the likelihood and the alternative Bayesian learning objective involving prior knowledge on the parameters $\theta$.

\paragraph{Restricted Boltzmann Machines}
Models described in the preceding paragraphs comprised only \emph{feed forward} neural networks. In feed forward neural networks, the state or value of successive layers is determined following the recursion \eqref{eq:chap1-dnn-rec1}-\eqref{eq:chap1-dnn-rec3}, in one pass from inputs to outputs.
Boltzmann machines instead involve \emph{undirected} neural networks which consist of stochastic neurons with symmetric interactions.
%  It means that they define a probability distributions instead of deterministic functions. 
The probability law associated with a neuron state is a function of neighboring neurons, themselves reciprocally function of the first neuron. Sampling a configuration therefore requires an equilibration in the place of a simple forward pass. 

A Restricted Boltzmann Machine (RBM) \cite{Ackley1985, Smolensky186} with M hidden neurons in practice defines a joint distribution over an input (or visible) layer $\x \in \{0,1\}^N$ and a hidden layer $\hidd  \in \{0,1\}^M$, 
\begin{gather}
\label{eq:chap1-rbm-meas}
p_\theta(\x, \hid) = \frac{1}{\cZ} e^{\vect{a}\T \x + \vect{b}\T \hidd + \x\T\W\hidd} \, , \qquad \theta = \{ \W, \vect{a}, \vect{b} \} \, , 
\end{gather}
where $\cZ$ is the normalization factor, similar to the partition function of statistical physics. The parametric density model over inputs is then the marginal $p_\theta(\x) = \sum_{\hidd \in \{0,1\}^M} p_\theta(\x, \hid)$. Although seemingly very similar to pairwise Ising models, the introduction of hidden units provides a greater representative power to RBMs as hidden units can mediate interactions between arbitrary groups of input units. Furthermore, they can be generalized to Deep Boltzmann Machines (DBMs) \cite{Salakhutdinov2009}, where several hidden layers are stacked on top of each other. 

Identically to VAEs, RBMs can represent sophisticated distributions at the cost of an intractable likelihood. Indeed the summation over $2^{M+N}$ terms in the partition function cannot be simplified by an analytical trick and is only realistically doable for small models. 
% This intractability remains a priori an issue for the learning by gradient ascent of the log-likelihood. Given an input training point $\x\kk$, the derivatives of $\log p(\x\kk)$ with respect to trainable parameters are
% \begin{subequations} \label{eq:chap1-rbm-grad}
% \begin{gather}
%     \nabla_{\vect{a}} \log p(\x\kk) = \x\kk - \langle \x \rangle \, , \\
%     \nabla_{\vect{b}} \log p(\x\kk) = \langle \hidd \rangle_{\x\kk} - \langle \hid\rangle \, ,\\
%     \nabla_{\W} \log p(\x\kk) = \langle \hidd \, \x\T \rangle_{\x\kk} - \langle \hid \, \x\T \rangle,
% \end{gather} 
% \end{subequations}
% where we introduce the notation $\langle . \rangle$ for the average over the RBM measure \eqref{eq:chap1-rbm-meas} and $\langle . \rangle_{\x\kk}$ for the average over the RBM measure conditioned on $\x=\x\kk$, also called the \emph{clamped} RBM measure.  To approximate the gradients, sampling over the clamped measure is actually easy as the hidden neurons are independent when conditioned on the inputs; in other words $p(\hidd |\x)$ is factorized.
%  Yet sampling over the complete RBM measure is more involved and requires for example the equilibration of a Markov Chain. 
RBMs are commonly trained through a gradient ascent of the likelihood using approximated gradients. As exact Monte Carlo evaluation is a costly operation that would need to be repeated at each parameter update in the gradient ascent, several more or less sophisticated approximations are preferred: contrastive divergence (CD) \cite{Hinton2002},  its persistent variant (PCD) \cite{Tieleman2008} or even parallel tempering \cite{Desjardins2010,Cho2010}.
% Instead it is commonly approximated by \emph{contrastive divergence} (CD) \cite{Hinton2002}, 
% which approximates the $\langle . \rangle$ in \eqref{eq:chap1-rbm-grad} by the final state of a Monte Carlo Markov chain initialized in $\x\kk$ and updated for a small number of steps.  

RBMs were the first effective generative models using neural networks. They found applications in various domains including dimensionality reduction \cite{Hinton2006a}, classification \cite{larochelle2008classification}, collaborative filtering \cite{salakhutdinov2007restricted}, feature learning \cite{coates2011analysis}, and topic modeling \cite{hinton2009replicated}.
Used for an unsupervised pre-training of deep neural networks layer by layer \cite{Hinton2006,Bengio2007}, they also played a crucial role in the take-off of supervised deep learning.

% The measure defined by a DBM of depth $L$ is
% \begin{gather}
%     \label{eq:chap1-dbm-meas}
%     p_\theta(\x, \hid) = \frac{1}{\cZ} e^{\vect{a}\T \x  + \x\T\W^{(1)}\hidd + {\vect{b}^{(L)}}\T \hidd^{(L)}  + \sum_{l=1}^{L-1} {\hid^{(l)}}\T\W^{(l+1)}\hidd^{(l+1)} + {\vect{b}^{(l)}}\T \hidd^{(l)} }\\
%     \theta = \{ \vect{a}\,, \; \vect{b}^{(l)}, \; \W^{(l)} \; ; \; l = 1 \cdots L \} .
% \end{gather}
% Similarly to \eqref{eq:chap1-rbm-grad}, the log-likelihood gradient ascent with respect to the parameters involves averages over the DBM measure and the corresponding clamped measure. In this case both of them involve an equilibration to be approximated by a Markov Chain. The contrastive divergence algorithm can be generalized to circumvent this difficulty, but the approximation of the gradients appear all the more less controlled.  

\paragraph{Open questions and challenges}
Generative models involving neural networks such as VAE, GANs and RBMs have great expressive powers at the cost of not being amenable to exact treatment. Their training, and sometimes even their sampling requires approximations. From a practical standpoint, whether these approximations can be either made more accurate or less costly is an open direction of research. 
Another important related question is the evaluation of the performance of generative models \cite{Sajjadi2018}. To start with the objective function of training is very often itself intractable (e.g. the likelihood of a VAE or a RBM), and beyond this objective, the unsupervised setting does not define a priori a test task for the performance of the model. 
Additionally, unsupervised deep learning inherits some of the theoretical puzzles already discussed in the supervised learning section. In particular, assessing the difficulty to represent a distribution and select a sufficient minimal model and/or training data set is an ongoing effort of research.

% In this thesis, unsupervised learning models will be discussed in particular in \citechap~\ref{sec:chap4}. A novel deterministic framework for RBMs and DBMs with a tractable learning objective is derived, allowing to exploit efficiently their representative power. 
% \review{In Chapter \ref{sec:chap6}, an algorithm for the optimal Bayesian learning of simple models of VAEs is proposed.}

% \newpage
\section{Statistical inference and the statistical physics approach
% \\ {\small (Fundamental theoretical frameworks)} 
}
\label{sec:chap2}

To tackle the open questions and challenges surrounding neural networks mentioned in the previous \citesec, we need to manipulate high-dimensional probability distributions. The generic concept of statistical inference refers to the extraction of useful information from these complicated objects.  Statistical physics, with its probabilistic interpretation of natural systems composed of many elementary components, is naturally interested in similar questions. 
% Before going into the details of the corresponding mean-field methods, 
We provide in this section a few concrete examples of inference questions arising in neural networks and explicit how statistical physics enters the picture. In particular, the theory of disordered systems appears here especially relevant.

\subsection{Statistical inference}

\label{sec:chap2-stat-inf}
 
\subsubsection{Some inference questions in neural networks for machine learning}
\label{sec:chap2-teacher-student}
% To fix the ideas, we present what kind of inference questions arising in the context of deep learning will be of particular interest to us.

\paragraph{Inference in generative models}
Generative models used for unsupervised learning are statistical models defining high-dimensional distributions with complex dependencies. As we have seen in \citesec~\ref{sec:chap1-unsupervised}, the most common training objective in unsupervised learning is the maximization of the log-likelihood, i.e. the log of the probability assigned by the generative model to the training set $\{\x\kk\}_{k=1}^{P}$. Computing the probability of observing a given sample $\x\kk$ is an inference question. It requires to marginalize over all the hidden representations of the problem. For instance in the RBM \eqref{eq:chap1-rbm-meas},
% \begin{align}
%     p(\x\kk) = \frac{1}{\cZ}\sum_{\hidd \in \{0,1\}^M} e^{{\x\kk}\T\W\hidd}p_x(\x\kk)p_t(\hidd). 
% \end{align}
\begin{gather}
    % \label{eq:chap1-rbm-meas}
    p_\theta(\x\kk) = \frac{1}{\cZ} \sum_{\hidd \in \{0,1\}^M} e^{\vect{a}\T \x\kk + \vect{b}\T \hidd + \x\kk\T\W\hidd}.
    \end{gather}
While the numerator will be easy to evaluate, the partition function has no analytical expression and its exact evaluation requires to sum over all possible states of the network.

\paragraph{Learning as statistical inference: Bayesian inference and the teacher-student scenario}

% \missing{Gardner reference in Sebastian's paper}

The practical problem of training neural networks from data as introduced in \citechap~\ref{sec:chap1} is not in general interpreted as inference. To do so, one needs to treat the learnable parameters as random variables, which is the case in Bayesian learning. For instance in supervised learning, an underlying prior distribution $p_\theta(\theta)$ for the weights and biases of a neural network \eqref{eq:chap1-dnn}-\eqref{eq:chap1-dnn-theta} is assumed, so that Bayes rule defines a posterior distribution given the training data $\mathcal{D}$,
\begin{align}
p(\theta | \mathcal{D}) & = \frac{p(\mathcal{D} | \theta) p_\theta(\theta)}{p(\mathcal{D})}. 
\end{align}
Compared to the single output of risk minimization, we obtain an entire distribution for the learned parameters $\theta$, which takes into account not only the training data but also some knowledge on the structure of the parameters (e.g. sparsity) through the prior. In practice, Bayesian learning and traditional empirical risk minimization may not be so different. On the one hand, the Bayesian posterior distribution is often summarized by a point estimate such as its maximum. On the other hand risk minimization is often biased towards desired properties of the weights through regularization techniques (e.g. promoting small norm) recalling the role of the Bayesian prior. 

However, from a theoretical point of view, Bayesian learning is of particular interest in the \emph{teacher-student} scenario. The idea here is to consider a toy model of the learning problem where parameters are effectively drawn from a prior distribution.
Let us use as an illustration the case of the supervised learning of the perceptron model \eqref{eq:chap1-perceptron}. We draw a weight vector $\vect{w}_0$, from a prior distribution $p_w(\cdot)$, along with a set of $P$ inputs $\{\x\kk\}_{k=1}^{P}$ i.i.d from a data distribution $p_x(\cdot)$. Using this \emph{teacher} perceptron model we also draw a set of possibly noisy corresponding outputs $y\kk$ from a teacher conditional probability $p(.| \vect{w}_0\T\x\kk)$. From the training set of the $P$ pairs $\mathcal{D} = \{\x\kk, y\kk\}$, one can attempt to rediscover the teacher rule by training a \emph{student} perceptron model.
The problem can equivalently be phrased as a reconstruction inference question: can we recover the value of $\vect{w}_0$ from the observations in $\mathcal{D}$? The Bayesian framework yields a posterior distribution of solutions
\begin{gather}
    p(\vect{w}| \mathcal{D}) = \prod_{k=1}^P p(y\kk| \vect{w}\T\x\kk)p_w(\vect{w}) \, / \, \prod_{k=1}^P p(y\kk | \x\kk).
\end{gather}

Note that the terminology of teacher-student applies for a generic inference problem of reconstruction: the statistical model used to generate the data along with the realization of the unknown signal is called the \emph{teacher}; the statistical model assumed to perform the reconstruction of the signal is called the \emph{student}. When the two models are identical or matched, the inference is \emph{Bayes optimal}. When the teacher model is not perfectly known, the statistical models can also be different (from slightly differing prior distributions to entirely different models), in which case they are said to be mismatched, and the reconstruction is suboptimal.

Of course in practical machine learning applications of neural networks, one has only access to an empirical distribution of the data and it is unclear whether there should exist a formal rule underlying the input-output mapping.
% data do not follow a known distribution and there is no mathematical ground truth rule underlying the input-output mapping. 
Yet the teacher-student setting is a modelling strategy of learning which offers interesting possibilities of analysis and we shall refer to numerous works resorting to the setup in  \citesec~\ref{sec:chapex-all}.

\subsubsection{Answering inference questions}
\label{sec:chap2-challenges-inf}

Many inference questions in the line of the ones mentioned in the previous \citesec~have no tractable exact solution. When there exists no analytical closed-form, computations of averages and marginals require summing over configurations. Their number typically scales exponentially with the size of the system, then becoming astronomically large for high-dimensional models. % Hence the computational hardness comes from the astronomically large number of configurations to be tracked to describe a high-dimensional joint probability distribution.
Hence it is necessary to design approximate inference strategies. They may require an algorithmic implementation but must run in finite (typically polynomial) time. 
An important cross-fertilization between statistical physics and information sciences have taken place over the past decades around the questions of inference. Two major classes of such algorithms are Monte Carlo Markov Chains (MCMC), and mean-field methods. The former is nicely reviewed in the context of statistical physics in \cite{Krauth2006}. The latter will be the focus of this short review, in the context of deep learning.

\begin{figure}[t]
    \centering
    \captionsetup{width=.3\linewidth}
    \subfloat[Restricted Boltzmann Machine]{\includegraphics[width=0.42\textwidth, valign=m]{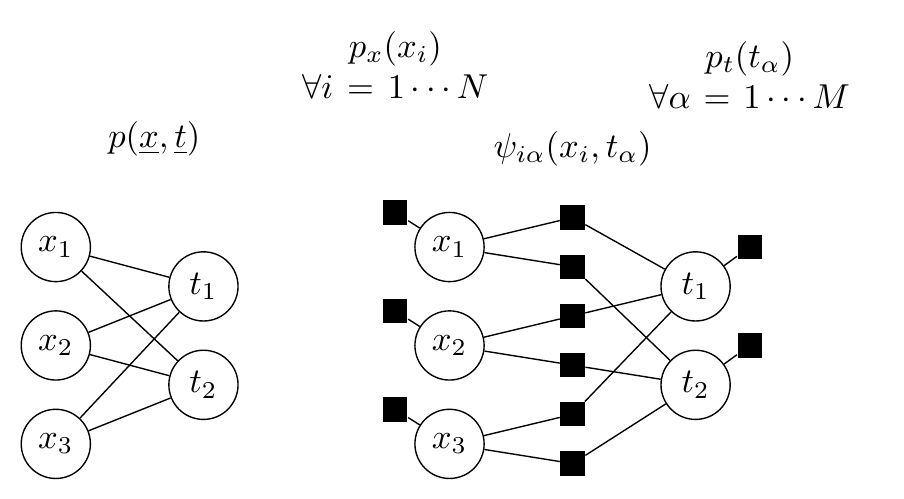}
    \label{fig:chap2-rbm-bis}}
    % \hspace{1cm}
    \captionsetup{width=.4\linewidth}
    \hspace{0.1\textwidth}
    \subfloat[Perceptron teacher and student. ]{\includegraphics[width=0.4\textwidth, valign=m]{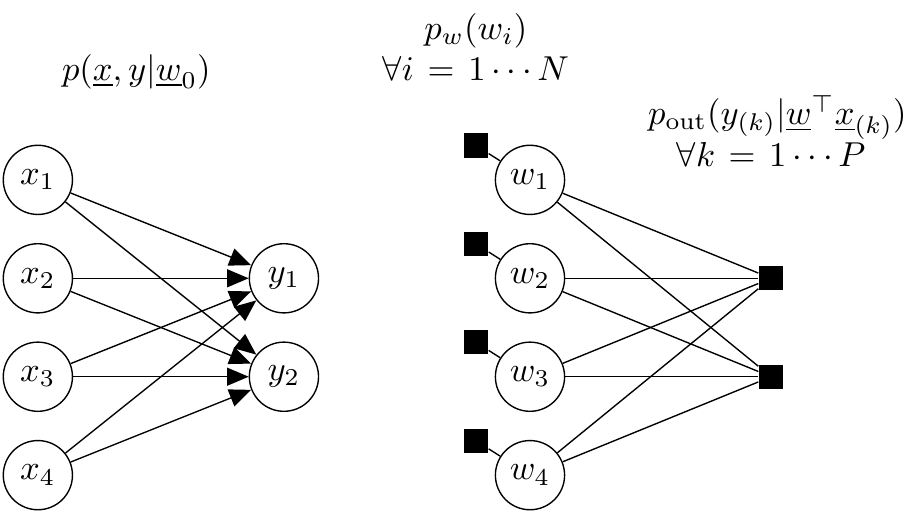}
    \label{fig:chap2-glm-bis}}
    \captionsetup{width=.9\linewidth}
    \caption{\textbf{(a)} Undirected probabilistic graphical model (left) and factor graph representation (right). \textbf{(b)} Left: Directed graphical model of the generative model for the training data knowing the teacher weight vector $\vect{w}_0$. Right: Factor graph representation of the posterior distribution for the student $p(\vect{w} | \X, \y )$, where the vector $\y \in \R^P$ gathers the outputs $y_{(k)}$ and the matrix $\X\in\R^{N \times P}$ gathers the inputs $\x_{(k)}$.\label{fig:chap2-graphs}}
\end{figure}

Note that representations of joint probability distributions through probabilistic graphical models and factor graphs are crucial tools to design efficient inference strategies. In \citeapp~\ref{app:chap2-graphs}, we quickly introduce for the unfamiliar reader these two formalisms that enable to encode and exploit independencies between random variables. As examples, \citefig~\ref{fig:chap2-graphs} presents graphical representations of the RBM measure \eqref{eq:chap1-rbm-meas} and the posterior distribution in the Bayesian learning of the perceptron as discussed in the previous \citesec.

%The graphical representations introduced in \citesec~\ref{sec:chap2-graphs} are of great help in doing so. 

% \emely loose when compared to practical examples.

% Michael Jordan 'Underlying much of the heightened interest in these links between statistical
% physics and the information sciences is the development (in the latter field) of a general framework for associating joint probability distributions with graphs, and for exploiting the structure of the graph in the computation of marginal probabilities and expectations. Probabilistic graphical models are graphs-directed or undirected-annotated with functions defined on local clusters of nodes that when taken together define families of joint probability distributions on the graph. Not only are the classical models of statistical physics instances of graphical models (generally involving undirected graphs), but many applied probabilistic models with no obvious connection to physics are graphical models as well-examples include phylogenetic trees in genetics, diagnostic systems in medicine, unsupervised learning models in machine learning, and error-control codes in information theory. The availability of the general framework has made it possible for ideas to flow more readily between these fields..'

\subsection{Statistical physics of disordered systems, first appearance on stage}
Here we re-introduce briefly fundamental concepts of statistical physics that will help to understand connections with inference and the origin of the methods presented in what follows.

\paragraph{The thermodynamic limit}
% To solve and study inference questions we will resort to concepts and methods originally developed by physicists. In a nutshell we remind some key concepts in statistical physics. 

% \paragraph{Boltzmann distribution and free energy}
% Statistical physics aims at explaining the behaviors of systems composed of a large number of degrees of freedom. Although the laws of classical physics governing microscopic interactions are deterministic, it is unrealistic to keep track of the evolution of each of the elementary component to account for the behavior of the whole. Remarkably, macroscopic quantities of interest to the physicist are well described by an averaging over the elementary states according to the \emph{Boltzmann distribution}. 
The equilibrium statistics of classical physical systems are described by the Boltzmann distribution.
For a system with $N$ degrees of freedom noted $\x \in \cX^N$ and an energy functional $E(\x)$, we have
\begin{align}
    \label{eq:chap2-boltzmann}
    p(\x) = \frac{e^{-\beta E(\x)}}{\cZ_N}, \quad \cZ_N = \sum_{\x \in \mathcal{X}^N} e^{-\beta E(\x)}, \quad \beta = 1/k_B T , %\\
\end{align}
where we defined the partition function $\cZ_N$ and the inverse temperature $\beta$.
% While the adoption of the probabilistic framework is already a great simplification to the complete microscopic description, it takes us back to the difficult task of handling high-dimensional joint probability distribution discussed in \citesec~\ref{sec:chap2-stat-inf}. Let us discuss key conceptual steps further taken by physicists to understand large interacting systems.
To characterize the macroscopic state of the system, an important functional is the free energy 
\begin{align}
    F_N & = - \log \cZ_N / \beta = - \frac 1 \beta \log \sum_{\x \in \mathcal{X}^N} e^{-\beta E(\x)}. 
    % \\
    % & = -\frac 1 \beta \left( \sum_{\x \in \mathcal{X}^N} -\beta E(\x)e^{-\beta E(\x)} /\cZ_N + \sum_{\x \in \mathcal{X}^N}  \left(\beta E(\x) + \log \cZ_N \right)e^{-\beta E(\x)} /\cZ_N \right) \\
    % & = \left\langle E(\x)\right\rangle + \left\langle \log p(\x) \right\rangle /\beta = U_N - H_N /\beta
\end{align} 
% where $\langle \cdot \rangle$ stands for the average over the Boltzmann distribution, $U_N$ is the average energy and $H_N$ is the entropy. 
% Note that the entropy of statistical physics and the entropy of information theory \cite{Shannon1948} have the same definition relatively to a distribution. 
While the number of available configurations $\cX^N$ grows exponentially with $N$, considering the \emph{thermodynamic} limit $N \to \infty$ typically simplifies computations due to concentrations. 
% Let $e_N = E /N$ be the energy per degree of freedom, $\Omega(E) = \sum_{\x \in \mathcal{X}^N} \delta\left(E - E(\x)\right) = e^{N h_N}$ be the number of configurations at energy $E = N e_N$, and $h_N(e_N)$ the entropy per degree of freedom. 
Let $e_N = E /N$ be the energy per degree of freedom, the partition function can be re-written as a sum over the configurations of a given energy $e_N$
\begin{align}
    \cZ_N = 
    % \sum_E \sum_{\x \in \mathcal{X}^N} \delta\left(E - E(\x)\right) e^{-N \beta e_N}  = \sum_{e_N} e^{-N (\beta e_N - h_N(e_N))} = 
    \sum_{e_N} e^{-N \beta f_N(e_N)} ,
\end{align}
where we define $f_N(e_N)$ the free energy density of states of energy $e_N$. This rewriting implies that 
at large $N$ the states of energy minimizing the free energy are exponentially more likely than any other states. Provided the following limits exist, the statistics of the system are dominated by the former states and we have the thermodynamic quantities
%  Using the definitions of the limits
\begin{gather}
    \cZ = \lim_{N \to \infty} \cZ_{N} =  e^{-\beta f}, \text{ and } f = \lim_{N \to \infty} F_N / N .
\end{gather}
% \missing{comment about the fact that there is a limit is the interaction are sufficienlty decreasing, extensivity}
The interested reader will also find a more detailed yet friendly presentation of the thermodynamic limit in \citesec~2.4 of \cite{Mezard2009}.

\paragraph{Disordered systems}

Remarkably, the statistical physics framework can be applied to inhomogeneous systems with \emph{quenched disorder}. In these systems, interactions are functions of the realization of some random variables. An iconic example is the Sherrington-Kirkpatrick (SK) model \cite{Sherrington1975}, a fully connected Ising model with random Gaussian couplings $\mat{J} = (J_{i j})$, that is where the $J_{ij}$ are drawn independently from a Gaussian distribution. As a result, the energy functional of disordered systems is itself function of the random variables. For instance here, the energy of a spin configuration $\x$ is then $E(\x; \mat{J}) = - \frac 1 2 \x\T \mat{J} \x$. 
In principle, system properties depend on a given realization of the disorder. In our example, the correlation between two spins $\langle x_i x_j \rangle_J$ certainly does. Yet some aggregated properties are expected to be \emph{self-averaging} in the thermodynamic limit, meaning that they concentrate on their mean with respect to the disorder as the fluctuations are averaged out.
It is the case for the free energy. As a result, here it formally verifies:
\begin{gather}
    \lim_{N\to\infty} F_{N; \mat{J}} / N = \lim_{N\to\infty} \E_{\mat{J}}[F_{N; \mat{J}} / N] = f.
\end{gather}
(see e.g. \cite{Mezard1986, Castellani2005} for discussions of self-averaging in spin glasses).
% \E_W[(F_{N; W} / N)^2] - \E_W[(F_{N; W} / N)]^2 = O(1/N)
Thus the typical behavior of complex systems is studied in the statistical physics framework by taking two important conceptual steps: averaging over the realizations of the disorder and considering the thermodynamic limit. These are starting points to design approximate inference methods. Before turning to an introduction to mean-field approximations, we stress the originality of the statistical physics approach to inference. 

% \missing{quenched avarange?}

% \missing{phase transitions}

% 'The free energy is self-averaging, and in particular,
% 2 F2 −F = O
%?? 1
% N (8)
% If a quantity has, for example, a bimodal distribution, it is not self-averaging. Indeed its average is a very poor indicator of the physical values of the quantity itself. A simple argument to work out equation (8) can be given in finite dimension. We
% divide our system in a number K of macroscopic sub-systems, with 1 ≪ K ≪ N. The total (extensive) free energy will be the sum of the free energies of the sub-systems, plus a contribution coming from the interactions at the interfaces between the sub-systems. Once we compute the free energy density, this surface contribution can be neglected in the limit N → ∞. Moreover, the sub-systems free energies are independent random variables and therefore we can apply the central limit theorem to the sum, and get (8).'

\paragraph{Statistical physics of inference problems}
Statistical inference questions are mapped to statistical physics systems by interpreting general joint probability distributions as Boltzmann distributions \eqref{eq:chap2-boltzmann}. 
Turning back to our simple examples of \citesec~\ref{sec:chap2-stat-inf}, the RBM is trivially mapped as it directly borrows its definition from statistical physics. We have
    \begin{gather}
        E(\x, \hidd ; \W) = - \vect{a}\T \x - \vect{b}\T \hidd - \x\T\W\hidd .
    \end{gather} 
The inverse temperature parameter can either be considered equal to 1 or as a scaling factor of the weight matrix $\W \leftarrow \beta \W$ and bias vectors $\vect{a} \leftarrow \beta \vect{a}$ and $\vect{b} \leftarrow \beta \vect{b} $. The RBM parameters play the role of the disorder. Here the computational hardness in estimating the log-likelihood comes from the estimation of the log-partition function, which is precisely the free energy. In our second example, the estimation of the student perceptron weight vector,
the posterior distribution is mapped to a Boltzmann distribution by setting 
\begin{gather}
    % $
    E(\vect{w} ; \y, \X) = - \log p(\y| \vect{w}\T\X)p_w(\vect{w})
    % $
    . 
\end{gather}The disorder is here materialized by the training data.
%  and the teacher weight vector $\vect{w}_0$. 
The difficulty is here to compute $p(\y | \X)$ which is again the partition function in the Boltzmann distribution mapping.
% \missing{planted ensemble}
Relying on the thermodynamic limit, mean-field methods will provide asymptotic results. Nevertheless, experience shows that the behavior of large finite-size systems are often well explained by the infinite-size limits. 

Also, the application of mean-field inference requires assumptions about the distribution of the disorder which is averaged over. Practical algorithms for arbitrary cases can be derived with ad-hoc assumptions, but studying a precise toy statistical model can also bring interesting insights. The simplest model in most cases is to consider uncorrelated disorder: in the example of the perceptron this corresponds to random input data points with arbitrary random labels. Yet, the teacher-student scenario offers many advantages with little more difficulty. It allows to create data sets with structure (the underlying teacher rule).
% Interestingly, the teacher-student perspective 
It also allows to formalize an analysis of the difficulty of a learning problem and of the performance in the resolution. Intuitively, the definition of a ground-truth teacher rule with a fixed number of degrees of freedom sets the minimum information necessary to extract from the observations, or training data, in order to achieve perfect reconstruction. This is an \emph{information-theoretic limit}. 

Furthermore, the assumption of an underlying statistical model enables the measurement of performance of different learning algorithms over the class of corresponding problems from an average viewpoint. This is in contrast with
the traditional approach of computer science in studying the difficulty of a class of problem based on the \emph{worst} case. This conservative strategy yields strong guarantees of success, yet it may be overly pessimistic compared to the experience of practitioners. Considering a distribution over the possible problems (a.k.a different realizations of the disorder), the average performances are sometimes more informative of \emph{typical} instances rather than worst ones.
For deep learning, this approach may prove particularly interesting as the traditional bounds, based on the VC-dimension \cite{Vapnik2000} and Rademacher complexity \cite{Bartlett2002,Shalev-Shwartz2014}, appear extremely loose when compared to practical examples.

Finally, we must emphasize that derivations presented here are not mathematically rigorous. They are based on `correct' assumptions allowing to push further the understanding of the problems at hand, while a formal proof of the assumptions is possibly much harder to obtain.

% \newpage
\section[
    Selected overview of mean-field treatments: free energies and algorithms 
    % \\ {\small (Techniques)}
    ]{
Selected overview of mean-field treatments: \\ Free energies and algorithms 
% \\{\small (Techniques)} 
}
\label{sec:chap3}

% \missing{what did we do with the temperature?}

Mean-field methods are a set of techniques enabling to approximate marginalized quantities of joint probability distributions by exploiting knowledge on the dependencies between random variables. 
They are usually said to be analytical - as opposed to numerical Monte Carlo methods. In practice they usually replace a summation exponentially large in the size of the system by an analytical formula involving a set of parameters, themselves solution of a closed set of non-linear equations. Finding the values of these parameters typically requires only a polynomial number of operations.  

In this \citechap, we will give a selected overview of mean-field methods as they were introduced in the statistical physics and/or signal processing literature. A key take away of what follows is that closely related results can be obtained from different heuristics of derivation. 
We will start by deriving the simplest and historically first mean-field method. We will then introduce the important broad techniques that are high-temperature expansions, message-passing algorithms and the replica method. In the following \citechap~\ref{sec:chap3further} we will additionally cover the most recent extensions of mean-field methods presented in the present \citechap~\ref{sec:chap3} that are relevant to study learning problems.
% \vspace{-0.3cm}
% for instance I will not talk about the bethe approximation - nor directly of the cavity formalism as introduced by physicist

% \missing{cite Advanced mean-field methods - + maybe review about the technique not discussed?}

% Quality of the approximation will be typically good, but some instance may be quite bad. 
% p73 of Mézard Montanari -> weakly correlated random variables
% \missing{Comment on the fact that sometimes there will not be any temperatures involved}

\subsection{Naive mean-field}
\label{sec:chap3-nmf}
The naive mean-field method is the first and somehow simplest mean-field approximation. It was introduced by the physicists Curie  \cite{Curie1895} and Weiss \cite{Weiss1907} and then adopted by the different communities interested in inference \cite{Wainwright2008}.

\subsubsection{Variational derivation}
The naive mean-field method consists in approximating the joint probability distribution of interest by a fully factorized distribution. Therefore, it ignores correlations between random variables. Among multiple methods of derivation, we present here the variational method: it is the best known method across fields and it readily shows that, for any joint probability distribution interpreted as a Boltzmann distribution, the rather crude naive mean-field approximation yields an upper bound on the free energy. 
For the purpose of demonstration we consider a Boltzmann machine without hidden units (Ising model) with variables (spins) $\x = (x_1, \cdots ,x_N) \in \mathcal{X} = \{0,1\}^N $, and energy function
\begin{gather}
    \label{eq:chap3-ising-energy}
    E(\x) = - \sum_{i=1}^N b_i x_i -  \sum_{(ij)} W_{ij}x_i x_j = - \vect{b}\T \x - \frac{1}{2} \x\T \W \x \, , \quad \vect{b} \in \R^N \, , \quad \W \in \R^{N\times N} \, ,
\end{gather}
where the notation $(ij)$ stands for pairs of connected spin-variables, and the weight matrix $\W$ is symmetric. 
The choices for $\{0,1\}$ rather than $\{-1,+1\}$ for the variable values, the notations $\W$ for weights (instead of couplings), $\vect{b}$ for biases (instead of local fields), as well as the vector notation, are leaning towards the machine learning conventions. We denote by $q_{\m}$ a fully factorized distribution on $\{0,1\}^N$, which is a multivariate Bernoulli distribution parametrized by the mean values $\m = (m_1, \cdots, m_N) \in [0,1]^N$ of the marginals (denoted by $q_{m_i}$):
\begin{gather}
    q_{\m} (\x) = \prod_{i=1}^N q_{m_i}(x_i) = \prod_{i=1}^N m_i \dirac(x_i - 1) + (1-m_i) \dirac(x_i). 
\end{gather}
We look for the optimal $q_{\m}$ distribution to approximate the Boltzmann distribution $p(\x) =  e^{-\beta E(\x)}/\cZ$ by minimizing the KL-divergence 
\begin{align}
    \minn{\m} \KL(q_{\m} || p) & = \minn{\m} \sum_{\x \in \mathcal{X}} q_{\m}(\x) \log \frac{q_{\m}(\x)}{p(\x)} \\
    & = \minn{\m}  \sum_{\x \in \mathcal{X}} q_{\m}(\x) \log q_{\m}(\x) +  \beta \sum_{\x \in \mathcal{X}} q_{\m}(\x) E(\x)  + \log \cZ \\
    & = \minn{\m} \; \beta G(q_{\m}) - \beta F \geq 0, \label{eq:chap3-variational-inequality}
\end{align}
where the last inequality comes from the positivity of the KL-divergence. For a generic distribution $q$, $G(q)$ is the \emph{Gibbs free energy} for the energy $E(\x)$, 
\begin{gather}
    G(q) =  \sum_{\x \in \mathcal{X}} q(\x) E(\x) + \frac{1}{\beta}\sum_{\x \in \mathcal{X}} q(\x) \log q(\x) = U(q) - H(q)/\beta \geq F ,
\end{gather}
involving the average energy $U(q)$ and the entropy $H(q)$.
It is greater than the true free energy $F$
except when $q = p$, in which case they are equal. Note that this fact also means that the Boltzmann distribution minimizes the Gibbs free energy. Restricting to factorized $q_{\m}$ distributions, we obtain the naive mean-field approximations for the mean value of the variables (or \emph{magnetizations}) and the free energy:
\begin{gather}
    \m^* = \argminn{\m} G(q_{\m}) = \langle \x \rangle_{q_{\m^*}} \, , \\
    % q^*(\x) = \prod_{i=1}^N m^*_i \dirac(x_i - 1) + (1-m^*_i) \dirac(x_i) \\
    F_{\rm NMF} = G(q_{\m^*}) \geq F.
\end{gather} 
The choice of a very simple family of distributions $q_{\m}$ limits the quality of the approximation but allows for tractable computations of observables, for instance the two-spins correlations $\langle x_i x_j \rangle_{q^*} = m^*_i  m^*_j$ or variance of one spin $\langle x_i^2 \rangle_{q^*} - \langle x_i \rangle_{q^*}^2 = m^*_i - {m^*_i}^2$. 

In our example of the Boltzmann machine, it is easy to compute 
% the entropic contribution $H_{\rm NMF}(m)$ and energetic contribution $U_{\rm NMF}(m)$ of 
the Gibbs free energy for the factorized ansatz, we define functions of the magnetization vector:
\begin{align}
    U_{\rm NMF}(\m) & = \langle E(\x) \rangle_{q_{\m}} = - \vect{b}\T \m - \frac{1}{2} \m\T\W\m \, ,\\
    \label{eq:chap3-hnmf}
    H_{\rm NMF}(\m) & = - \langle \log q_{\m}(\x) \rangle_{q_{\m}} = - \sum_{i=1}^N m_i \log m_i + (1-m_i) \log (1-m_i) \, ,\\
    G_{\rm NMF}(\m) &= G(q_{\m})  = U_{\rm NMF}(\m) - H_{\rm NMF}(\m) / \beta.
\end{align}
Looking for stationary points we find a closed set of non linear equations for the $m^*_i$,
\begin{gather}
    \label{eq:chap3-nmf-eq}
    \left. \frac{\partial G_{\rm NMF}}{\partial m_i} \right|_{\m^*} = 0
    \quad \Rightarrow \quad m^*_i = \sigm(\beta b_i + \sum_{j \in \partial i} \beta W_{ij} m^*_j) \quad \forall i = 1 \cdots N\,,% \text{ where } \sigm(x) = \frac{1}{1  + e^{-x}} .
\end{gather}
where $\sigm(x) = (1  + e^{-x})^{-1}$.
The solutions can be computed by iterating these relations from a random initialization until a fixed point is reached.

To understand the implication of the restriction to factorized distributions,
it is instructive to compare this naive mean-field equation with the exact identity 
\begin{align}
    \label{eq:chap3-mf-identity}
    \langle x_i \rangle_p = \langle \sigm(\beta b_i + \sum_{j \in \partial i}\beta W_{ij} x_j) \rangle_p\,,
\end{align}
derived in a few lines in \citeapp~\ref{app:chap3-mf-identity}.
Under the Boltzmann distribution $p(\x) = e^{-\beta E(\x)}/\cZ$, these averages are difficult to compute. The naive mean-field method is neglecting the fluctuations of the effective field felt by the variable $x_i$:  $\sum_{j \in \partial i} W_{ij} x_j$, keeping only its mean $\sum_{j \in \partial i} W_{ij} m_j$. This incidentally justifies the name of mean-field methods.   

\subsubsection{When does naive mean-field hold true?}
The previous derivation shows that the naive mean-field approximation allows to bound the free energy. While this bound is expected to be rough in general, the approximation is reliable when the fluctuations of the local effective fields $\sum_{j \in \partial i} W_{ij} x_j$ are small. This may happen in particular in the thermodynamic limit $N\to \infty$ in \emph{infinite range} models, that is when weights or couplings are not only local but distributed in the entire system, or if each variable interacts directly with a non-vanishing fraction of the whole set of variables (e.g. \cite{opper2001advanced} \citechap~2). The influence on one given variable of the rest of the system can then be treated as an average background. Provided the couplings are weak enough, the naive mean-field method may even become asymptotically exact. This is the case of the \emph{Curie-Weiss} model, which is the fully connected version of the model \eqref{eq:chap3-ising-energy} with all $W_{ij} = 1/N$ (see e.g. \citesec~2.5.2 of \cite{Mezard2009}). The sum of weakly dependent variables then concentrates on its mean by the central limit theorem. 
We stress that it means that for finite dimensional models (more representative of a physical system, where for instance variables are assumed to be attached to the vertices of a lattice with nearest neighbors interactions), mean-field methods are expected to be quite poor. By contrast, infinite range models (interpreted as infinite-dimensional models by physicists) are thus traditionally called \emph{mean-field models}. 

In the next \citesec~we will recover the naive mean-field equations through a different method. The following derivation will also allow to compute corrections to the rather crude approximation we just discussed by taking into account some of the correlations it neglects.

\subsection{Thouless Anderson and Palmer equations}

\label{sec:chap3-tap}
The TAP mean-field equations \cite{Thouless1977, Morita1976} were originally derived as an exact mean-field theory for the Sherrington-Kirkpatrick (SK) model \cite{Sherrington1975}. The emblematic \emph{spin glass} SK model we already mentioned corresponds to a fully connected Ising model with energy \eqref{eq:chap3-ising-energy} and disordered couplings $W_{ij}$ drawn independently from a Gaussian distribution with zero mean and variance $W_0 / N$. The derivation of \cite{Thouless1977} followed from arguments specific to the SK model. Later, it was shown that the same approximation could be recovered from a second order Taylor expansion at high temperature by Plefka \cite{Plefka1982} and that it could be further corrected by the systematic computation of higher orders by Georges and Yedidia \cite{Georges1999}. We will briefly present this last derivation, having again in mind the example of the generic Boltzmann machine \eqref{eq:chap3-ising-energy}. 

\subsubsection{Outline of the derivation}

\label{sec:chap3-GY}
% \missing{definition of the Boltzmann bracket?}
Going back to the variational formulation \eqref{eq:chap3-variational-inequality}, we shall now perform a minimization in two steps. Consider first the family of distributions $q_{\m}$ enforcing $\langle \x \rangle_{q_{\m}} = \m$ for a fixed vector of magnetizations $\m$, but without any factorization constraint. The corresponding Gibbs free energy is
\begin{gather}
    G(q_{\m})  = U(q_{\m}) - H(q_{\m}) / \beta .% \\
    % U(q_{\m})  = \sum_{\x \in \mathcal{X}} q_{\m}(\x) E(\x), \quad
    % H(q_{\m})  = - \sum_{\x \in \mathcal{X}} q_{\m}(\x) \log q_{\m}(\x),
\end{gather}  
A first minimization at fixed $\m$ over the $q_{\m}$ defines another auxiliary free energy
\begin{align}
    G_{\rm TAP}(\m) = \minn{q_{\m}} G(q_{\m}).
\end{align} 
A second minimization over $\m$ would recover the overall unconstrained minimum of the variational problem \eqref{eq:chap3-variational-inequality} which is the exact free energy
\begin{align}
    F = -\log \cZ / \beta = \minn{\m} G_{\rm TAP}(\m).
\end{align}
Yet the actual value of $G_{\rm TAP}(\m)$ turns out as complicated to compute as $F$ itself. Fortunately, $\beta G_{\rm TAP}(\m)$ can be easily approximated by a Taylor expansion around $\beta = 0$ due to interactions vanishing at high temperature, as noticed by Plefka, Georges and Yedidia \cite{Plefka1982, Georges1999}. 
After expanding, the minimization over $G_{\rm TAP}(\m)$ yields a set of self consistent equations on the magnetizations $\m$, called the \emph{TAP equations}, reminiscent of the naive mean-field equations \eqref{eq:chap3-nmf-eq}. Here again, the consistency equations are typically solved by iterations. Plugging the solutions $\m^*$ back into the expanded expression yields the \emph{TAP free energy} $F_{\rm TAP}=G_{\rm TAP}(\m^*)$.
Note that ultimately the approximation lies in the truncation of the expansion. At first order the naive mean-field approximation is recovered. Historically, the expansion was first stopped at the second order. This choice was model dependent, it results from the fact that the mean-field theory is already exact at the second order for the SK model \cite{Morita1976, Thouless1977, Plefka1982}.

% \missing{We now unroll the different steps. - or do we leave it to the computation for the GRBM where we need to have two Lagrange mutlplipliers?
% reference to a derivation later in the text?} 
% LOOK AT DERIVATION CHAP 2 ADVANCED MEAN FIELD OPPER <3 <3 <3
% \paragraph{Minimization at fixed $\m$}
% The constrained minimization problem can be 

% \paragraph{Expansion}

% \paragraph{Minimization over $\m$}

\subsubsection{Illustration on binary Boltzmann machines and important remarks}
For the Boltzmann machine \eqref{eq:chap3-ising-energy}, the TAP equations and TAP free energy (truncated at second order) are \cite{Thouless1977},
\begin{gather}
    m^*_i = \sigm\left(\beta b_i + \sum_{j \in \partial i} \beta W_{ij} m^*_j - \beta^2 W_{ij}^2(m^*_j - \frac 1 2)(m^*_i - {m^*_i}^2 )\right) \; \forall i \label{eq:chap3-tap-eq}\\
    \beta G_{\rm TAP}(\m^*) = - H_{\rm NMF}(\m^*) - \beta \sum_{i=1}^N b_i m^*_i - \beta \sum_{(ij)} m_i^*W_{ij}m^*_j\\
    \qquad \qquad \qquad \qquad \qquad \qquad \qquad \qquad \qquad \qquad - \frac{\beta^2}{2} \sum_{(ij)} W_{ij}^2 (m^*_i - {m^*_i}^2)(m^*_j - {m^*_j}^2)\, , \notag
\end{gather}
where the naive mean-field entropy $H_{\rm NMF}$ was defined in \eqref{eq:chap3-hnmf}. For this model, albeit with $\{+1, -1\}$ variables instead of $\{0,1\}$, several references pedagogically present the details of the derivation sketched in the previous paragraph. The interested reader should check in particular \cite{opper2001advanced, Zamponi2010}. We also present a more general derivation in \citeapp~\ref{app:chap3-real-GY}, see \citesec~\ref{sec:chap3-GY-generalized}.

\subparagraph{Onsager reaction term}
Compared to the naive mean-field approximation the TAP equations include a correction to the effective field called the \emph{Onsager reaction term}. The idea is that, in the effective field at variable $i$, we should consider corrected magnetizations of neighboring spins $j \in \partial i$, that would correspond to the absence of variable $i$. 
% The correction term then measures the reaction of the system to the deletion of variable $i$. 
This intuition echoes at two other derivations of the TAP approximation: the cavity method \cite{Mezard1986} that will not be covered here and the message passing which will be discussed in the next \citesec.

% This correction is enough to make the mean-field description exact in the thermodynamic limit for the SK model for some region of the $W_0, \beta$ plane
% \missing{What does it tell us about the SK model?
As far as the SK model is concerned,
this second order correction is enough in the thermodynamic limit as the statistics of the weights imply that higher orders will typically be subleading. Yet in general, the correct TAP equations for a given model will depend on the statistics of interactions and there is no guarantee that there exists a finite order of truncation leading to an exact mean-field theory. 
In \citesec~\ref{sec:chap3-ortho-invariant} we will discuss models beyond SK where a conjectured exact TAP approximation can be derived. 

% \missing{ref Hopefield?}
% . Two models with the same connectivities but different dis- tributions for the couplings, like, e.g., the SK model and the Hopfield model  have different expressions for the On- sager corrections see, e.g.,  Chap. XIIIMANFRED PR E ADATPIVE TAP - end of first poage - 

\subparagraph{Single instance}
Although the selection of the correct TAP approximation relies on the statistics of the weights, the derivation of the expansion outlined above does not require to average over them, i.e. it does not require an average over the disorder. Consequently, the approximation method is well defined for a single instance of the random disordered model and the TAP free energy and magnetizations can be computed for a given (realization of the) set of weights $\{W_{ij}\}_{(ij)}$ as explained in the following paragraph. 
In other words, it means that the approximation can be used to design practical inference algorithms in finite-sized problems and not only for theoretical predictions on average over the disordered class of models. Crucially, these algorithms may provide approximations of disorder-dependent observables, such as correlations, and not only of self averaging quantities. 
% , as will be made clear in \citesec~\ref{sec:chap3-bp-to-amp} where the connection between TAP equations and message passing algorithms is discussed.  
% Note that this fact further allowed the designing of the adaptive TAP approximation \cite{Opper2001, Opper2001prl, Opper2005}, that we will discuss in \citesec~\ref{sec:chap3-ortho-invariant}, requiring no prior knowledge on the statistics of the weights. The Onsager correction is here computed directly for a given $\W$. 

\subparagraph{Finding solutions}
The self-consistent equations on the magnetizations \eqref{eq:chap3-tap-eq} are usually solved by turning them into an iteration scheme and looking for fixed points. This generic recipe leaves nonetheless room for interpretation: which exact form should be iterated? How should the updates for the different equations be scheduled? Which time indexing should be used? While the following scheme may seem natural
\begin{align}
    {m_i}^{(t+1)} \leftarrow \sigm\left(\beta b_i + \sum_{j \in \partial i} \beta W_{ij} {m_j}^{(t)} - W_{ij}^2\left({m_j}^{(t)} - \frac 1 2\right)\left({m_i}^{\mathbf{(t)}} - {{m_i}^{\mathbf{(t)}}}^2 \right)\right),
\end{align}
it typically has more convergence issues than the following alternative scheme including the time index $t-1$
\begin{align}
    {m_i}^{(t+1)} \leftarrow \sigm\left(\beta b_i + \sum_{j \in \partial i} \beta W_{ij} {m_j}^{(t)} - W_{ij}^2\left({m_j}^{(t)} - \frac 1 2\right)\left({m_i}^{\mathbf{(t-1)}} - {{m_i}^{\mathbf{(t-1)}}}^2 \right)\right).
\end{align}
This issue was discussed in particular in \cite{Kabashima2003,bolthausen2014iterative}.
Remarkably, this last scheme, or algorithm, is actually the one obtained by the approximate message passing derivation that will be discussed in the upcoming \citesec~\ref{sec:chap3-bp-to-amp}.

\subparagraph{Solutions of the TAP equations}
The TAP equations can admit multiple solutions with either equal or different TAP free energy. 
While the true free energy $F$ corresponds to the minimum of the Gibbs free energy, reached for the Boltzmann distribution, the TAP derivation consists in performing an effectively unconstrained minimization in two steps, but with an approximation through a Taylor expansion in between. The truncation of the expansion therefore breaks the correspondence between the discovered minimizer and the unique Boltzmann distribution, hence the possible multiplicity of solutions. For the SK model for instance, the number of solutions of the TAP equations increases rapidly as $\beta$ grows \cite{Mezard1986}. While the different solutions can be accessed using different initializations of the iterative scheme, it is notably hard in phases where they are numerous to find exhaustively all the TAP solutions. In theory, they should be weighted according to their free energy density and averaged to recover the thermodynamics predicted by the replica computation \cite{Dominicis1983}, another mean-field approximation discussed in \citesec~\ref{sec:chap3-replica}.

% . First, one considers the ensemble of distributions with fixed moments $\am$ and $\cm$ and finds the corresponding minimizer $G(\am, \cm)$, which is in turn minimized with respect to $\am$ and $\cm$ in a second step. Yet before the second step, $G(\am, \cm)$ is approximated by a Taylor expansion. The truncation of the expansion therefore breaks the correspondence between the discovered minimum and the true free energy. In particular, the TAP equations possess multiple solutions whose number increases rapidly as $\beta$ grows  \cite{Mezard1986}.}

\subsubsection{Generalizing the Georges-Yedidia expansion}
\label{sec:chap3-GY-generalized}
In the derivation outlined above for binary variables, $x_i = 0$ or $1$, the mean of each variable $m_i$ was fixed. This is enough to parametrize the corresponding marginal distribution $q_{m_i}(x_i)$. Yet the expansion can actually be generalized to Potts variables (taking multiple discrete values) or even real valued variables by introducing appropriate parameters for the marginals. A general derivation fixing arbitrary real valued marginal distribution was proposed in \citeapp~B of \cite{Lesieur2017} for the problem of low rank matrix factorization. Alternatively, another level of approximation can be introduced for real valued variables by restricting the set of marginal distributions tested to a parametrized family of distributions. By choosing a Gaussian parametrization, one recovers TAP equations equivalent to the approximate message passing algorithm that will be discussed in the next \citesec. In \citeapp~\ref{app:chap3-real-GY}, we present a derivation for real-valued Boltzmann machines with a Gaussian parametrization as proposed in \cite{Tramel2018}.

\subsection{Belief propagation and approximate message passing}
\label{sec:chap3-bp-to-amp}
Another route to rediscover the TAP equations is through the approximation of message passing algorithms. Variations of the latter were discovered multiple times in different fields. In physics they were written in a restricted version  as soon as 1935 by Bethe \cite{Bethe1935}. In statistics, they were developed by Pearl as methods for probabilistic inference \cite{Pearl1988}. 
In this section we will start by introducing a case-study of interest, the Generalized Linear Model. We will then proceed by steps to outline the derivation of the Approximate Message Passing (AMP) algorithm from the Belief Propagation (BP) equations.

\subsubsection{Generalized linear model}
\label{sec:chap3-glm}
\begin{figure}
    \centering
    \includegraphics[width=\textwidth]{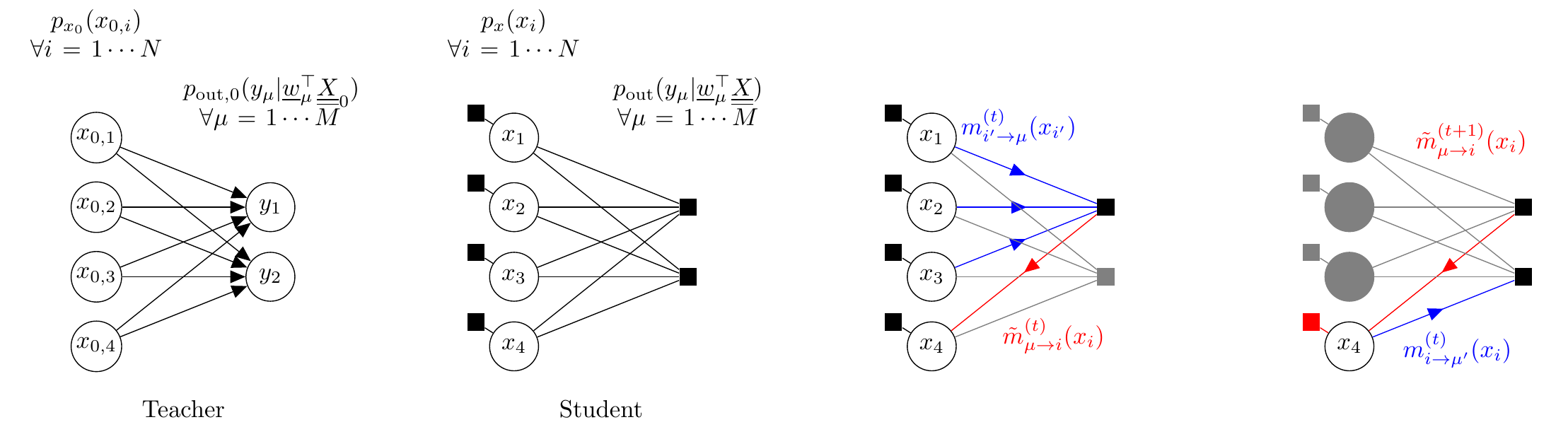}
    \caption{Graphical representations of the Generalized Linear Model. \textbf{Left:} Probabilistic graphical model of the teacher. \textbf{Middle left:} Factor graph representation of the posterior distribution on the signal $\x$ under the student statistical model. \textbf{Middle right and right:} Belief propagation updates \eqref{eq:chap3-bp-glm-1} - \eqref{eq:chap3-bp-glm-2} for approximate inference.   \label{fig:chap3-glm}}
\end{figure}

\subparagraph{Definition} We introduce the \emph{Generalized Linear Model} (GLM) which is a fairly simple model to illustrate message passing algorithms and which is also an elementary brick for a large range of interesting inference questions on neural networks. It falls under the teacher-student set up: a student model is used to reconstruct a signal from a teacher model producing indirect observations. 
In the GLM, the product of an unknown signal $\x_0 \in \R^N$ and a known weight matrix $\W \in \R^{N \times M}$ is observed as $\y$ through a noisy channel $\pouto$,
\begin{gather}
    \left\{
    \begin{array}{l}
    \W \sim p_W(\W)
    \\
    % \quad 
    \x_0 \sim p_{x_0}(\x_0) = \prod\limits_{i=1}^N p_{x_0}(x_{0,i})
    \end{array}
    \right. 
     \quad 
    % \\
    \Rightarrow
     \y \sim \pouto(\y | \W\x_0) = \prod_{\mu=1}^M \pouto(y_\mu | \vect{w}_\mu\T\x_0). 
\end{gather}
The probabilistic graphical model corresponding to this teacher is represented in \citefig~\ref{fig:chap3-glm}. The prior over the signal $p_{x_0}$ is supposed to be factorized, and the channel $\pouto$ likewise. The inference problem is to produce an estimator $\xh$ for the unknown signal $\x_0$ from the observations $\y$. Given the prior $p_x$ and the channel $\pout$ of the student, not necessarily matching the teacher, the posterior distribution is 
\begin{align}
    % p(\x|\y, \W) = \frac{1}{\cZ(\y, \W)} \pout(\y | \x, \W) p_x(\x) \, \\
    p(\x|\y, \W) &= \frac{1}{\cZ(\y, \W)} \, \prod_{\mu=1}^M \pout(y_\mu | \sum_{i=1}^N W_{\mu i} x_i)\, \prod_{i=1}^N p_x(x_i) \, ,  \label{eq:chap3-glm-meas}\\
    % = \frac{1}{\cZ(\y, \W)} e^{\log \pout(\y | \x, \W) p_x(\x) } \\
    \cZ(\y, \W) &= \int \dd{\x} \pout(\y | \x, \W) p_x(\x), \label{eq:chap3-glm-Z}
\end{align}
represented as a factor graph also in \citefig~\ref{fig:chap3-glm}. The difficulty of the reconstruction task  of $\x_0$ from $\y$ is controlled by the measurement ratio $\alpha = M/N$ and the amplitude of the noise possibly present in the channel. 

% \missing{notations with the g function}

\subparagraph{Applications}
The generic GLM underlies a number of applications. 
In the context of neural networks of particular interest in this technical review, the channel $\pout$ generating observations $\y \in \R^M$ can equivalently be seen as a stochastic activation function $g(\cdot; \eps)$ incorporating a noise $\eps \in \R^M$ component-wise to the output,
\begin{gather}
    y_\mu = g(\vect{w}_\mu\T\x \,; \, \eps_\mu).
\end{gather}
The inference of the teacher signal in a GLM has then two possible interpretations. 
On the one hand, it can be interpreted as the reconstruction of the input $\x$ of a stochastic single-layer neural network from its output $\y$. For example, this inference problem can arise in the maximum likelihood training of a one-layer VAE (see corresponding paragraph in \citesec~\ref{sec:chap1-vae}).  On the other hand, the same question can also correspond to the Bayesian learning of a single-layer neural network with a single output - the perceptron - where this time $\{\W, \y\}$ are interpreted as the collection of training input-output pairs and $\x_0$ plays the role of the unknown weight vector of the teacher (as cited as an example in \citesec~\ref{sec:chap2-teacher-student}). 
However, note that one of the most important applications of the GLM, Compressed Sensing (CS) \cite{Donoho2006}, does not involve neural networks.

% It consist in recovering a $\rho$-sparse signal $\x_0$, i.e. with $\rho N$ $(< N)$ non zero entries, from  $M$ noisy linear measurements gathered in $\y$. At zero noise, the information theoretic limit for the reconstruction is $M > \rho N$, or equivalently $\alpha > \rho$. Yet at this threshold the reconstruction requires the knowledge of the position of the non-zeros entries, or an exponential number of operations to test all possibilities. Hence the problem is non-trivial as soon as $M<N$. Studying the performance of algorithms in between the impossible and trivial regime is of primal practical and theoretical interest.

\subparagraph{Statistical physics treatment, random weights and scaling}
% For the 
% \missing{$\W$ of order $1/\sqrt{N}$}
% \missing{thermodynamic limit with M and N at fixed ratio}
% \missing{clearer definition of the equivalent g function? Could also be the moment to define the mismatched priors - }
From the statistical physics perspective, the effective energy functional is read from the posterior \eqref{eq:chap3-glm-Z} seen as a Boltzmann distribution with energy
\begin{gather}
    E(\x) = - \log \pout(\y | \x, \W) p_x(\x) = - \sum_{\mu =1}^M\log \pout(y_\mu | \sum_{i=1}^N W_{\mu i} x_i) - \sum_{i=1}^
    N \log p_x(x_i) .
\end{gather} 
The inverse temperature $\beta$ has here no formal equivalent and can be thought as being equal to 1. The energy is a function of the random realizations of $\W$ and $\y$, playing the role of the disorder. Furthermore, the validity of the approximation presented below require additional assumptions.  Crucially, the weight matrix is assumed to have i.i.d. Gaussian entries with zero mean and variance $1/N$, much like in the SK model. The prior of the signal is chosen so as to ensure that the $x_i$-s (and consequently the $y_\mu$-s) remain of order 1.
Finally, the thermodynamic limit $N \to \infty$ is taken for a fixed measurement ratio $\alpha=M/N$.

% \begin{figure}[t]
%     \centering
%     \subfloat[]{\includegraphics[width=0.22\textwidth]{chap3_glm_teacher.pdf}
%     \label{fig:chap3-glm-teacher}}
%     % \hspace{1cm}
%     \subfloat[]{\includegraphics[width=0.22\textwidth]{chap3_glm_student.pdf}
%     \label{fig:chap3-glm-student}}
%     \subfloat[]{\includegraphics[width=0.44\textwidth]{chap3_glm_bp.pdf}
%     \label{fig:chap3-glm-bp}}
%     \caption{}
% \end{figure}

\subsubsection{Belief Propagation} 
\label{sec:chap3-bp}
Recall that inference in high-dimensional problems consists in marginalizations over complex joint distributions, typically in the view of computing partition functions, averages or marginal probabilities for sampling. Belief Propagation (BP) is an inference algorithm, sometimes exact and sometimes approximate as we will see, leveraging the known factorization of a distribution, which encodes the precious information of (in)depencies between the random variables in the joint distribution. For a generic joint probability distribution $p$ over $\x \in \R^N$ factorized as
\begin{gather}
    \label{eq:chap3-factorization}
    p(\x) = \frac 1 \cZ \prod_{\mu = 1}^M \psi_\mu(\x_{\partial \mu}),
\end{gather}
$\psi_\mu$ are called potential functions taking as arguments the variables $x_i$-s involved in the factor $\mu$ 
% in the factor graph 
shortened as $\x_{\partial \mu}$.

% The Belief Propagation algorithm (BP) is an exact inference algorithm for joint probability distributions with acyclic (or tree) underlying factor graph. It allows to compute marginals of the random variables using a set of auxiliary functions called \emph{messages}, attached to the edges of the factor graph. Starting at a leaf of the tree, the messages communicate beliefs of a given node variable taking a given value based on the nodes and factors already visited along the tree.

\subparagraph{Definition of messages}
Let us first write the BP equations and then explain the origin of these definitions. 
The underlying factor graph of \eqref{eq:chap3-factorization}
has $N$ nodes carrying variables $x_i$-s and $M$ factors associated with the potential functions $\psi_\mu$-s (see \citeapp~\ref{app:chap2-graphs} for a quick reminder). BP acts on \emph{messages} variables which are tied to the edges of the factor graph.
% taking as arguments the neighboring variables $x_i$-s of factor $\mu$ 
% in the factor graph 
% shortened as $\x_{\partial \mu}$, 
Specifically, the sum-product version of the algorithm (as opposed to the max-sum, see e.g. \cite{Mezard2009}) consists in the update equations
\begin{align}
    \label{eq:chap3-bp1}
     \msg{\tilde{m}^{(t)}}{\mu}{i}(x_i) & =  \frac{1}{\msg{\cZ}{\mu}{i}} \int \prod_{i'\in \partial \mu \setminus i} \dd{x_{i'}} \psi_\mu(\x_{\partial \mu})  \prod_{i'\in \partial \mu \setminus i} \msg{m^{(t)}}{i'}{\mu}(x_{i'}), \\
    \label{eq:chap3-bp2}
    \msg{m^{(t+1)} }{i}{\mu}(x_i)  & = \frac{1}{\msg{\cZ}{i}{\mu}} p_x(x_i)\prod_{\mu'\in \partial i \setminus  \mu} \msg{\tilde{m}^{(t)}}{\mu'}{i}(x_i) 
\end{align}
where again the $i$-s index the variable nodes and the $\mu$-s index the factor nodes. 
% The $\psi_\mu$ are the potential functions taking as arguments the neighboring variables $x_i$-s of factor $\mu$ 
% % in the factor graph 
% shortened as $\x_{\partial \mu}$. 
The notation $\partial \mu \setminus i$ designate the set of neighbor variables of the factor $\mu$ except the variable $i$ (and reciprocally for $\partial i \setminus \mu$).
The partition functions $\msg{\cZ}{i}{\mu}$ and $\msg{\cZ}{\mu}{i}$ are normalization factors ensuring that the messages can be interpreted as probabilities.

\begin{figure}[t]
    \centering
    \includegraphics[width=0.75\textwidth]{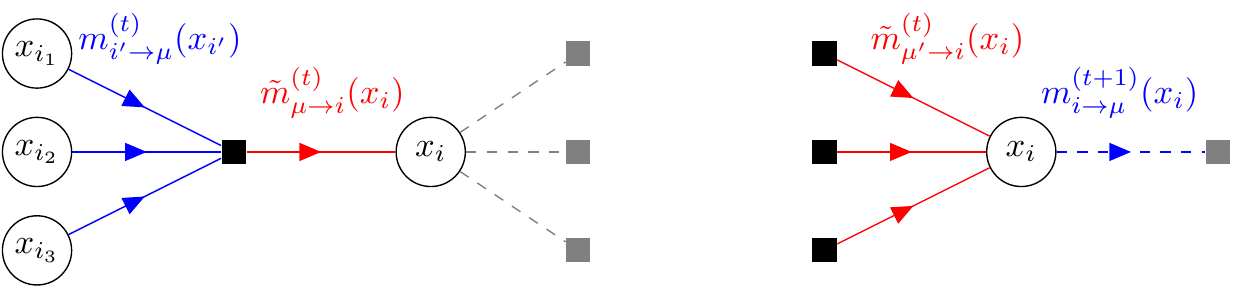}
    \caption{Representations of the neighborhood of edge $i$-$\mu$ in the factor graph and corresponding local BP updates. Factors are represented as squares and variable nodes as circles. \textbf{Left:} In the factor graph where all factors around $x_i$ are removed (in gray) except for the factor $\mu$, the marginal of $x_i$ (in red) is updated from the messages incoming at factor $\mu$ (in blue) \eqref{eq:chap3-bp1}. \textbf{Right:} In the factor graph where factor $\mu$ is deleted (in gray), the marginal of $x_i$ (in blue) is updated with the incoming messages (in red) from the rest of the factors \eqref{eq:chap3-bp2}. \label{fig:chap3-bp}}
\end{figure}

For acyclic (or tree-like) factor graphs, the BP updates are guaranteed to converge to a fixed point, that is a set of time independent messages $\{\msg{m}{i}{\mu} , \msg{\tilde{m}}{\mu}{i}\}$ solution of the system of equations \eqref{eq:chap3-bp1}-\eqref{eq:chap3-bp2}. Starting at a leaf of the tree, these messages communicate beliefs of a given node variable taking a given value based on the nodes and factors already visited along the tree. More precisely, $\msg{\tilde{m}}{\mu}{i}(x_i)$ is the marginal probability of $x_i$ in the factor graph before visiting the factors in $\partial i$ except for $\mu$, and $\msg{m}{i}{\mu}(x_i)$ is equal the marginal probability of $x_i$ in the factor graph before visiting the factor $\mu$, see \citefig~\ref{fig:chap3-bp}. 

Thus, at convergence of the iterations, the marginals can be computed as
\begin{gather}
    \label{eq:chap3-bp-marginal}
    m_i(x_i) = \frac{1}{\cZ_i} p_x(x_i) \prod_{\mu \in \partial i} \msgt{m}{\mu}{i}(x_i),
\end{gather}
which can be seen as the main output of the BP algorithm. 
These marginals will only be exact on trees where incoming messages, computed from different part of the graph, are independent. Nonetheless, the algorithm \eqref{eq:chap3-bp1}-\eqref{eq:chap3-bp2}, occasionally then called \emph{loopy-BP}, can sometimes be converged on graphs with cycles and in some cases will still provide high quality approximations. For instance, graphs with no short loops are locally tree like and BP is an efficient method of approximate inference, provided correlations decay with distance (i.e. incoming messages at each node are still effectively independent). 
BP will also appear principled for some infinite range mean-field models previously discussed; an example of which being our case-study the GLM discussed below. 
While this is the only example that will be discussed here in the interest of conciseness, getting fluent in BP generally requires more than one application. The interested reader could also consult \cite{Yedidia2002} and \cite{Mezard2009} \citesec~14.1. for simple concrete examples.

\paragraph{The Bethe free energy}
The BP algorithm can also be recovered from a variational argument. Let's consider both the single variable marginals $m_i(x_i)$ and the marginals of the neighborhood of each factor $\tilde{m}_\mu(\x_{\partial \mu}$). On tree graphs, the joint distribution \eqref{eq:chap3-factorization} can be re-expressed as
\begin{gather}
    p(\x) = \frac{\prod_{\mu=1}^M \tilde{m}_\mu(\x_{\partial \mu})}{\prod_{i=1}^Nm_i(x_i)^{n_i-1}},
\end{gather}
where $n_i$ is the number of neighbor factors of the $i$-th variable. Abusively, we can use this form as an ansatz for loopy graph and plug it in the Gibbs free energy to derive an approximation of the free energy, similarly to the naive mean-field derivation of \citesec~\ref{sec:chap3-nmf}. This time the variational parameters will be the distributions $m_i$ and $\tilde{m}_\mu$ (see e.g. \cite{Yedidia2002, Mezard2009} for additional details). The corresponding functional form of the Gibbs free energy is called the Bethe free energy:
\begin{gather}
    F_{\rm Bethe}(m_i, \tilde{m}_\mu) =  - \int \dd{\x_{\partial \mu}} \tilde{m}_\mu(\x_{\partial \mu}) \ln \psi_\mu(\x_{\partial \mu}) + (n_i -1) H(m_i) - H(\tilde{m}_\mu), 
\end{gather}
where $H(q)$ is the entropy of the distribution $q$. Optimization of the Bethe free energy with respect to its arguments under the constraint of consistency
\begin{gather}
    \int \dd{\x_{\partial \mu \setminus i}} \tilde{m}_\mu(\x_{\partial \mu }) = m_i(x_i)
\end{gather}
involves Lagrange multipliers which can be shown to be related to the messages defined in \eqref{eq:chap3-bp1}-\eqref{eq:chap3-bp2}. Eventually, one can verify that marginals defined as \eqref{eq:chap3-bp-marginal} and 
\begin{gather}
    \tilde{m}_\mu(\x_{\partial_\mu}) = \frac{1}{\cZ_\mu} \psi(\x_{\partial \mu}) \prod_{i \in \partial \mu}\msg{m}{i}{\mu}(x_i),
\end{gather}
are stationary point of the Bethe free energy for messages that are BP solutions. In other words, the BP fixed points are consistent with the stationary point of the Bethe free energy. Using the normalizing constants of the messages, the Bethe free energy can also be re-written as
% and the Bethe approximation of the free energy is given by
\begin{gather}
    \label{eq:chap3-bethe-fe}
    F_{\rm Bethe} = - \sum_{i \in V} \log \cZ_i  - \sum_{\mu \in F} \log \cZ_\mu + \sum_{(i \mu) \in E} \log \cZ_{\mu i} \, , 
\end{gather}
with
\begin{gather}
    \cZ_i = \int \dd{x_i} p_x(x_i) \prod_{\mu \in \partial i} \msgt{m}{\mu}{i}(x_i) \, ,\\
    \cZ_\mu = \int \prod_{i \in \partial \mu} \dd{x_i} \psi(\x_{\partial \mu}) \prod_{i \in \partial \mu}\msg{m}{i}{\mu}(x_i) \, ,\\
    \cZ_{\mu i } = \int \dd{x_i} \msgt{m}{\mu}{i}(x_i)  \msg{m}{i}{\mu}(x_i)\, .
 \end{gather}

%  \review{The Belief Propagation algorithm (BP) is an exact inference algorithm for joint probability distributions with acyclic (or tree) underlying factor graph. It allows to compute marginals of the random variables using a set of auxiliary functions called \emph{messages}, attached to the edges of the factor graph. Starting at a leaf of the tree, the messages communicate beliefs of a given node variable taking a given value based on the nodes and factors already visited along the tree.}

As for the marginals, the Bethe free energy, will only be exact if the underlying factor graph is a tree. Otherwise it is an approximation of the free energy, that is not generally an upper bound.

% \missing{with diosorered couplings to answer some decoherence between messages -> see florent and lenka introductory paragraph on compressed sensing}

\subparagraph{Belief propagation for the GLM} The writing of the BP-equations for our case-study is schematized on the right of \citefig~\ref{fig:chap3-glm}. There are $2 \times N \times M$ updates:
\begin{align}
    \label{eq:chap3-bp-glm-1}
     \msg{\tilde{m}^{(t)}}{\mu}{i}(x_i) 
     &  = \frac{1}{\msg{\cZ}{\mu}{i}} \int \prod_{i'\neq i} \dd{x_{i'}} \pout(y_\mu | {w_\mu}\T\x)  \prod_{i'\neq i} \msg{m^{(t)}}{i'}{\mu}(x_{i'}),\\
     \label{eq:chap3-bp-glm-2}
     \msg{m^{(t+1)}}{i}{\mu}(x_i)  
    & = \frac{1}{\msg{\cZ}{i}{\mu}} p_x(x_i)\prod_{\mu'\neq \mu} \msg{\tilde{m}^{(t)}}{\mu'}{i}(x_i),
\end{align}
for all $i-\mu$ pairs. 
Despite a relatively concise formulation, running BP in practice turns out intractable since for a signal $\x$ taking continuous values it would entail keeping track of distributions on continuous variables. In this case, BP is approximated by the (G)AMP algorithm presented in the next section.

% \missing{messages = fixed points of Bethe?}

% Bethe free energy = Gibbs like free energy with constraint on the correlations -> factorization of the distribution with pairwise potential is then exact for trees. Bethe free energy is not generally an upper bound.

% \subparagraph{Max-sum algorithm}
% \missing{mention or descitption to the max-sum algorithm}

\subsubsection{(Generalized) approximate message passing}
\label{sec:chap3-gamp}
The name of approximate message passing (AMP) was fixed by Donoho, Maleki and Montanari \cite{Donoho2009} who derived the algorithm in the context of Compressed Sensing. Several works from statistical physics had nevertheless already proposed related algorithmic procedures and made connections with the TAP equations for different systems \cite{Kabashima1998, opper2001advanced, Kabashima2003}. The algorithm was derived systematically for any channel of the GLM by Rangan \cite{Rangan2011} and became Generalized-AMP (GAMP), yet again it seems that \cite{Kabashima2004} proposed the first generalized derivation.

The systematic procedure to write AMP for a given joint probability distribution consists in first writing BP on the factor graph, second project the messages on a parametrized family of functions to obtain the corresponding \emph{relaxed-BP} and third close the equations on a reduced set of parameters by keeping only leading terms in the thermodynamic limit. We will quickly review and justify these steps for the GLM. 
Note that here a relevant algorithm for approximate inference will be derived from message passing on a fully connected graph of interactions. As it tuns out, the high connectivity limit and the introduction of short loops does not break the assumption of independence of incoming messages in this specific case thanks to the small scale $O(1/\sqrt{N})$ and the independence of the weight entries. The statistics of the weights are here crucial.

% \missing{make the connection of why and when would be nice}
\paragraph{Relaxed Belief Propagation}
In the thermodynamic limit $M, N \to + \infty$, one can show that the scaling $1/\sqrt{N}$ of the $W_{ij}$ and the extensive connectivity of the underlying factor graph imply that messages are approximately Gaussian.
Without giving all the details of the computation which can be cumbersome, let us try to provide some intuitions. We drop the time indices for simplicity and start with \eqref{eq:chap3-bp-glm-1}. Consider the intermediate reconstruction variable $z_\mu = \vect{w}_\mu\T\x = \sum_{i'\neq i}W_{\mu i'}x_{i'} + W_{\mu i}x_i$. Under the statistics of the messages $\msg{m}{i'}{\mu}(x_{i'})$, the $x_{i'}$ are independent such that by the central limit theorem  $z_\mu - W_{\mu i}x_i$ is a Gaussian random variable with respectively mean and variance
\begin{gather}
    \label{eq:chap3-rbp-om}
    \msg{\omega}{\mu}{i} = \sum_{i'\neq i}W_{\mu i'}\msg{\hat{x}}{i'}{\mu},\\
    \label{eq:chap3-rbp-V}
    \msg{V}{\mu}{i} = \sum_{i'\neq i}W^2_{\mu i'}\msg{C^x}{i'}{\mu},
\end{gather}
where we defined the mean and the variance of the messages $\msg{m}{i'}{\mu}(x_{i'})$,
\begin{gather}
    \label{eq:chap3-rbp-xhat}
    \msg{\hat{x}}{i'}{\mu} = \int \dd{x_{i'}} x_{i'} \, \msg{m}{i'}{\mu}(x_{i'}), \\
    \label{eq:chap3-rbp-cx} 
    \msg{C^x}{i'}{\mu} = \int \dd{x_{i'}} x^2_{i'} \, \msg{m}{i'}{\mu}(x_{i'}) - \msg{\hat{x}}{i'}{\mu}^2.
\end{gather}
Using these new definitions, \eqref{eq:chap3-bp-glm-1} can be rewritten as 
\begin{gather}
    \label{eq:chap3-rbp-01}
    \msg{\tilde{m}}{\mu}{i}(x_i) \propto \int \dd{z_\mu} \pout(y_\mu | z_\mu)e^{-\frac{(z_\mu - W_{\mu i} x_i - \msg{\omega}{\mu}{i})^2}{2 \msg{V}{\mu}{i}}} ,
\end{gather}
where the notation $\propto$ omits the normalization factor for distributions. Considering that $W_{\mu i}$ is of order $1/\sqrt{N}$, the development of \eqref{eq:chap3-rbp-01} shows that at leading order $\msg{\tilde{m}}{\mu}{i}(x_i)$ is Gaussian:
\begin{gather}
    \label{eq:chap3-rbp-02}
    \msg{\tilde{m}}{\mu}{i}(x_i) \propto e^{\msg{B}{\mu}{i}x_i + \frac 1 2 \msg{A}{\mu}{i}x_i^2 }
\end{gather}
where the details of the computations yield
\begin{gather}
    \label{eq:chap3-rbp-B}
    \msg{B}{\mu}{i} = W_{\mu i} \, \gouts(y_\mu, \msg{\omega}{\mu}{i}, \msg{V}{\mu}{i}) \\
    \label{eq:chap3-rbp-A}
    \msg{A}{\mu}{i} = - W_{\mu i}^2 \, \dgouts(y_\mu, \msg{\omega}{\mu}{i}, \msg{V}{\mu}{i})
\end{gather}
using the \emph{output update functions}
\begin{gather}
\label{eq:chap3-gout}
\gouts(y, \omega, V) = \frac{1}{\Zout}\int \dd{z} \frac{(z - \omega)}{V} \pout(y|z)  \cN(z; \omega, V), \\
\label{eq:chap3-dgout}
\dgouts(y, \omega, V) = \frac{1}{\Zout}\int \dd{z} \frac{(z - \omega)^2}{V^2} \pout(y|z)  \cN(z; \omega, V) - \frac 1 V - \gouts(y, \omega, V)^2,\\ 
\label{eq:chap3-Zout}
\Zout( y, \omega, V) = \int \dd{z}  \pout(y|z)  \cN(z; \omega, V).
\end{gather}
These arguably complicated functions, again coming out of the development of \eqref{eq:chap3-rbp-01}, can be interpreted as the estimation of the mean and the variance of the gap between two different estimate of $z_\mu$ considered by the algorithm: the mean estimate $\msg{\omega}{\mu}{i}$ given incoming messages $\msg{m}{i'}{\mu}(x_{i'})$ and the same mean estimate updated to incorporate the information coming from the channel $\pout$ and observation $y_\mu$. 
Finally, the Gaussian parametrization \eqref{eq:chap3-rbp-02} of $\msg{\tilde{m}}{\mu}{i}(x_i)$ serves to rewrite the other type of messages $\msg{m}{i}{\mu}(x_i)$ \eqref{eq:chap3-bp-glm-2},
\begin{gather}
    \label{eq:chap3-rbp-2}
    \msg{m}{i}{\mu}(x_i) \propto p_x(x_i)e^{-\frac{(\msg{\lambda}{i}{\mu}- x_i)^2}{2 \msg{\sigma}{i}{\mu}}}, 
\end{gather}
with
\begin{gather}
    \label{eq:chap3-rbp-sig}
    \msg{\sigma}{i}{\mu} = \left(\sum_{\mu' \neq \mu}\msg{A}{\mu'}{i} \right)^{-1} \\ 
    \label{eq:chap3-rbp-lbd}
    \msg{\lambda}{i}{\mu} = \msg{\sigma}{i}{\mu} \left(\sum_{\mu' \neq \mu} \msg{B}{\mu'}{i}\right).
\end{gather}
The set of equations can finally be closed by recalling the definitions \eqref{eq:chap3-rbp-xhat}-\eqref{eq:chap3-rbp-cx}:
\begin{gather}
    \label{eq:chap3-rbp-xhat-f1}
    \msg{\hat{x}}{i}{\mu} = f^x_1(\msg{\lambda}{i}{\mu}, \msg{\sigma}{i}{\mu})\\
    \label{eq:chap3-rbp-xhat-f2}
    \msg{C^x}{i}{\mu} = f^x_2(\msg{\lambda}{i}{\mu}, \msg{\sigma}{i}{\mu})
\end{gather}
with now the \emph{input update functions}
\begin{gather}
    \label{eq:chap3-Zx}
    \cZ^x = \int \dd{x} p_x(x)e^{-\frac{(x-\lambda)^2}{2\sigma}}, \\
    \label{eq:chap3-f1x}
    f^x_1(\lambda, \sigma) = \frac{1}{\cZ^x}\int \dd{x} x \, p_x(x)e^{-\frac{(x-\lambda)^2}{2\sigma}}, \\
    \label{eq:chap3-f2x}
    f^x_2(\lambda, \sigma) =  \frac{1}{\cZ^x} \int \dd{x} x^2 \, p_x(x)e^{-\frac{(x-\lambda)^2}{2\sigma}} - f^x_1(\lambda, \sigma)^2.
\end{gather}
The input update functions can be interpreted as updating the estimation of the mean and variance of the signal $x_i$ based on the information coming from the incoming messages grasped by $\msg{\lambda}{i}{\mu}$ and $\msg{\sigma}{i}{\mu}$ with the information of the prior $p_x$.

To sum-up, by considering the leading order terms in the thermodynamic limit, the BP equations can be self-consistently re-written as a closed set of equations over mean and variance variables \eqref{eq:chap3-rbp-om}-\eqref{eq:chap3-rbp-V}-\eqref{eq:chap3-rbp-B}-\eqref{eq:chap3-rbp-A}-\eqref{eq:chap3-rbp-sig}-\eqref{eq:chap3-rbp-lbd}-\eqref{eq:chap3-rbp-xhat-f1}-\eqref{eq:chap3-rbp-xhat-f2}. Eventually, r-BP can equivalently be thought of as the projection of BP onto the following parametrizations of the messages
\begin{gather}
    \label{eq:chap3-rbp-1}
    \msg{\tilde{m}^{(t)}}{\mu}{i}(x_i) \propto e^{\msg{B^{(t)}}{\mu}{i}x_i + \frac 1 2 \msg{A^{(t)}}{\mu}{i}x_i^2} \propto \int \dd{z_\mu} \pout(y_\mu | z_\mu)e^{-\frac{(z_\mu - W_{\mu i} x_i - \msg{\omega^{(t)}}{\mu}{i})^2}{2 \msg{V^{(t)}}{\mu}{i}}} ,\\
    \label{eq:chap3-rbp-2}
    \msg{m^{(t+1)}}{i}{\mu}(x_i) \propto e^{-\frac{(\msg{\hat{x}^{(t+1)}}{i}{\mu}- x_i)^2}{2 \msg{{C^x}^{(t+1)}}{i}{\mu}}} \propto p_x(x_i)e^{-\frac{(\msg{\lambda^{(t)}}{i}{\mu}- x_i)^2}{2 \msg{\sigma^{(t)}}{i}{\mu}}}. 
\end{gather}
Note that, at convergence, an approximation of the marginals is recovered from the projection on the parametrization \eqref{eq:chap3-rbp-2} of \eqref{eq:chap3-bp-marginal},
\begin{gather}
    m_i(x_i) = \frac{1}{\cZ_i}p_x(x_i)e^{-(\lambda_i-x_i)^2/2\sigma_i} = f^x_1(\lambda_i, \sigma_i), \\
    \msg{\sigma}{i}{\mu} = \left(\sum_{\mu}\msg{A}{\mu}{i} \right)^{-1}, \\ 
    \msg{\lambda}{i}{\mu} = \msg{\sigma}{i}{\mu} \left(\sum_{\mu} \msg{B}{\mu}{i}\right).
\end{gather} 
% where we introduced the dummy variable $\z = \W\x$, an important intermediate step in the reconstruction of $\x$ from $\y$, and the notation $\propto$ that omits the normalization factor for distributions.
% The BP algorithm projected on this parametrization makes up the r-BP inference algorithm 

Nonetheless, r-BP is scarcely used as such as the computational cost can be readily reduced with little more approximation. Because the parameters in \eqref{eq:chap3-rbp-1}-\eqref{eq:chap3-rbp-2} take the form of messages on the edges of the factor graph there are still $O(M \times N)$ quantities to track to solve the self-consistency equations by iterations. Yet, in the thermodynamic limit, the messages are closely related to the marginals as the contribution of 
the missing message between \eqref{eq:chap3-bp2} and \eqref{eq:chap3-bp-marginal} is to a certain extent negligible. Careful book keeping of the order of contributions of these small differences leads to a set of closed equations on parameters of the marginals, i.e. $O(N)$ variables, corresponding to the GAMP algorithm.

A detailed derivation and developed algorithm of r-BP for the GLM can be found for example in \cite{Zdeborova2016} (\citesec~6.3.1). In \citesec~\ref{sec:chap3-multivalue} of the present paper, we also present the derivation in a slightly more general setting where the variables $x_i$ and $y_\mu$ are possibly vectors instead of scalars.

\paragraph{Generalized approximate message passing}

The GAMP algorithm with respect to marginal parameters, analogous to the messages parameters introduced above (summarized in  \eqref{eq:chap3-rbp-1}-\eqref{eq:chap3-rbp-2}), is given in \citealg~\ref{alg:chap3-amp}. 
The origin of GAMP is again the development of the r-BP message-like equations around marginal quantities. The details of this derivation for the GLM can be found for instance in \cite{Zdeborova2016} (\citesec~6.3.1).
%lines~\eqref{alg:chap3-amp-V}, \eqref{alg:chap3-amp-om}, \eqref{alg:chap3-amp-sig} and \eqref{alg:chap3-amp-lbd} 
For a random initialization, the algorithm can be decomposed in 4 steps per iteration which refine the estimate of the signal $\x$ and the intermediary variable $\z$ by incorporating the different sources of information.
Steps 2) and 4) involve the \emph{update functions} relative to the prior and output channel defined above. 
Steps 1) and 3) are general for any GLM with a random Gaussian weight matrix, as they result from the consistency of the two alternative parametrizations introduced for the same messages in \eqref{eq:chap3-rbp-1}-\eqref{eq:chap3-rbp-2}.%lines~\eqref{alg:chap3-amp-dg}, \eqref{alg:chap3-amp-g}, \eqref{alg:chap3-amp-cx} and \eqref{alg:chap3-amp-xh} 
% Conversely, steps 2) and 4) involve \emph{update functions}.
% These special functions are defined respectively to choices of a prior on the signal $\x$ and an output channel producing the observations $\y$.
% \missing{They are also the only steps to be modified when deriving an AMP version of the max-sum BP as we will show in \citechap~\ref{sec:chap6}}.

% \subparagraph{Input update functions relative to the prior}
% \begin{gather}
%     \label{eq:chap3-Zx}
%     \cZ^x = \int \dd{x} p_x(x)e^{-\frac{(x-\lambda)^2}{2\sigma}} \\
%     \label{eq:chap3-f1x}
%     f^x_1(\lambda, \sigma) = \frac{1}{\cZ^x}\int \dd{x} x \, p_x(x)e^{-\frac{(x-\lambda)^2}{2\sigma}} \\
%     \label{eq:chap3-f2x}
%     f^x_2(\lambda, \sigma) =  \frac{1}{\cZ^x} \int \dd{x} x^2 \, p_x(x)e^{-\frac{(x-\lambda)^2}{2\sigma}} - f^x_1(\lambda, \sigma)^2
% \end{gather}
% \subparagraph{Output update functions relative to the channel}
% \begin{gather}
%     \label{eq:chap3-Zout}
%     \Zout( y, \omega, V) = \int \dd{z}  \pout(y|z)  \cN(z; \omega, V) \\
%     \label{eq:chap3-gout}
%     \gouts(y, \omega, V) = \frac{1}{\Zout}\int \dd{z} \frac{(z - \omega)}{V} \pout(y|z)  \cN(z; \omega, V) \\
%     \label{eq:chap3-dgout}
%     \dgouts(y, \omega, V) = \frac{1}{\Zout}\int \dd{z} \frac{(z - \omega)^2}{V^2} \pout(y|z)  \cN(z; \omega, V) - \frac 1 V - \gouts(y, \omega, V)^2
% \end{gather}

\begin{algorithm*}[t]
\caption{Generalized Approximate Message Passing\label{alg:chap3-amp}}   
\begin{algorithmic}
    \State {\bfseries Input:} vector $\y \in \R^{M}$ and matrix $\W \in \R^{M \times N}$:
    \State \emph{Initialize}: $\hat{x}_i$, $C^x_i \quad \forall i$ and $\gouts_\mu$, $\dgouts_\mu \quad \forall \mu$
    \Repeat   
    \Statex 1) Estimate mean $\omega_{\mu}^{(t)}$ and variance $V_{\mu}^{(t)}$ of $z_\mu$ given current $\hat{\x}^{(t)}$
    \vspace{-0.3cm}
    \begin{align}
        V_{\mu}^{(t)} &= \sum\limits_{i=1}^N W_{\mu i}^2 {C^x_{i}}^{(t)} \label{alg:chap3-amp-V} \\
        \omega_{\mu}^{(t)} &= \sum\limits_{i = 1}^N W_{\mu i} \hat{x}^{(t)}_i - \sum\limits_{i = 1}^N W_{\mu i}^2 {C^x}^{(t)}_i \gouts_\mu^{(t-1)} \label{alg:chap3-amp-om}
    \end{align} 
    \vspace{-0.3cm}
    \Statex 2) Estimate mean $\gouts^{(t)}_\mu$ and variance $\dgouts^{(t)}_\mu$ of the gap between optimal $z_\mu$ and $\omega_\mu^{(t)}$ given $y_\mu$
    \vspace{-0.3cm}
    \begin{align}
        \dgouts^{(t)}_\mu &=  \dgouts( y_\mu, \omega^{(t)}_\mu, V^{(t)}_\mu) \label{alg:chap3-amp-dg} \\
        \gouts^{(t)}_\mu &= \gouts( y_\mu, \omega^{(t)}_\mu, V^{(t)}_\mu)  \label{alg:chap3-amp-g}
    \end{align}
    \vspace{-0.3cm}
    \Statex 3) Estimate mean $\vect{\lambda}$ and variance $\vect{\sigma}$ of $\x$ given current gap between optimal $\z$ and $\vect{\omega}$
    \vspace{-0.3cm}
    \begin{align}
        \sigma_i^{(t)} &= \left(- \sum\limits_{\mu=1}^{M}W_{\mu i}^2\dgouts_\mu^{(t)}\right)^{-1} \label{alg:chap3-amp-sig} \\
        \lambda_i^{(t)} &=  \hat{x}^{(t)}_i + \sigma_i^{(t)}\left( \sum\limits_{\mu=1}^{M}W_{\mu i}\gouts_\mu^{(t)}\right) \label{alg:chap3-amp-lbd}
    \end{align}
    \vspace{-0.3cm}
    \Statex 4) Estimate of mean $\hat{\x}^{(t+1)}$ and  ${\vect{C}^x}^{(t+1)}$ variance of $\x$ updated with information from the prior
    \vspace{-0.3cm}
    \begin{align}
        {C^x_i}^{(t+1)} &= f^x_2(\lambda^{(t)}_i, \sigma_i^{(t)}) \label{alg:chap3-amp-cx} \\
        \hat{x}^{(t+1)}_i &= f^x_1(\lambda^{(t)}_i, \sigma_i^{(t)})  \label{alg:chap3-amp-xh}
    \end{align}

    \vspace{0.05cm}
    \Until{convergence} 

\end{algorithmic}
\end{algorithm*}

% \begin{gather}
%     m_i^{(t+1)}(x_i) \propto e^{-\frac{(\hat{x}^{(t+1)}_i- x_i)^2}{2 {C_i^x}}^{(t+1)}} \propto p_x(x_i)e^{-\frac{(\lambda^{(t)}_i - x_i)^2}{2 \sigma^{(t)}_i}},
% \end{gather}

\subparagraph{Relation to TAP equations}
Historically the main difference between the AMP algorithm and the TAP equations is that the latter was first derived for binary variables with $2$-body interactions (SK model) while the former was proposed for continuous random variables with $N$-body interactions (Compressed Sensing). The details of the derivation (described in \cite{Zdeborova2016} or in a more general case in \citesec~\ref{sec:chap3-multivalue}), rely on the knowledge of the statistics of the disordered variable $\W$ but do not require a disorder average, as in the Georges-Yedidia expansion yielding the TAP equations.
By
% presented the derivation for 
focusing on
the GLM with a random Gaussian weight matrix scaling as $O(1/\sqrt{N})$ (similarly to the couplings of the SK model) we naturally obtained TAP equations at second order, with an Onsager term in the update \eqref{alg:chap3-amp-om} of $\omega_\mu$. 
%The TAP free energy can also be recovered by plugging-in the approximate parametrized messages in the definition of the Bethe free energy \eqref{eq:chap3-bethe-fe}.
Yet an advantage of the AMP derivation from BP over the high-temperature expansion is that it explicitly provides `correct' time indices in the iteration scheme to solve the self consistent equations \cite{bolthausen2014iterative}. 

\subparagraph{Reconstruction with AMP}
AMP is therefore a practical reconstruction algorithm which can be run on a single instance (the disorder is not averaged) to estimate an unknown signal $\x_0$. Note that the prior $p_x$ and channel $\pout$ used in the algorithm correspond to the student statistical model and they may be different from the true underlying teacher model that generates $\x_0$ and $\y$. In other words, the AMP algorithm may be used either in the Bayes optimal or in the mismatched setting defined in \citesec~\ref{sec:chap2-teacher-student}.
Remarkably, it is also possible to consider a disorder average in the thermodynamic limit to study the average case computational hardness, here of the GLM inference problem, in either of these matched or mismatched configurations.

\paragraph{State Evolution}
The statistical analysis of the AMP equations for Compressed Sensing in the average case and in the thermodynamic limit $N\to \infty$ lead to another closed set of equations that was called State Evolution (SE) in \cite{Donoho2009}. Such an analysis can be generalized to other problems of application of approximate message passing algorithms. The derivation of SE starts from the r-BP equations and relies on the assumption of independent incoming messages to invoke the Central Limit Theorem. It is therefore only necessary to follow the evolution of a set of means and variances parametrizing Gaussian distributions. When the different variables and factors are statistically equivalent, as it is the case of the GLM, SE reduces to a few scalar equations. The interested reader should refer to \citeapp~\ref{app:chap6-vect-amp} for a detailed derivation in a more general setting.

\subparagraph{Mismatched setting} In the general mismatched setting we need to carefully differentiate the teacher and the student. We note $p_{x_0}$ the prior used by the teacher. We also rewrite its channel $\pouto(y|\vect{w}\T\x)$ as the explicit function $y = g_0(\vect{w}\T\x; \epsilon)$ assuming the noise $\epsilon$ to be distributed according to the standard normal distribution.
The tracked quantities are the \emph{overlaps},
\begin{gather}
    q =\lim_{N\to \infty}\frac{1}{N} \sum_{i=1}^N \hat{x}_i^2\,, \quad m = \lim_{N\to \infty}\frac{1}{N} \sum_{i=1}^N \hat{x}_i x_{0,i} \,, \quad q_0 = \lim_{N\to \infty}\frac{1}{N} \sum_{i=1}^N x_{0,i}^2 = \E_{p_{x_0}}[x_0^2] ,
\end{gather}
along with the auxiliary $V$, $\hat{q}$, $\hat{m}$ and $\hat{\chi}$:
\begin{align}
\label{eq:chap3-se-nonishi-out-q}
\hat{q}^{(t)} & = \int \D{\epsilon} \int \dd{\omega} \dd{z} \cN(z, \omega ; 0, \mat{Q}^{(t)}) 
	\gouts(\omega, g_0\left( z ; \epsilon\right) , V^{(t)})^2 \, ,\\
	\label{eq:chap3-se-nonishi-out-m}
\hat{m}^{(t)} & = \int \D{\epsilon} \int \dd{\omega} \dd{z} \cN(z, \omega ; 0, \mat{Q}^{(t)})
		\partial_{z} \gouts(\omega, g_0\left( z ; \epsilon\right) , V^{(t)}) \, ,\\
\label{eq:chap3-se-nonishi-out-xi}
\hat{\chi}^{(t)} & = - \int \D{\epsilon} \int \dd{\omega}\dd{z} \cN(z, \omega ; 0, \mat{Q}^{(t)}) 
	 \partial_{\omega} \gouts(\omega, g_0\left( z ; \epsilon_\mu\right) , V^{(t)}) \, ,
\end{align}
\begin{align}
\label{eq:chap3-se-nonishi-in-q}
q^{(t+1)} & = \int \dd{x_0} p_{x_0}(x_0) \int \D{\xi} 
	f^x_1 \left( (\alpha \hat{\chi}^{(t)})^{-1}\left({\sqrt{\alpha \hat{q}^{(t)}} \xi + \alpha \hat{m}^{(t)}\x_0}\right); (\alpha \hat{\chi}^{(t)})^{-1}  \right) ^2  \, ,\\
\label{eq:chap3-se-nonishi-in-m}
m^{(t+1)} & = \int \dd{x_0} p_{x_0}(x_0) \int \D{\xi} x_0 
	f^x_1 \left( (\alpha \hat{\chi}^{(t)})^{-1}\left({\sqrt{\alpha \hat{q}^{(t)}} \xi + \alpha \hat{m}^{(t)}\x_0}\right); (\alpha \hat{\chi}^{(t)})^{-1} \right) \, ,\\
\label{eq:chap3-se-nonishi-in-V}
V^{(t+1)} & = \int \dd{x_0} p_{x_0}(x_0) \int \D{\xi} 
f^x_2 \left( (\alpha \hat{\chi}^{(t)})^{-1}\left({\sqrt{\alpha \hat{q}^{(t)}} \xi + \alpha \hat{m}^{(t)}x_0}\right); (\alpha \hat{\chi}^{(t)})^{-1} \right) \, ,
\end{align}
where we use the notation $\cN(\cdot;\cdot,\cdot)$ for the normal distribution, $\D{\xi}$ for the standard normal measure and the covariance matrix $\mat{Q}^{(t)}$ is given at each time step by
\[
\mat{Q}^{(t)} = 
\begin{bmatrix}
q_0 & m^{(t)} \\
\\
{m^{(t)}} & q^{(t)} \\
\end{bmatrix}.
\]

Due to the self-averaging property, the performance of the reconstruction by the AMP algorithm on an instance of size $N$ can be tracked along the iterations given 
\begin{gather}
    \label{eq:chap3-se-mse}
    \MSE(\hat{x})= \frac{1}{N}\sum_{i=1}^N (\hat{x}_i - x_{0,i})^2 =  q - 2 m + q_0,
\end{gather}
with only minor differences coming from finite-size effects.
State Evolution also provides an efficient procedure to study from the theoretical perspective the AMP fixed points for a generic model, such as the GLM, as a function of some control parameters. It reports the average results for running the complete AMP algorithm on $O(N)$ variables
%  for a representatively large number of instances, 
using a few scalar equations. Furthermore, the State Evolution equations simplify further in the Bayes optimal setting.

\subparagraph{Bayes optimal setting}
When the prior and channel are identical for the student and the teacher, the true unknown signal $\x_0$ is in some sense statistically equivalent to the estimate $\xh$ coming from the posterior. More precisely one can prove the Nishimori identities \cite{Opper1991, Iba1999, Nishimori2001} (or \cite{Kabashima2016} for a concise demonstration and discussion) implying that $q  = m$, $ V  = q_0 - m$ and $\hat{q} = \hat{m} = \hat{\chi}$. Only two equations are then necessary to track the performance of the reconstruction:
\begin{align}
    \label{eq:chap3-se-bo-qh}
    \hat{q}^{(t)} & = \int \dd{\epsilon} p_{\epsilon_0}(\epsilon) \int \dd{\omega} \dd{z} \cN(z, \omega ; 0, \mat{Q}^{(t)}) 
    \gouts(\omega, g_0\left( z ; \epsilon\right) , V^{(t)})^2 \\
    \label{eq:chap3-se-bo-q}
    q^{(t+1)} & = \int \dd{x_0} p_{x_0}(x_0) \int \D{\xi} 
	f^x_1 \left( (\alpha \hat{\chi}^{(t)})^{-1}\left({\sqrt{\alpha \hat{q}^{(t)}} \xi + \alpha \hat{m}^{(t)}\x_0}\right); (\alpha \hat{\chi}^{(t)})^{-1}  \right) ^2 \, .
\end{align}

\subsection{Replica method}
\label{sec:chap3-replica}
Another powerful technique from the statistical physics of disordered systems to examine models with infinite range interactions is the replica method. It enables an analytical computation of the quenched free energy via non-rigorous mathematical manipulations. More developed introductions to the method can be found in \cite{Mezard1986, Nishimori2001, Castellani2005}. 

\subsubsection{Steps of a replica computation}
The basic idea of the replica computation is to compute the average over the disorder of $\log \cZ$ by considering the identity $\log \cZ = \lim_{n\to 0} (\cZ^n - 1)/n$. First the expectation of $\cZ^n$ is evaluated for $n\in\mathbb{N}$, then the $n\to 0$ limit is taken by `analytic continuation'. Thus the method takes advantage of the fact that the average of a power of $\cZ$ is sometimes easier to compute than the average of a logarithm. We illustrate the key steps of the calculation for the partition function of the GLM \eqref{eq:chap3-glm-Z}.

\subparagraph{Disorder average for the replicated system: coupling of the replicas}
The average of $\cZ^n$ for $n\in \mathbb{N}$ can be seen as the partition function of a system with $n + 1$ non interacting replicas of $\x$ indexed by $a \in \{0, \cdots, n\}$, where the first replica $a=0$ is representative of the teacher and the $n$ other replicas are identically distributed as the student:
\begin{align}
    \E_{\W, \y, \x_0}\left[ \cZ^n \right] &  = \E_{\W}\left[
        \int \dd{\y} \dd{\x_0} 
        \pouto(\y|\W\x_0)
        p_{\x_0}(\x_0)
        \left( \int \dd{\x} \pout(\y|\W\x)p_x(\x)
        \right)^n
    \right] \\
    & = \E_{\W}\left[ 
        \int \dd{\y} \prod_{a=0}^{n} \left( \dd{\x_a} \pouta(\y|\W\x_a)p_{x_a}(\x_a) \right)
        \right] \\
    & = \E_{\W}\left[ \int \dd{\y} \prod_{a=0}^{n} \left( \dd{\x_a} \dd{\z_a} \delta(\z_a - \W\x_a)\pouta(\y|\z_a)p_{x_a}(\x_a) \right) 
    \right] \;.
\end{align}
To perform the average over the disordered interactions $\W$ we consider the statistics of $\z_a = \W\x_a$. Recall that $W_{\mu i} \sim \cN(W_{\mu i} ;0,1/N)$, independently for all $\mu$ and $i$. Consequently, 
% by the CLT, 
the $\z_a$ are jointly Gaussian in the thermodynamic limit with means and covariances
\begin{gather}
    \E_{\W}[z_{a,\mu}] = \E_{\W}\left[\sum_{i=1}^N W_{\mu i}x_{a,i}\right]  =  0\, , \quad E_{\W}\left[ z_{a,\mu} z_{b, \nu}\right]  = \sum_{i=1}^N x_{a, i}x_{b, i} / N = q_{ab}.
\end{gather}
The overlaps, that we already introduced in the SE formalism, naturally re-appear. We introduce the notation $\q$ for the $(n+1) \times (n+1)$ overlap matrix. Integrating out the disorder $\W$ shared by the $n+1$ replicas will therefore leave us with an effective system of now coupled replicas:
\begin{align}
    \E_{\W, \y, \x_0} \left[ \cZ^n \right] = 
    \int \prod_{a,b} \dd{N q_{ab}}  & \int \dd{\y} \prod_{a=0}^{n} \dd{\z_a} \pouta(\y|\z_a) \\
     \exp&\left(-\frac{1}{2} \displaystyle \sum_{\mu=1}^M \sum_{a,b}z_{a,\mu}z_{b,\mu} (\q^{-1})_{ab} - M C(\q,n)\right)\notag \\
    & \int \prod_{a=1}^n\dd{\x_a} p_{x_a}(\x_a) \delta(N q_{ab} - \sum_{i=1}^N x_{a,i}x_{b,i}). \notag
\end{align}

\subparagraph{Change of variable for the overlaps: decoupling of the variables}
We consider the Fourier representation of the Dirac distribution fixing the consistency between overlaps and replicas,
\begin{align}
    \delta(N q_{ab} - \sum_{i=1}^N x_{a,i}x_{b,i}) =
   \int \frac{{\mathrm{d}\hat{q}_{ab}}}{2 i \pi } \, e^{\hat{q}_{ab}(N q_{ab} - \sum_{i=1}^N x_{a,i}x_{b,i})},
\end{align}
where $\hat{q}_{ab}$ is purely imaginary, which yields
\begin{align}
    \E_{\W, \y, \x_0} \left[ \cZ^n \right] = & \int  \prod_{a,b} \dd{Nq_{ab}}  \int \prod_{a,b} \frac{{\mathrm{d}\hat{q}_{ab}}}{2 i \pi } \, \exp\left(N \hat{q}_{ab}q_{ab}\right) \\
    & \int \dd{\y} \prod_{a=0}^{n} \dd{\z_a} \pouta(\y|\z_a) \exp\left(-\frac{1}{2} \displaystyle \sum_{\mu=1}^M \sum_{a,b}z_{a,\mu}z_{b,\mu} (\q^{-1})_{ab} - M C(\q,n)\right)\notag \\
    & \int \prod_{a=1}^n\dd{\x_a} p_{x_a}(\x_a) \exp\left(- \hat{q}_{ab} \displaystyle \sum_{i=1}^N x_{a,i}x_{b,i}\right) \notag
\end{align}
where $C(\q,n)$ is related to the normalization of the Gaussian distributions over the $\z_a$ variables, and the integrals can be factorized over the $i$-s and $\mu$-s. Thus we obtain
\begin{gather}
    \label{eq:chap3-replica-Nscaling}
    \E_{\W, \y, \x_0}\left[ \cZ^n \right] = \int \prod_{a,b} \dd{N q_{ab}}  \int \prod_{a,b} \dd{ \hat{q}_{ab}} e^{N \hat{q}_{ab}q_{ab}} e^{M \log \hat{\mathcal{I}}_z(\q)} e^{N \log \hat{\mathcal{I}}_x(\hat{\q})} \, , 
\end{gather}
with
\begin{gather}
    \hat{\mathcal{I}}_z(\q) = \int \dd{y} \prod_{a=0}^{n} \dd{z_a} \pouta(y|z_a) \exp\left(-\frac{1}{2} \displaystyle \sum_{a,b}z_{a}z_{b} (q_{ab})^{-1} - C(\q,n)\right) \, ,\\
    \hat{\mathcal{I}}_x(\hat{\q}) = \int \prod_{a=1}^n\dd{x_a} p_{x_a}(x_a) \exp\left(- \hat{q}_{ab} \displaystyle x_{a}x_{b}\right) \, ,
\end{gather}
where we introduce the notation $\hat{\q}$ for the auxiliary overlap matrix with entries $(\hat{\q})_{ab} = \hat{q}_{ab}$ and we omitted the factor $2i\pi$ which is eventually subleading as $N\to + \infty$.
The decoupling of the $x_i$ and the $z_\mu$ of the infinite range system yields pre-factors $N$ and $M$ in the exponential arguments. In the thermodynamic limit, we recall that both $N$ and $M$ tend to $+\infty$ while the ratio $\alpha=M/N$ remains fixed. Hence, the integral for the replicated average is easily computed in this limit by the saddle point method: 
\begin{gather}
   \log \E_{\W, \y, \x_0}\left[ \cZ^n \right] \simeq N \mathrm{extr}_{\q \hat{\q}}\left[\mathcal{\phi}(\q, \hat{\q})\right]  \, , \quad  \mathcal{\phi}(\q, \hat{\q}) = \sum_{a,b} \hat{q}_{ab}q_{ab} + \alpha \hat{\mathcal{I}}_z(\q) + \hat{\mathcal{I}}_x(\hat{\q}),
\end{gather}
where we defined the replica potential $\mathcal{\phi}$.

\subparagraph{Exchange of limits: back to the quenched average}
The thermodynamic average of the log-partition is recovered through an a priori risky mathematical manipulation: (i) perform an analytical continuation from $n \in \mathbb{N}$ to $n \to 0$ 
\begin{gather}
    % - f = \lim_{N\to \infty} 
    \frac{1}{N}\E_{\W, \y, \x_0}\left[ \log \cZ \right] 
    = 
    % \lim_{N\to \infty}
    \lim_{n\to 0} \frac{1}{nN} \E_{\W, \y, \x_0}\left[ \cZ^n -1\right] 
    = 
    % \lim_{N\to \infty}
    \lim_{n\to 0} \frac{1}{nN} \log \E_{\W, \y, \x_0}\left[ \cZ^n \right] 
\end{gather}
and (ii) exchange limits 
\begin{gather}
    -f   
    = \lim_{N\to \infty} \lim_{n\to 0}\frac{1}{n}  \frac{1}{N} \log \E_{\W, \y, \x_0}\left[ \cZ^n \right] = \lim_{n\to 0} \frac{1}{n} \mathrm{extr}_{\q \hat{\q}}\left[\mathcal{\phi}(\q, \hat{\q})\right].
\end{gather}
Despite the apparent lack of rigour in taking these last steps, the replica method has been proven to yield exact predictions in the thermodynamic limit for different problems and in particular for the GLM \cite{Reeves2016, Barbier2017a}.

\subparagraph{Saddle point solution: choice of a replica ansatz}
At this point, we are still left with the problem of computing the extrema of $\mathcal{\phi}(\q, \hat{\q})$. To solve this optimization problem over $\q$ and $\hat{\q}$, a natural assumption is that replicas, that are a pure artefact of the calculation, are equivalent. This is reflected in a special structure for overlap matrices between replicas that only depend on three parameters each,
\begin{align}
    \q = 
\begin{bmatrix}
q_0 & m & m & m \\
m & q & q_{12} & q_{12}\\
m & q_{12} & q & q_{12}\\ 
m & q_{12} & q_{12} & q \\
\end{bmatrix} \, , \quad
\hat{\q} = 
\begin{bmatrix}
\hat{q_0} & \hat{m} & \hat{m} & \hat{m} \\
\hat{m} & \hat{q} & \hat{q}_{12} & \hat{q}_{12}\\
\hat{m} & \hat{q}_{12} & \hat{q} & \hat{q}_{12}\\ 
\hat{m} & \hat{q}_{12} & \hat{q}_{12} & \hat{q} \\
\end{bmatrix},
\end{align} 
here given as an example for $n=3$ replicas. 
Plugging this \emph{replica symmetric} (RS) ansatz in the expression of $\mathcal{\phi}(\q, \hat{\q})$, taking the limit $n\to0$ and looking for the stationary points as a function of the parameters $q$, $m$, $q_{12}$ and $\hat{m}$, $\hat{q}$, $\hat{q}_{12}$ recovers a set of equations equivalent to SE \eqref{eq:chap1-dnn-rec1}, albeit without time indices. Hence the two a priori different heuristics of BP and the replica method are remarkably consistent under the RS assumption.

Nevertheless, the replica symmetry can be spontaneously broken in the large $N$ limit and the dominating saddle point does not necessarily correspond to the RS overlap matrix. This replica symmetry breaking (RSB) corresponds to substantial changes in the structure of the examined Boltzmann distribution. It is among the great strengths of the replica formalism to naturally capture it. 
Yet for inference problems falling under the teacher-student scenario, the correct ansatz is always replica symmetric in the Bayes optimal setting \cite{Nishimori2001, Castellani2005, Zdeborova2016}, and we will not investigate here this direction further. The interested reader can refer to the classical references for an introduction to replica symmetry breaking \cite{Mezard1986, Nishimori2001, Castellani2005} in the context of the theory of spin-glasses.

\subparagraph{Bayes optimal setting} As in SE the equations simplify in the matched setting, where the first replica corresponding to the teacher becomes equivalent to all the others. The replica free energy of the GLM is then given as the extremum of a potential over two scalar variables:
\begin{gather}
    \label{eq:chap3-replica-fe-glm}
    - f = \mathrm{extr}_{q \hat{q}}\left[ - \frac{1}{2} q \hat{q} +  \mathcal{I}_x(\hat{q}) + \alpha \mathcal{I}_z(q_0, q)\right]\\
    \label{eq:chap3-replica-fe-glm_Ix}
    \mathcal{I}_x(\hat{q}) = \int \D{\xi} \dd{x} p_x(x)e^{-\hat{q}\frac{x^2}{2} + \sqrt{\hat{q}}\xi x} \log\left( \int \dd{x'} p_x(x')e^{-\hat{q}\frac{x'^2}{2} + \sqrt{\hat{q}}\xi x'}\right) \\
    \mathcal{I}_z(q , q_0) = \int \D{\xi} \dd{y} \dd{z} \pout(y|z)\cN(z;\sqrt{q}\xi,q_0-q) \notag \\
    \label{eq:chap3-replica-fe-glm_Iz}
    \qquad \qquad \qquad \qquad \qquad \qquad \qquad \qquad \log\left( \int \dd{z'} \pout(y|z')\cN(z';\sqrt{q}\xi,q_0-q)  \right) .
\end{gather}
The saddle point equations corresponding to the extremization \eqref{eq:chap3-replica-fe-glm}, fixing the values of $q$ and $\hat{q}$, would again be found equivalent to the Bayes optimal SE \eqref{eq:chap3-se-bo-qh} - \eqref{eq:chap3-se-bo-q}. This Bayes optimal result is derived in \cite{Krzakala2012} for the case of a linear channel and Gauss-Bernoulli prior, and can also be recovered as a special case of the low-rank matrix factorization formula (where the measurement matrix is in fact known) \cite{Kabashima2016}.

\subsubsection{Assumptions and relation to other mean-field methods}
A crucial point in the above derivation of the replica formula is the extensivity of the interactions of the infinite range model that allowed the factorization of the $N$ scaling of the argument of the exponential integrand in \eqref{eq:chap3-replica-Nscaling}. The statistics of the disorder $\W$ and in particular the independence of all the $W_{\mu i}$ was also necessary. While this is an important assumption for the technique to go through, it can be possible to relax it for some types of correlation statistics, as we will see in \citesec~\ref{sec:chap3-ortho-invariant}.

Note that the replica method directly enforces the disorder averaging and does not provide a prediction at the level of the single instance. Therefore it cannot be turned into a practical algorithm of reconstruction. Nonetheless, we have seen that the saddle point equations of the replica derivation, under the RS assumption, matches the SE equations derived from BP. This is sufficient to theoretically study inference questions under a teacher-student scenario in the Bayes optimal setting, and in particular predict the MSE following \eqref{eq:chap3-se-mse}.

In the mismatched setting however, the predictions of the replica method under the RS assumption and the equivalent BP conclusions can be wrong. By introducing the symmetry breaking between replicas, the method can sometimes  be corrected. It is an important endeavor of the replica formalism to grasp the importance of the overlaps and connect the form of the replica ansatz to the properties of the joint probability distribution examined. When BP fails on loopy graphs, correlations between variables are not decaying with distance, which manifests into an RSB phase. Note that there also exists message passing algorithms operating in this regime \cite{Mezard2001, Mezard2002, Mezard2009, Saglietti2019, Antenucci2019, Antenucci2019a}.
% To tackle this case, a message passing formalism taking into account (one step of) replica symmetry breaking was first proposed in \cite{Mezard2001} (nicely formalized and generalized in \cite{Mezard2009}); it maintains the consistency of the replica method with the corresponding ansatz but can only be solved for the ensemble with disorder averaged out. Survey propagation as algorithm \cite{Mezard2002} Recent works incorporate replica symmetry braking into AMP like algorithms, for instance \cite{Saglietti2019,Antenucci2019,Antenucci2019a}.}

% The replica method 
% Among its most interesting aspects is the role played by ‘overlaps’ among replicas. It turns out that the subtle probabilistic structure of the systems under study are often most easily phrased in terms of such variables

% Not so obviously correlated at first sight - replica method is another method, that in some of the cases of the one given above will allow approximation averaged over the quenched disorder. 

% Mean-field models = fully connected -> according to Spin glass for pedestrians conditions for the replica ansatz to allow for decoupling sites (-> partition to the power N) and then coupled replicas when averaging over the disorder.
% Saddle point computation - 

% The replica symmetric solution and the RSB - 
% The RSB = predictions 

% In Bayes optimal - always replica symmetric - 

% \subparagraph{Several level of approximations yet reigorous proofs}
% Also justifies AMP

\section{Further extensions of interest for learning}
\label{sec:chap3further}
In the previous \citesec~we presented the classical mean-field approximations focusing on the simple and original examples of the Boltzmann machine (a.k.a. SK model) and the GLM with Gaussian i.i.d weight matrices. Along the way, we tried to emphasize how the procedures of approximation rely on structural (e.g. connectivity) and statistical properties of the model under scrutiny. In the present \citesec, we will see that extensions of the message passing and replica methods have now broadened the span of applicability of mean-field approximations. We focus on a selection of recent developments of particular interest to study learning problems.

\subsection{Streaming AMP for online learning}
\label{sec:chap3-streaming-amp}
In learning applications, it is sometimes advantageous for speed or generalization concerns to only treat a subset of examples at the time - making
for instance the SGD algorithm the most popular training algorithm in deep learning. Sometimes also, the size of the current data sets may exceed the available memory. Methods implementing a step-by-step learning as the data arrives are referred to as \emph{online}, \emph{streaming} or \emph{mini-batch} learning, as opposed to \emph{offline} or \emph{batch} learning. 

% TO ADD: Andre's introduction - In this work, we treat streaming inference within a Bayesian framework where, as new data arrives,
% posterior beliefs are updated according to Bayes’ rule. One well known approach in this direction is assumed density filtering (ADF) [3, 4], which processes a single data point at a time, a procedure to which we refer to as fully online. A number of other works analyzed various related fully online algorithms [5, 6], especially in the statistical physics literature [7–11]. We are instead interested in the case where multiple samples – a mini-batch – arrive at once. Tuning the size of these mini-batches allows us to to explore the trade-off between the precision and efficiency

In \cite{Manoel2018}, a mini-batch version of the AMP algorithm is proposed. It consists in a generalization of Assumed Density Filtering \cite{Opper1999, Rossi2016} that are fully-online, meaning that only a single example is received at once, or mini-batches have size 1. The general derivation principle is the same. On the example of the GLM, one imagines receiving at each iteration a subset of the components of $\y$ to reconstruct $\x$. We denote by $\y\kk$ these successive mini-batches. Bayes formula gives the posterior distribution over $\x$ after seeing $k$ mini-batches
\begin{align}
    p(\x|\y\kk, \{\y_{(k-1)}, \cdots \y_{(1)}\}) = \frac{p(\y\kk|\x)p(\x|\{\y_{(k-1)}, \cdots \y_{(1)}\})}{\int \dd{\x}p(\y\kk|\x)p(\x|\{\y_{(k-1)}, \cdots \y_{(1)}\})} .
\end{align}
This formula suggests the iterative scheme of using as a prior on $\x$ at iteration $k$ the posterior obtained at iteration $k-1$. This idea can be implemented in different approximate inference algorithms, as also noticed by \cite{Broderick2013} using a variational method. In the regular version of AMP an effective factorized posterior is given at convergence by the input update functions \eqref{eq:chap3-Zx}-\eqref{eq:chap3-f2x}:
\begin{align}
p(\x|\y, \W) \simeq \prod_{i=1}^N \frac{1}{\cZ_x(\lambda_i, \sigma_i)}p_x(x_i)e^{-\frac{(\lambda_i-x_i)^2}{2 \sigma_i}}.     
\end{align}
Plugging this posterior approximation in the iterative scheme yields the mini-AMP algorithm using the converged values of ${\lambda_{(\ell)}}_i$ and ${\sigma_{(\ell)}}_i$ at each anterior mini-batch $\ell < k$ to compute the prior
\begin{align}
    p_{x}^{(k)}(\x) = p(\x|\{\y_{(k-1)}, \cdots \y_{(1)}\}, \W) \simeq \prod_{i=1}^N \frac{1}{\cZ_{x,i}} \; p_x(x_i)  \; e^{-\sum\limits_{\ell=1}^{k-1}\frac{(\lambda_{(\ell), i}-x_i)^2}{2 \sigma_{(\ell), i}}},     
\end{align}
where the $\cZ_{x,i}$ normalize each marginal factor.
Compared to a naive mean-field variational approximation of the posterior, AMP takes into account more correlations and is indeed found to perform better in experiments reported by \cite{Manoel2018}. Another advantage of the AMP based online inference is that it is amenable to theoretical analysis by a corresponding State Evolution \cite{Opper1999, Rossi2016, Manoel2018}.
% In \citechap~\ref{sec:chap6}, we follow the lines of \cite{Manoel2018} to derive a streaming version of the Cal-AMP algorithm (see next section), for the training of minimal models of neural networks. 

% \missing{cite committee paper for the vector derivation?}

\subsection{Algorithms and free energies beyond i.i.d. matrices} 
\label{sec:chap3-ortho-invariant}
The derivations outlined in the previous \citesecs~of the equivalent replica, TAP and AMP equations required the weight matrices to have Gaussian i.i.d. entries. In this case, rigorous proofs of asymptotic exactness of the mean-field solutions were found, for the SK model \cite{Talagrand2006} and the GLM \cite{Reeves2016, Barbier2017a}. Mean-field inference with different weight statistics is a priori feasible if one finds a way either to perform the corresponding disorder average in the replica computation, to evaluate the corresponding Onsager correction in the TAP equations, or to write a message passing where messages remain uncorrelated (even in the high-connectivity limit we may be interested in).  

Efforts to broaden in practice the class of matrices amenable to such mean-field treatments lead to a series of works in statistical physics and signal processing with related propositions.
Parisi and Potters pioneered this direction by deriving mean-field equations for orthogonal weight matrices using a high-temperature expansion \cite{Parisi1995}.
The adaptive TAP approach proposed by Opper and Winther \cite{Opper2001, Opper2001prl} further allowed for inference in densely connected graphical models without prior knowledge on the weight statistics. The Onsager term of these TAP equations was evaluated using the cavity method for a given weight sample. The resulting equations were then understood to be a particular case of the Expectation Propagation (EP) \cite{Minka2001} - belonging to the class of message passing algorithms for approximate inference - yet applied in densely connected models \cite{Opper2005}. An associated approximation of the free energy called Expectation Consistency (EC) was additionally derived from the EP messages. Subsequently, Kabashima and collaborators \cite{Shinzato2008, Shinzato2009, Kabashima2008} focused on the perceptron and the GLM to propose TAP equations and a replica derivation of the free energy for the ensemble of orthogonally invariant random weight matrices. In the singular value decomposition of such weight matrices, $\W=\mat{U}\,\mat{S}\,\mat{V}\T \in \R^{M\times N}$, the orthogonal basis matrices $\mat{U}$ and $\mat{V}$ are drawn uniformly at random from respectively $\mathrm{O}(M)$ and $\mathrm{O}(N)$, while the diagonal matrix of singular values $\mat{S}$ has an arbitrary spectrum. The consistency between the EC free energy and the replica derivation for orthogonally invariant matrices was verified by \cite{Kabashima2014} for signal recovery from linear measurements (the GLM without G). From the algorithmic perspective, Fletcher, Rangan and Schniter \cite{Rangan2016, Schniter2016} applied the EP to the GLM to obtain the (Generalized) Vector-Approximate Message Passing (G-VAMP) algorithm. Remarkably, these authors proved that the behavior of the algorithm could be characterized in the thermodynamic limit, provided the weight matrix is drawn from the orthogonally invariant ensemble, by a set of scalar State Evolution equations similarly to the AMP algorithm. These equations are again related to the saddle point equations of the replica free energy. Concurrently, Opper, Cakmak and Winther proposed an alternative procedure for solving TAP equations with orthogonally invariant weight matrices in Ising spin systems relying on an analysis of iterative algorithms \cite{Opper2016, Cakmak2019}. Finally, \cite{Maillard2019} revisits the above cited contributions and provides detailed considerations on their connections. 

Below we present the aforementioned free energy as proposed by \cite{Shinzato2008, Shinzato2009, Kabashima2008}, and the G-VAMP algorithm of \cite{Schniter2016}.

\subsubsection{Replica free energy for the GLM in the Bayes Optimal setting}
Consider the ensemble of orthogonally invariant weight matrices $\W$ with spectral density $\sum_{i=1}^N\dirac(\lambda - \lambda_i) / N$ of their `square' $\W \W\T$ converging in the thermodynamic limit $ N \to + \infty$ to a given density $\rho_\lambda(\lambda)$. The quenched free energy of the GLM in the Bayes optimal setting derived in \cite{Kabashima2008, Shinzato2009} writes
\begin{gather}
    \label{eq:chap3-kaba-fe} 
    -f = \mathrm{extr}_{q \hat{q}} \left[ - \frac{1}{2} q \hat{q} +  \mathcal{I}_x(\hat{q}) + \mathcal{J}_z(q_0, q, \alpha, \rho_\lambda) \right] \, ,\\
    \mathcal{J}_z(q_0, q, \alpha, \rho_\lambda) =  \mathrm{extr}_{u \hat{u}} \left[ F_{\rho_\lambda, \alpha}(q_0-q, \hat{u}/\lambda_0) +\frac{\hat{u}q_0}{2} - \frac{\alpha \hat{u}u}{2 \lambda_0} + \alpha \mathcal{I}_z(q_0\lambda_0/\alpha, u) \right],
\end{gather}
where $\mathcal{I}_x$ and $\mathcal{I}_z$ were defined as \eqref{eq:chap3-replica-fe-glm_Ix}-\eqref{eq:chap3-replica-fe-glm_Iz} and the spectral density $\rho_\lambda(\lambda)$  appears via its mean $\lambda_0=\E_{\lambda}[\lambda]$ and in the definition of 
\begin{gather}
    F_{\rho_\lambda, \alpha}(q, u) = \frac{1}{2} \mathrm{extr}_{\Lambda_q, \Lambda_u} \left[ -(\alpha-1)\log\Lambda_u - \E_{\lambda}\log(\Lambda_u\Lambda_q + \lambda) + \Lambda_q q + \alpha\Lambda_u u \right] \notag \\
    \qquad \qquad \qquad  - \frac{1}2 (\log q + 1) + \frac{\alpha}2 (\log u + 1) .
\end{gather}
Gaussian random matrices are a particular case of the considered ensemble. Their singular values are characterized asymptotically by the Marcenko-Pastur distribution \cite{Marcenko1967}. In this case, one can check that the above expression reduces to \eqref{eq:chap3-replica-fe-glm}. More generally, note that $\mathcal{J}_z$ generalizes $\mathcal{I}_z$.

\subsubsection{Vector Approximate Message Passing for the GLM}
\begin{figure}
    \centering
    \includegraphics[width=0.5\textwidth]{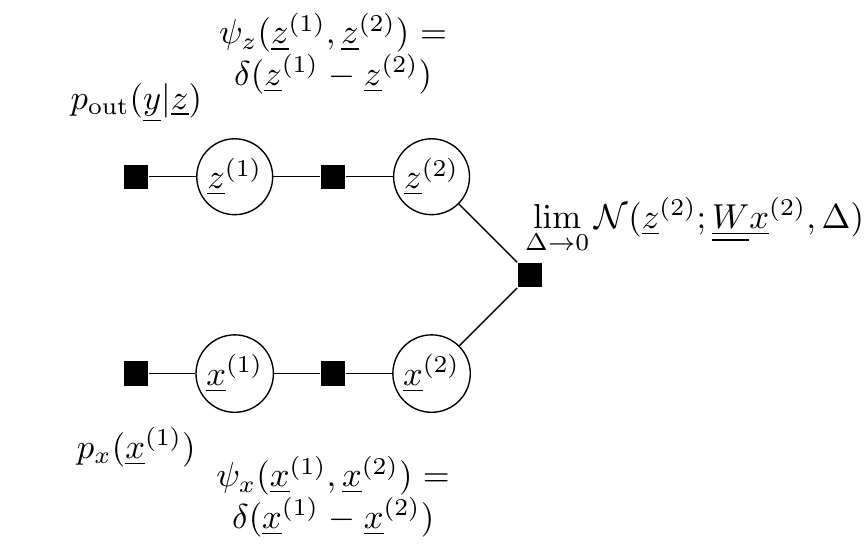}
    \caption{Factor graph representation of the GLM for the derivation of VAMP \label{fig:chap3-vamp-glm}}
\end{figure}

The VAMP algorithm consists in writing EP \cite{Minka2001} with Gaussian messages on the factor graph representation of the GLM posterior distribution given in \citefig~\ref{fig:chap3-vamp-glm}. The estimation of the signal $\x$ is decomposed onto four variables, two duplicates of $\x$ itself and two duplicates of the linear transformation $\z = \W\x$. The potential functions $\psi_x$ and $\psi_z$ of factors connecting copies of the same variable are Dirac distributions enforcing their equality. The factor node linking $\z^{(2)}$ and $\x^{(2)}$ is assumed Gaussian with variance going to zero.
The procedure of derivation, equivalent to the projection of the BP algorithm on Gaussian messages, is recalled in \citeapp~\ref{sec:app-chap3-vamp} and leads to \citealg~\ref{alg:chap3-vamp}. Like for AMP, the algorithm features some auxiliary variables introduced along the derivation. At convergence the duplicated $\xh_1$, $\xh_2$ (and $\hat{\z}_1$, $\hat{\z}_2$) are equal and either can be returned by the algorithm as an estimator.
For readability, we omitted the time indices in the iterations that here simply follow the indicated update.
%  schedule without calling from previous updates conversely to the AMP \citealg~\ref{alg:chap3-amp}.

\begin{algorithm*}[t]
\caption{Vector Approximate Message Passing\label{alg:chap3-vamp}}   
\begin{algorithmic}
    \State {\bfseries Input:} vector of observations $\y \in \R^{M}$ and weight matrix $\W \in \R^{M \times N}$:
    \State \emph{Initialize}: $\A_1^{x}$,  $\B_1^{x}$, $\A_1^z$,  $\B_1^z$
    \Repeat   
    \vspace{-0.3cm}
    \begin{gather}
        \xh_1 = f_1^x( {\B_1^{x}}, \A_1^{x}) 
        %\\
        \, , \qquad 
        {\Cx}_1 =  f_2^x( {\B_1^{x}}, \A_1^{x}) 
    \\ 
    % \end{gather}
    % \begin{gather}
        \label{alg:chap3-vamp-Ax2}
        \A_2^x = {\Cx_1}^{-1} - \A_1^{x} 
        %\\
        \, , \qquad
        % \label{alg:chap3-vamp-Bx2}
        \B_2^x = {\Cx_1}^{-1}\hat{\x}_2 - \B_1^{x}
    % \\ 
    \end{gather}
    \begin{gather}
        \hat{\z}_1 = f_1^z( {\B_1^z}, \A_1^z) 
        %\\
        \, , \qquad
        {\Cz}_1 =  f_2^z( {\B_1^z}, \A_1^z) 
    \\ 
    % \end{gather}
    % \begin{gather}
        \label{alg:chap3-vamp-Az2}
        \A_2^z = {\Cx_1}^{-1} - \A_1^z 
        %\\
        \, , \qquad 
        % \label{alg:chap3-vamp-Bz2}
        \B_2^z = {\Cx_1}^{-1}\hat{\x}_2 - \B_1^z
    % \\ 
    \end{gather}
    \begin{gather}
        \xh_2 = g^x_1({\B_2^x}, A_2^x, {\B_2^z}, A_2^z)  
        %\\
        \, , \qquad 
        \Cx_2 = g^x_2({\B_2^x}, A_2^x, {\B_2^z}, A_2^z)
    \\ 
    % \end{gather}
    % \begin{gather}
        \label{alg:chap3-vamp-Ax1}
        \A_1^x = {\Cx_2}^{-1} - \A_2^x 
        %\\
        \, , \qquad 
        % \label{alg:chap3-vamp-Bx1}
        \B_1^x = {\Cx_2}^{-1}\hat{\x}_1 - \B_2^x
    % \\ 
    \end{gather}
    \begin{gather}
    \hat{\z}_2 = g^z_1({\B_2^x}, A_2^x, {\B_2^z}, A_2^z)  
    %\\
    \, , \qquad 
    \Cz_2 = g^z_2({\B_2^x}, A_2^x, {\B_2^z}, A_2^z)
    \\ 
    % \end{gather}
    % \begin{gather}
    \label{alg:chap3-vamp-Az1}
    \A_1^z = {\Cx_2}^{-1} - \A_2^z 
    %\\
    \, , \qquad 
    % \label{alg:chap3-vamp-Bz1}
    \B_1^z = {\Cx_2}^{-1}\hat{\x}_1 - \B_2^z
    \end{gather}
    \Until{convergence} 
    \State {\bfseries Output:} signal estimate $\xh_1 \in \R^N$, and estimated covariance $\Cx_1 \in \R^{N \times N} $
\end{algorithmic}
\end{algorithm*}

For a given instance of the GLM inference problem, i.e. a given weight matrix $\W$, one can always launch either the AMP algorithm or the VAMP algorithm to attempt the reconstruction. If the weight matrix has i.i.d. zero mean Gaussian entries, the two strategies are conjectured to be equivalent and GAMP can be provably convergent for certain settings \cite{Rangan2014}. If the weight matrix is not Gaussian but orthogonally invariant, then only VAMP is expected to always converge. More generally, even in cases where none of these assumptions are verified, VAMP has been observed to have less convergence issues than AMP. 

Like for AMP, a State Evolution can also be derived for VAMP (which was actually directly proposed for the multi-layer GLM \cite{Fletcher2018a}). It rigorously characterizes the behavior of the algorithm when $\W$ is orthogonally invariant. One can also verify that the SE fixed points can be mapped to the solutions of the saddle point equations of the replica free energy \eqref{eq:chap3-kaba-fe} (see \citesec~1 of Supplementary Material of \cite{Gabrie2018}); so that the robust algorithmic procedure can advantageously be used to compute the fixed points to be plugged in the replica potential to approximate the free energy.

\subsection{Multi-value AMP}
\label{sec:chap3-multivalue}
A recent extension of AMP consists in treating the simultaneous reconstruction of multiple signals undergoing the same mixing step from the corresponding multiple observations. This is a situation of particular interest for learning appearing for instance in the teacher-student set-up of committee machines. The authors of \cite{Aubin2018} showed that the different weight vectors of these neural networks can be inferred from the knowledge of training input-output pairs introducing this extended version of AMP. Here the same matrix of training input data mixes the teacher weight vectors to produce the training output data. For a matter of consistency with the examples used in the previous sections, we here formalize the algorithm for the GLM. Nevertheless this is just a matter of rewriting of the committee algorithm of \cite{Aubin2018}.

Concretely let's consider a GLM with $P$ pairs of signal and observations $\{\x\kk, \y\kk\}_{k=1}^P$, gathered in matrices $\X \in \R^{N\times P}$ and $\Y \in \R^{M\times P}$. We are interested in the posterior distribution 
\begin{gather}
    \label{eq:chap6-glm-vec-meas}
    p(\X | \Y, \W) = \frac{1}{\cZ(\Y, \W)} \prod_{i=1}^N p(\x_i)\prod_{\mu=1}^M \pout(\y_\mu | \vect{w}_\mu\T\X), \quad \x_i \in \R^P, \quad \y_\mu \in \R^P. 
\end{gather}
Compared to the traditional GLM measure \eqref{eq:chap3-glm-meas}, scalar variables are here replaced by vectors in $\R^P$. In \citeapp~\ref{app:chap6-vect-amp} we present a derivation starting from BP of the corresponding AMP presented in \citealg~\ref{alg:chap6-vect-amp}.  
The major difference with the scalar GLM consists in the necessity of tracking covariance matrices between the $P$ different variables instead of simple variances.

\begin{algorithm*}[t]
\caption{Approximate Message Passing for multi-value GLM \label{alg:chap6-vect-amp}}   
\begin{algorithmic}
    \State {\bfseries Input:} matrix $\Y \in \R^{M  \times P}$ and matrix $\W \in \R^{M \times N}$:
    \State \emph{Initialize}: $\xh_i$, $\Cx_i \quad \forall i$ and $\gout_\mu$, $\dgout_\mu \quad \forall \mu$
    \Repeat   
    \Statex 1) Estimate mean and variance of $\z_\mu$ given current $\xh_i$
    \vspace{-0.3cm}
    \begin{align}
        \V_{\mu}^{(t)} &= \sum\limits_{i=1}^N \frac{W_{\mu i}^2}{N} {\Cx_{i}}^{(t)} \label{alg:chap6-vect-amp-V} \\
        \w_{\mu}^{(t)} &= \sum\limits_{i = 1}^N \frac{W_{\mu i}}{\sqrt{N}} \xh^{(t)}_i - \sum\limits_{i = 1}^N \frac{W_{\mu i}^2}{N} (\sig_i^{(t)})^{-1} {\Cx}^{(t)}_i \sig_i \gout_\mu^{(t-1)} \label{alg:chap6-vect-amp-om}
    \end{align} 
    \vspace{-0.3cm}
    \Statex 2) Estimate mean and variance of the gap between optimal $\z_\mu$ and $\w_\mu$ given $\y_\mu$
    \vspace{-0.3cm}
    \begin{align}
        \dgout^{(t)}_\mu &=  \dgout( \y_\mu, \w^{(t)}_\mu, \V^{(t)}_\mu) \label{alg:chap6-vect-amp-dg} \\
        \gout^{(t)}_\mu &= \gout( \y_\mu, \w^{(t)}_\mu, \V^{(t)}_\mu)  \label{alg:chap6-vect-amp-g}
    \end{align}
    \vspace{-0.3cm}
    \Statex 3) Estimate mean and variance of $\x_i$ given current optimal $\z_\mu$
    \vspace{-0.3cm}
    \begin{align}
        \sig_i^{(t)} &= \left(- \sum\limits_{\mu=1}^{M}\frac{W_{\mu i}^2}{N}\dgout_\mu^{(t)}\right)^{-1} \label{alg:chap6-vect-amp-sig} \\
        \lbd_i^{(t)} &=  \xh^{(t)}_i + \sig_i^{(t)}\left( \sum\limits_{\mu=1}^{M}\frac{W_{\mu i}}{\sqrt{N}}\gout_\mu^{(t)}\right) \label{alg:chap6-vect-amp-lbd}
    \end{align}
    \vspace{-0.3cm}
    \Statex 4) Estimate of mean and variance of  $\x_i$ augmented of the information about the prior
    \vspace{-0.3cm}
    \begin{align}
        {\Cx_i}^{(t+1)} &= \mat{f}^x_2(\lbd^{(t)}_i, \sig_i^{(t)}) \label{alg:chap6-vect-amp-cx} \\
        \xh^{(t+1)}_i &= \vect{f}^x_1(\lbd^{(t)}_i, \sig_i^{(t)})  \label{alg:chap6-vect-amp-xh}
    \end{align}

    \vspace{0.05cm}
    \Until{convergence} 
\end{algorithmic}
\end{algorithm*}

This AMP algorithm can also be analyzed by a State Evolution. 
In \cite{Aubin2018}, the teacher-student matched setting of the committee machine is examined through the replica approach and the Bayes optimal State Evolution equations are obtained as the saddle point equations of the replica free energy.
In \citeapp~\ref{app:chap6-vect-amp} we present the alternative derivation of the State Evolution equations from the message passing and without assuming a priori matched teacher and student, as done in \cite{Gabrie2019}.

\begin{figure}[t]
    \centering
    {\includegraphics[width=0.5\textwidth]{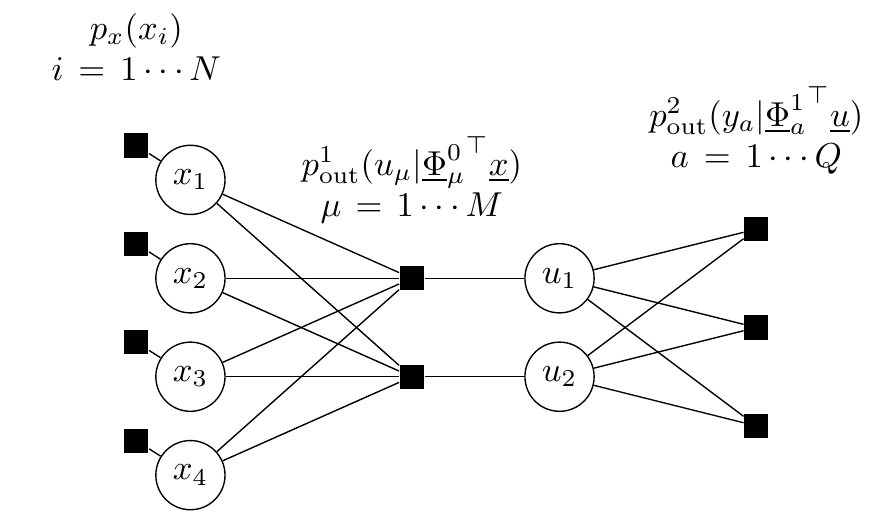}
    }
    % \subfloat[blu]{\includegraphics[width=0.35\textwidth]{}
    \caption{Factor graph representation of a generic 2-layer GLM. \label{fig:chap6-mlglm}}
\end{figure}

\subsection{Model composition and multi-layer inference}
% \missing{Add the reference to multi-layer free energy of our papers and MLVAMP}
Another recent and ongoing direction of extension of mean-field methods is the combination of solutions of elementary models to tackle more sophisticated inference questions. The graphical representations of probability distributions (reintroduced briefly in \citeapp~\ref{app:chap2-graphs}) are here of great help. In a complicated joint probability distribution, it is sometimes possible to identify well-known sub-models, such as the GLM or the RBM. Understanding how and when it is justified to plug-in different solutions is of course non-trivial and a very promising direction of research. 

A particularly relevant extension in this direction is the treatment of multi-layer GLMs, or in other words multi-layer neural networks. With depth $L$,  hidden layers noted $\vect{u}^\ell \in \R^{N_\ell}$, and weight matrices $\mat{\Phi}^\ell \in \R^{N_{\ell +1} \times N_\ell}$, it formally corresponds to the statistical model
\begin{gather}
    \vect{u}^0 = \x_0 \sim p_{x_0}(\x_0) \, ,\\
    \vect{u}^\ell \sim \pout^\ell(u^\ell | \mat{\Phi}^{\ell-1} \vect{u}^{\ell -1}) \quad \forall \ell = 1 \cdots L -1 \, , \\
    \y \sim \pout^{L}(\y | \mat{\Phi}^{L-1} \vect{u}^{L -1} ).
\end{gather}
In \cite{Manoel2017b} a multi-layer version of AMP is derived, assuming Gaussian i.i.d weight matrices, along with a State Evolution and a replica free energy. Remarkably, the asymptotic replica prediction was mathematically proven correct in \cite{Gabrie2018}.  In \cite{Fletcher2018a}, the multi-layer version of the VAMP algorithm is derived with the corresponding State Evolution for orthogonally invariant weight matrices. The matching free energies were obtained independently in \cite{Gabrie2018} by the generalization of a replica result and by \cite{Reeves2016} from a different argument.  

In the next paragraph we sketch a derivation of the 2-layer AMP presented in \citealg~\ref{alg:chap6-amp-2layer}, it provides a good intuition of the composition ability of mean-field inference methods.

\paragraph{Heuristic derivation of 2-layer AMP}

% For instance, in \citechap~\ref{sec:chap4}, we perform an ad hoc combination of the RBM and GLM model to design a practical algorithm operating on arbitrary (real) data sets, albeit not yet analyzable. 
% An instead theoretically grounded extension in this direction is the treatment of multi-layer GLMs, or in other words multi-layer neural networks. In \cite{Manoel2017b} a multi-layer version of AMP is derived, assuming Gaussian i.i.d weight matrices, along with a State Evolution and a free energy. In \cite{Fletcher2018a}, the multi-layer version of the VAMP algorithm is derived with the corresponding State Evolution for orthogonally invariant weight matrices.
% In \citechap~\ref{sec:chap5}, we derive the corresponding replica free energy.
% Yet another example, in \citechap~\ref{sec:chap6}, we derive AMP and SE while combining a calibration problem with a GLM. The derivation presented there will follow the lines of ML-AMP although in a slightly more general case.
The derivation of the multi-layer AMP follows identical steps to the derivation of the single layer presented in \citesec~\ref{sec:chap3-bp-to-amp},  yet for a more complicated factor graph and consequently a larger collection of messages. Without conducting the lengthy procedure, one can form an intuition for the resulting algorithm starting from the single-layer AMP. 
The corresponding factor graph is given on \citefig~\ref{fig:chap6-mlglm}. Compared to the single-layer case (see \citefig~\ref{fig:chap3-glm}), an interface with a set of $M=N_1$ hidden variables $u_\mu$ is inserted between the $N=N_0$ signals $x_i$ and the $Q=N_2$ observations $y_a$. In the neighborhood of the inputs $x_i$ the factor graph is identical to the single-layer and the input functions can be defined from a normalization partition identical to \eqref{eq:chap3-Zx}, 
\begin{gather}
    \cZ^x(\lambda_i, \sigma_i) = \int \dd{x_i} p_x(x_i)e^{-\frac{ (x_i-\lambda_i)^2}{2 \sigma_i} },
    % \xh_i %& 
    % = f^x_1\left( \lbdo_i, \sigo_i   \right) ,\\
    % \Cx_i%&
    % = f_2^x\left( \lbdo_i, \sigo_i   \right) \\
\end{gather}
yielding updates \eqref{alg:chap6-vect-amp-2layer-xi}-\eqref{alg:chap6-vect-amp-2layer-Cxi}. Similarly, the neighborhood of the observations $y_a$ is also unchanged and the updates \eqref{alg:chap6-vect-amp-2layer-goutii} and \eqref{alg:chap6-vect-amp-2layer-dgoutii} are following from the definition of
\begin{gather}
    \Zout^y(\omega^2_a, V^2_a) =  \int \dd{z_a} \; \pout^2(\y_a|z_a) \;  e^{- \frac{\left(z_a - \omega^2_a\right)^2}{2V^2_a }} ,
\end{gather}
identical to the single layer \eqref{eq:chap3-Zout}. At the interface however, the variables $u_\mu$ play the role of outputs for the first GLM and of inputs for the second GLM, which translates into a normalization partition function of mixed form
\begin{gather}
    \label{eq:chap6-2layer-Zout-t}
    \Zout^u(\omega^1_\mu, V^1_\mu, \lambda^1_\mu, \sigma^1_\mu )  = \int \dd{z_\mu} \int \dd{u_\mu} \pout^1(u_\mu | z_\mu) \; \times 
     e^{- \frac{\left(u_\mu - \lambda^1_\mu\right)^2}{ 2 \sigma^1_\mu} } e^{-\frac{\left(z_\mu - \omega^1_\mu\right)^2}{2 V^1_\mu}}. \notag
    % & =  \int \dd{z} \; {\pout^1}^*(z)} \;  e^{\frac 1 2 \left(z - \omega^1_\mu\right)\T{V^1_\mu}^{-1}\left(z - \omega^1_\mu\right)} \\
    % & =  \int \dd{u_\mu}\; \mathbf{p_t^*(u)} \; e^{- \frac 1 2 \left(u_\mu - \lambda^1_\mu\right)\T {\sigma^1_\mu}^{-1} \left(u_\mu - \lambda^1_\mu\right)}
\end{gather}
Updates \eqref{alg:chap6-vect-amp-2layer-gouti} and \eqref{alg:chap6-vect-amp-2layer-dgouti} are obtained by considering that the second layer acts as an effective channel for the first layer, i.e. from the normalization interpreted as
\begin{gather}
    \Zout^u(\omega^1_\mu, V^1_\mu, \lambda^1_\mu, \sigma^1_\mu ) =  \int \dd{z_\mu} \; \pout^{\rm eff}(z_\mu) \;  e^{- \frac{\left(z_\mu - \omega^1_\mu\right)^2}{2 V^1_\mu}} .
    % \\
    % \gout^1(\omega^1_\mu, V^1_\mu, \lambda^1_\mu, \sigma^1_\mu ) = \partial_{\w} \log \Zout^u \quad \text{ and } \quad
    % \dgout^1 = \partial_{\w} \gout.
\end{gather}
Finally, update equations \eqref{alg:chap6-vect-amp-2layer-t} and \eqref{alg:chap6-vect-amp-2layer-Cti} are in turn derived considering the first layer defines an effective prior for the hidden variables and rewriting the normalization as
\begin{gather}
    \Zout^u=  \int \dd{u_\mu}\; p_u^{\rm eff}(u_\mu) \; e^{- \frac{\left(u_\mu - \lambda^1_\mu\right)^2} {2 \sigma^1_\mu}}.
\end{gather}
The rest of the algorithm updates follows as usual from the self-consistency between the different variables introduced as they correspond to different parametrization of the same marginals. The schedule of updates and the time indexing reported in \citealg~\ref{alg:chap6-amp-2layer} results from the entire derivation starting from the BP messages. The generalization of the algorithm to an arbitrary number of layers is easily obtained repeating the heuristic arguments presented here. 

\begin{algorithm*}[t]
\caption{Generalized Approximate Message Passing for the 2-layer GLM \label{alg:chap6-amp-2layer}}   
\begin{algorithmic}
    \State {\bfseries Input:} vector $\y \in \R^{M}$ and matrices $\Phio \in \R^{M \times N}$, $\Phii \in \R^{Q \times M}$:
    \State \emph{Initialize}: $\xh \in \R^N$, $\vect{C}^x\in \R^N$, $\vect{\hat{u}}\in \R^M$, $\vect{C}^u\in \R^M$, $\gouti\in \R^M$, $\vect{\partial \gouts}^1\in \R^M$, $\goutii\in \R^Q$, $\vect{\partial \gouts}^2 \in \R^Q$ and $t=0$.
    \Repeat 

    \Statex 1) Update auxiliary variables of second layer:
    \vspace{-0.3cm}
    \begin{gather}
        \label{alg:chap6-vect-amp-2layer-wii}
        {\omega^{2}_a}^{(t)}%& 
        = \sum_{\mu} \frac{{\Phi^{1}_{a \mu}}}{\sqrt{N}}\hat{u}_\mu^{(t)} - \sum_{\mu} \frac{{({\Phi^{1}_{a \mu}})^2}}{N} {C^u_\mu}^{(t)}{{\gouts^2_a}^{(t-1)}} \\
        \label{alg:chap6-vect-amp-2layer-Vii}
        {V^2_a}^{(t)}%& 
        = \sum_{\mu} \frac{{({\Phi^{1}_{a \mu}})^2}}{N} {C^u_\mu}^{(t)}\\
        \label{alg:chap6-vect-amp-2layer-goutii}
        {\gouts^2_a}^{(t)} = \gouts^2(\y_a, {\omega^{2}_a}^{(t)}, {V^2_a}^{(t)} ) \\
        \label{alg:chap6-vect-amp-2layer-dgoutii}
        {\dgouts^2_a}^{(t)} = \dgouts^2(\y_a, {\omega^{2}_a}^{(t)}, {V^2_a}^{(t)} )\\
        % \msg{\Bii}{a}{\mu} %& 
        % = \frac{{\Phi^{1}_{a \mu}}}{\sqrt{N}} \gouts^2(\x_a, \msg{\omega^{2}}{a}{\mu}, \msg{{V^2}^{-1}}{a}{\mu}) \\
    % \end{gather}
    %
    % \begin{gather}
        \label{alg:chap6-vect-amp-2layer-lbdi}
        {\lambda^1_\mu}^{(t)}  %&
        = \sigma^1_\mu\left( \sum_{a} \frac{{\Phi^{1}_{a \mu}}}{\sqrt{N}} {\gouts^2_a}^{(t)} - \frac{{({\Phi^{1}_{a \mu}})^2}}{N} {\dgouts^2_a}^{(t)} \hat{u}_\mu \right)\\
        \label{alg:chap6-vect-amp-2layer-sigi}
        {\sigma^1_\mu}^{(t)}%&
        = \left(- \sum_{a} \frac{{({\Phi^{1}_{a \mu}})^2}}{N} {\dgouts^2_a}^{(t)} \right)^{-1}    
    \end{gather}
    \Statex 1) Update auxiliary variables of first layer:
    \vspace{-0.3cm}
    \begin{gather}
        \label{alg:chap6-vect-amp-2layer-wi}
        {\omega^1_\mu}^{(t)} %& 
        = \sum_{i} \frac{{\Phi^0_{\mu i}}}{\sqrt{N}}\hat{x}_i^{(t)} - \sum_{i} \frac{{{(\Phi^{0}_{\mu i})}^2}}{N}  {C_x}_i  {\gouts^1_\mu}^{(t-1)} \\
        \label{alg:chap6-vect-amp-2layer-Vi}
        {V^1_\mu}^{(t)}%& 
        = \sum_{i} \frac{{{(\Phi^{0}_{\mu i})}^2}}{N} {{C_x}_i}^{(t)} \\
        % \msg{\Bi}{\mu}{i}  %& 
        % = \frac{{\Phi^0_{\mu i}}}{\sqrt{N}} \gouti(\msg{\wi}{\mu}{i}, \msg{\Vi}{\mu}{i}, \lambda^1_\mu, \sigma^1_\mu) \\
        \label{alg:chap6-vect-amp-2layer-gouti}
        {\gouts^1_\mu}^{(t)} = \gouts^1({\omega^1_\mu}^{(t)}, {V^1_\mu}^{(t)}, {\lambda^1_\mu}^{(t)}, {\sigma^1_\mu}^{(t)} ) \\
        \label{alg:chap6-vect-amp-2layer-dgouti}
        {\dgouts^1_\mu}^{(t)} = \dgouts^1({\omega^1_\mu}^{(t)}, {V^1_\mu}^{(t)}, {\lambda^1_\mu}^{(t)}, {\sigma^1_\mu}^{(t)} )\\
        \label{alg:chap6-vect-amp-2layer-sigo}
        {\sigma^0_i}^{(t)}%&
        = \left(- \sum_{i} \frac{{{(\Phi^{0}_{\mu i})}^2}}{N} {\dgouts^1_\mu}^{(t)} \right)^{-1}  \\
        \label{alg:chap6-vect-amp-2layer-lbdo}
        \lambda^0_i  %&
        = \sigma^0_i\left( \sum_{\mu} \frac{{\Phi^0_{\mu i}}}{\sqrt{N}} {\gouts^1_\mu}^{(t)} - \frac{{{(\Phi^{0}_{\mu i})}^2}}{N} {\dgouts^1_\mu}^{(t)} \hat{x}_i \right) %\\
    \end{gather}
    \Statex 3) Update means and variances of variables of both layers, $\x$ and $\tu$:
    \vspace{-0.3cm}
    \begin{gather}
        \label{alg:chap6-vect-amp-2layer-xi}
        \hat{x}_i^{(t+1)} %& 
        = f^x_1\left( {\lambda^0_i}^{(t)}, {\sigma^0_i}^{(t)}   \right) \\
        \label{alg:chap6-vect-amp-2layer-Cxi}
        {{C^x}_i}^{(t+1)}%&
        = f_2^x\left({\lambda^0_i}^{(t)}, {\sigma^0_i}^{(t)}     \right) \\
    % \end{gather}
    %
    % \begin{gather}
        \label{alg:chap6-vect-amp-2layer-t}
        \hat{u}_\mu^{(t+1)} %&
         = f^u_1 \left({\omega^1_\mu}^{(t)}, {V^1_\mu}^{(t)}, {\lambda^1_\mu}^{(t)}, {\sigma^1_\mu}^{(t)}  \right) \\
         \label{alg:chap6-vect-amp-2layer-Cti}
        {C^u_\mu}^{(t+1)}%&
        = f^u_2\left( {\omega^1_\mu}^{(t)}, {V^1_\mu}^{(t)}, {\lambda^1_\mu}^{(t)}, {\sigma^1_\mu}^{(t)}   \right) %\\
    \end{gather}

    \vspace{0.05cm}
    $t = t+1$
    \vspace{0.05cm}
    \Until{convergence} 
    \State {\bfseries Output:} signal estimate $\xh \in \R^N$, and estimated  variance $\vect{C}^x \in \R^{N} $
\end{algorithmic}
\end{algorithm*}

\section{Some applications}
\label{sec:chapex-all}
\subsection{A brief pre-deep learning history}
\label{sec:chap1-nn-and-mf}
% En gros ici on veut justifier que c'est pertinent d'utiliser ce genre de technique a.k.a. 
% With these first elements of context, 

The application of mean-field methods of inference to machine learning, and in particular to neural networks, already have a long history and significant contributions to their records. Here we briefly review some historical connections anterior to the deep learning revival of neural networks in the 2010s.

\paragraph{Statistical mechanics of learning} In the 80s and 90s, a series of works pioneered the analysis of learning with neural networks through the statistical physics lense.
By focusing on simple models with simple data distributions, and relying on the mean-field method of replicas, these papers managed to predict quantitatively important properties such as \emph{capacities}: the number of training data point that could be memorized by a model, or \emph{learning curves}: the generalization error (or population risk) of a model as a function of the size of the training set. This effort was initiated by the study of the Hopfield model \cite{Amit1985}, an undirected neural network providing associative memory \cite{Hopfield1982}. The analysis of feed forward networks with simple architectures followed (among which \cite{Gardner1987, Gardner1988, Opper1991,  Monasson1995, Opper1996,  Monasson2004}, see also the reviews \cite{Seung1992,Watkin1993, Opper1995, Saad1999a, Engel2001}). The dynamics of simple learning problems was also analyzed through a mean-field framework (not covered in the previous sections) initially in the simplifying case of online learning with infinite training set \cite{Saad1995, Saad1995a, Biehl1995, Saad1999a} but also with finite data \cite{Sollich1997, Li1999}.

Physicists, accustomed to studying natural phenomena, fruitfully brought the tradition of modelling to their investigation of learning, which translated into assumptions of random data distributions or teacher-student scenarios. Their approach was in contrast to the focus of the machine learning theorists on worst case guarantees: bounds for an hypothesis class that hold for any data distribution (e.g. Vapnik-Chervonenkis dimension and Rademacher complexity). 
The originality of the physicists approach, along with the heuristic character of the derivations of mean-field approximations, may nonetheless explain the minor impact of their theoretical contributions in the machine learning community at the time.

\paragraph{Mean-field algorithms for practictioners} 
Along with these contributions to the statistical mechanics theory of learning, new practical training algorithms based on mean-field approximations were also proposed at the same period (see e.g.\cite{Wong1995,Opper1996,Wong1997}). 
Yet, before the deep learning era, mean-field methods probably had a greater influence in the practice of unsupervised learning through density estimation, where we saw that approximate inference is almost always necessary. In particular the simplest method of naive mean-field, our first example in \citechap~\ref{sec:chap3}, was easily adopted and even extended by statisticians (see e.g. \cite{Wainwright2008} for a recent textbook and \cite{Blei2017} for a recent review).  The belief propagation algorithm is another example of a well known mean-field methods by machine learners, as it was actually discovered in both communities. 
% Impossible to cover all the bridges and applications 
Yet, for both methods, early applications rarely involved neural networks and rather relied on simple probabilistic models such as mixtures of elementary distributions.
They also did not take full advantage of the latest simultaneous developments in statistical physics of the mean-field theory of disordered systems.

\paragraph{Transferring advanced mean-field methods}
In this context, the inverse Ising problem has been a notable exception. The underlying question, rooted in theoretical statistical physics, is to infer the parameters of an Ising model given a set of equilibrium configurations. This is related to the unsupervised learning of the parameters of a Boltzmann machine (without hidden units) in the machine learning jargon, while it does not necessarily rely on a maximum likelihood estimation using gradients. The corresponding Boltzmann distribution, with pairwise interactions, is remarkable, not only to physicists. It is the least biased model under the assumption of fixed first and second moments in the sense that it maximizes the entropy. For this problem, physicists proposed dedicated developments of advanced mean-field methods for applications in other fields, and in particular in bio-physics (see \cite{Nguyen2017} for a recent review). A few works even considered the case of Boltzmann machines with hidden units, more common in the machine learning community \cite{Peterson1987,Galland1993}.

Beyond the specific case of Boltzmann machines, the language barrier between communities is undoubtedly a significant hurdle delaying the global transfer of developments in one field to the other. 
% A second potential issue is the apparent restricted span of applicability of some of the mean-field methods.  originally derived following model-specific procedures. 
In machine learning, the potential of the most recent progress of mean-field approximations was advocated for in a pioneering workshop mixing communities in 1999 \cite{opper2001advanced}. Yet the first widely-used application is possibly the Approximate Message Passing (AMP) algorithm for compressed sensing in 2009 \cite{Donoho2009}.
%  - related to the belief propagation algorithm and even more closely to the TAP equations known in the physics of disordered systems.
% (as clarified in \citechap~\ref{sec:chap3}). 
Meanwhile, in the different field of Constraint Satisfaction Problems (CSPs), there have been much tighter connections between developments in statistical physics and algorithmic solutions. The very first popular application of advanced mean-field methods outside of physics, beyond naive mean-field and belief propagation, is probably the survey propagation algorithm \cite{Mezard2002} in 2002. It borrows from the 1RSB cavity method (not treated in the present paper) to solve efficiently certain types of CSPs.

\subsection{Some current directions of research}
\label{sec:chapex}

The great leap forward in the performance of machine learning with neural networks brought by deep learning algorithms, along with the multitude of theoretical and practical challenges it has opened, has re-ignited the interest of physicists for the theory of neural networks. 
% While the power of advanced mean-field theories was mostly unexploited in this context a few years ago, there has been over the last couple of years a number of theoretical and practical contributions. 
In this \citesec, far from being exhaustive, we review some current directions of research leveraging mean-field approximations. Another relevant review is \cite{Carleo}, which provides references both for machine learning research helped by physics methods and conversely research in physics using machine learning.

Works presented below do not necessarily implement one of the classical inference methods presented in \citesecs~\ref{sec:chap3} and \ref{sec:chap3further}. In some cases, the mean-field limit corresponds to some asymptotic setting where the problem simplifies: typically some correlations weaken, fluctuations are averaged out by concentration effects and, as a result, ad-hoc methods of resolution can be designed. Thus, in the following contributions, different assumptions are considered to serve different objectives. For instance some take an infinite size limit, some assume random (instead of learned) weights or vanishing learning rates. Hence, there is no such a thing as one mean-field theory of deep neural networks. The below cited works are rather complementary pieces of solving a great puzzle.

% \missing{Not only the mean-field methods presented above, also some self averaging argument in the mean field limit Heuristic arguments in the mean field limits - understood as some concentration in some high dimensional limit. Can be very powerful, and not require the methods we introduced above. Inspired as they follow overlaps. And finally, also some more complicated theory DMFT}.

\subsubsection{Neural networks for unsupervised learning}

\paragraph{Fundamental study of learning}
Given their similarity with the Ising model, Restricted Boltzmann Machines have unsurprisingly attracted a lot of interest.  Studying an ensemble of RBMs with random parameters using the replica method, Tubiana and Monasson \cite{Tubiana2017} evidenced different regimes of typical pattern of activations in the hidden units and identified control parameters as the sparsity of the weights, their strength (playing the role of an effective temperature) or the type of prior for the hidden layer. Their study contributes to the understanding of the conditions under which the RBMs can represent high-order correlations between input units, albeit without including data and learning in their model. 
Barra and collaborators \cite{Barra2017,Barra2018}, exploited the connections between the Hopfield model and RBMs to characterize RBM learning understood as an associative memory.  Relying again on replica calculations, they characterize the retrieval phase of RBMs.
Mézard \cite{Mezard2017} also re-examined retrieval in the Hopfield model using its RBM representation and message passing, showing in particular that the addition of correlations between memorized patterns could still allow for a mean-field treatment at the price of a supplementary hidden layer in the Boltzmann Machine representation. This result remarkably draws a theoretical link between correlations in the training data and the necessity of depth in neural network models.

While the above results do not include characterization of the learning driven by data, a few others were able to discuss the dynamics of training. Huang \cite{Huang2017} studied with the replica method and TAP equations the Bayesian leaning of a RBM with a single hidden unit and binary weights.
Barra and collaborators \cite{Barra2017} empirically studied a teacher-student scenario of unsupervised learning by maximum likelihood on samples of an Hopfield model which they could compare to their theoretical characterization of the retrieval phase. 
Decelle and collaborators \cite{Decelle2017,Decelle2018} introduced an ensemble of RBMs characterized by the spectral properties of the weight matrix and derived the typical dynamics of the corresponding order parameters during learning driven by data. Beyond RBMs, analyses of the learning in other generative models are starting to appear \cite{Wang2018}.

\paragraph{Training algorithm based on mean-field methods}
Beyond bringing theoretical insights, mean-field methods are also found useful to build tractable estimators of the likelihood in generative models, which in turn serves to design novel training algorithms. 

For Boltzmann machines, this direction was already investigated in the 80s and 90s, \cite{Peterson1987,Hinton1989,Galland1993,Kappen1998}, albeit in small models with binary units and for artificial data sets very different from modern machine learning benchmarks. More recently, a deterministic training based on naive mean-field was tested on RBMs \cite{Welling2002,Tieleman2008}. On toy deep learning data sets, the algorithm was found to perform poorly when compared to both CD and PCD, the commonly employed approximate Monte Carlo methods.
However going beyond naive mean-field, considering the second order TAP expansion, allows to bridge the gap in efficiency \cite{Gabrie2015, Tramel2018} . Additionally, the deterministic mean-field framework offers a tractable way of evaluating the learning success by exploiting the mean-field observables to visualize the representation learned by RBMs. Interestingly, high temperature expansions of estimators different from the maximum likelihood have also been recently proposed as efficient inference method for the inverse Ising problem \cite{Lokhov2018}.

% In the machine learning community fast approximate Monte Carlo methods, specifically contrastive divergence (CD)\cite{Hinton2002} and persistent CD (PCD) \cite{Neal1992b,Tieleman2008}, have made large-scale training of RBMs possible. These rather crude methods have popularized RBMs even though the reason of their efficiency remains unclear. We propose to examine an alternative strategy based on mean-field approximations.

% Meanwhile, the recent craze for deep learning also motivated theoretical studies of RBMs following the statistical physics tradition and using similar tools.  

By construction, variational auto-encoders (VAEs) %\cite{Kingma2014, Rezende2014} 
rely on a variational approximation of the likelihood. In practice, the posterior distribution of the latent representation given an input (see \citesec~\ref{sec:chap1-unsupervised}) is typically approached by a factorized Gaussian distribution with mean and variance parametrized by neural networks. The factorization assumption relates the method to a naive mean-field approximation. 

\paragraph{Structured Bayesian priors}
With the progress of unsupervised learning, the idea of using generative models as expressive priors has emerged. 

For reconstruction tasks, in the event where a set of typical signals is available a priori, the latter can serve as a training set to learn a model of the underlying data distribution with a generative model. Subsequently, in the reconstruction of a new signal of the same type, the generative model can serve as a Bayesian prior.
In particular, the idea to exploit RBMs in CS applications was pioneered by \cite{Dremeau2012} and \cite{Tramel2015}, who trained binary RBMs using Contrastive Divergence (to locate the support of the non-zero entries of sparse signals) and combined it with an AMP reconstruction. They demonstrated drastic improvements in the reconstruction with structured learned priors compared to the usual sparse unstructured priors. The approach, requiring to combine the AMP reconstruction for CS and the RBM TAP inference, was further generalized in \cite{Tramel2016, Tramel2018} to real valued distributions. In the line of these applications, several works have also investigated using feed forward generative models for inference tasks.  Using this time multi-layer VAMP inference, Rangan and co-authors \cite{Pandit2019} showed that VAEs could help for in-painting partially observed images. 
Note also that a different line of works, mainly considering GANs, examined the same type of applications without resorting to mean-field algorithms \cite{Bora2017, Hand2018, Hand2018a, Mixon2018}. Instead they performed the inference via gradient descent and back-propagation.

Another application of generative priors is to model synthetic data sets with structure. In \cite{Gabrie2018, Aubin2019, Goldt2019a}, the authors designed learning problems amenable to a mean-field theoretical treatment  by assuming the inputs to be drawn from a generative prior (albeit with untrained weights so far). This approach goes beyond the vanilla teacher-student scenario where input data is typically unstructured with i.i.d. components. This is a crucial direction of research as the role of structure in data appears as an important component to understand the puzzling efficiency of deep learning. 

\subsubsection{Neural networks for supervised learning}

\paragraph{New results in the replica analysis of learning}
The classical replica analysis of learning with simple architectures, following bases set by Gardner and Derrida 30 years ago, continues to be explored. Among the most prominent results, Kabashima and collaborators \cite{Kabashima2008, Shinzato2008, Shinzato2009} extended the mean-field treatment of the perceptron from data matrices with i.i.d entries to random orthogonal matrices. It is a much larger class of random matrices where matrix entries can be correlated.
More recently, a series of works explored in depth the specific case of the perceptron with binary weight values for classification on random inputs. 
Replica computations showed that the space of solutions is dominated in the thermodynamic limit by isolated solutions \cite{Huang2013, Huang2014}, but also that subdominant dense clusters of solutions exist with good generalization properties in the teacher-student scenario case \cite{Baldassi2015, Baldassi2016, Baldassi2018}. This observation inspired a novel training algorithm \cite{Chaudhari2017a}. 
The simple two-layer architecture of the committee machine was also reexamined recently \cite{Aubin2018}. In the teacher-student scenario, a computationally hard phase of learning was evidenced by comparing a message passing algorithm (believed to be optimal) and the replica prediction. In this work, the authors also proposed a strategy of proof of the replica prediction. 

\paragraph{Signal propagation in depth}
Mean-field approximations can also help understand the role and implications of depth by characterizing signal propagation in neural networks. The following papers consider the width of each layer to go to infinity. In this limit, Sompolinsky and collaborators characterized how neural networks manage to progressively separate data manifolds fed as inputs \cite{Kadmon2016, Chung2018a, Cohen2019}. Another line of works focused on the initialization of neural networks (i.e. with random weights), and found an order-to-chaos transition in the signal propagation as a function of hyperparameters of training \cite{Poole2016,Schoenholz2017}. As a result, the authors could formulate recommendations for combinations of hyperparameters to practitioners. This type of analysis could furthermore be generalized to convolutional networks \cite{Novak2019}, recurrent networks \cite{Gilboa2019} and networks with batch-normalization regularization \cite{Yang2019}. The space of functions spanned by deep random networks in the infinite-size limit was also studied by \cite{Li2018,Li2019}, using the different but related approach of the generating functional analysis.
Yet another mean-field argument, this time relying on a replica computation, allowed to compute the mutual information between layers of large non-linear deep neural networks with orthogonally invariant weight matrices \cite{Gabrie2018}. Using this method, mutual informations can be followed precisely along the learning for an appropriate teacher-student scenario. The strategy offers an experimental test bed to characterize possible links between the generalization ability of deep neural networks and information compression phases in the training (see \cite{Tishby2015, Shwartz2017, Saxe2018}).

\paragraph{Dynamics of SGD learning in simple networks and generalization}
% \review{Un peu trop rien à voir avec le schmilblic ?}
A number of different mean-field limits led to interesting analyses of the dynamics of gradient descent learning. In particular, the below mentioned works contribute to shed light on the generalization power of neural networks in the so-called overparametrized regimes, that is where the number of parameters exceeds largely either the number of training points or the underlying degrees of freedom of the teacher rule.
In linear networks first, an exact description in the high-dimensional limit was obtained for the teacher-student setup by \cite{Advani2017} using random matrix theory. The generalization was predicted to improve with the overparametrization of the student. 
Non-linear networks with one infinitely wide hidden layer were considered by \cite{Mei2018, Rotskoff2018, Chizat2018a, Sirignano2018} who showed that gradient descent converges to a finite generalization error. 
Their results are related to others obtained in a slightly different limit of infinitely large neural networks \cite{Jacot2018}. 
For arbitrarily deep networks, Jacot and collaborators \cite{Jacot2018} showed that, in a certain setting, gradient descent was effectively performing a kernel regression with a kernel function converging to a fixed value for the entire training as the size of the layers increases. 
In both related limits, the absence of divergence is accounting for generalization not deteriorating despite of the explosion of the number of parameters. The relationship between the two above limits was discussed in \cite{Chizat2018, Mei2019, Geiger2019a}. 
Subsequent works, leveraged the formalism introduced in \cite{Jacot2018}. Scaling for the generalization error as a function of network sizes were derived by \cite{Geiger2019}. Other authors focused on the characterization of the network output function in this limit, which takes the form of a Gaussian process \cite{Lee2019}. This fact was probably first noticed by Opper and Winther with one hidden layer \cite{Opper99}, to whom it inspired a TAP based Bayesian classification method using Gaussian processes. 
Finally, yet another limit was analyzed by \cite{Goldt2019}, considering a finite number of hidden units with an infinitely wide input. Following classical works on the mean-field analysis of online learning (not covered in the previous sections \cite{Saad1995, Saad1995a, Biehl1995, Saad1999a}), a closed set of equations can be derived and analyzed for a collection of overlaps. Note that these are the same order parameters as in replica computations. The resulting learning curves evidence the necessity of multi-layer learning to observe the improvement of generalization with overparametrization. An interplay between optimization, architecture and data sets seems necessary to explain the phenomenon.

\section{Conclusion}
\label{sec:conclu}
% \addcontentsline{toc}{section}{Conclusion}%
% \lettrine[lraise=0.05, nindent=0.2em, slope=-.5em]{D}{eep} Learning}

% \vspace{1cm}
This review aimed at presenting in a pedagogical way a selection of inference methods coming from statistical physics. In past and current lines of research that were also reviewed, these methods are sometimes turned into practical and efficient inference algorithms, or sometimes the angle stone in theoretical computations.

\textbf{What is missing}
There are more of these methods beyond what was covered here. In particular the cavity method \cite{Mezard1986}, closely related to message passing algorithms and the replica formalism, played a crucial role in the physics of spin glasses. Note also that we assumed replica symmetry, which is only guaranteed to be correct in the Bayes optimal case. References of introductions to replica symmetry breaking are \cite{Mezard1986, Castellani2005}, and newly proposed message passing algorithms with RSB are \cite{Saglietti2019, Antenucci2019, Antenucci2019a}.
The methods of analysis of online learning algorithms pioneered by \cite{Saad1995, Saad1995a, Biehl1995} and reviewed in \cite{Saad1999a} also deserve the name of classical mean-field analysis. They are currently actively serving research efforts in deep learning theory \cite{Goldt2019}. Another important method is the off-equilibrium mean-field theory \cite{Crisanti1988, Crisanti1993,Cugliandolo1993}, recently used for example to characterize a specific type of neural networks called graph neural networks \cite{Kawamoto2018} or to study properties of gradient flows \cite{Mannelli2019}.

\textbf{On the edge of validity}
We have also touched upon the limitations of the mean-field approach. To start with, the thermodynamic limit is ignoring finite-size effects. Moreover, different ways of taking the thermodynamic limit for the same problem sometimes lead to different results. Also, necessary assumptions of randomness for weights or data matrices are sometimes in clear contrast with real applications. 

% In \citechap~\ref{sec:chap5}, we had to consider a restricted setting of learning in order to guarantee the accuracy of the replica formula of the entropy. Only the spectrum of the weight matrices was learned, which effectively reduced the number of degrees of freedom. A similar restriction is also included in the proposed direction to study multi-layer Bayesian learning in \citechap~\ref{sec:chap6}. Additionally, for both of these analyses, we can only consider synthetic data sets with specific distributions.

Thus, the temptation to apply abusively results from one field to the other can be a dangerous pitfall of the interdisciplinary approach. We could mention here the characterization of the dynamics of optimization. While physicists have extensively studied Langevin dynamics with Gaussian white noise, the continuous time limit of SGD is unfortunately not an equivalent in the general case. While some works attempt to draw insights from this analogy using strong assumptions (e.g. \cite{Choromanska2015, Jastrzebski2017}), others seek precisely to understand the differences between the two dynamics in neural networks optimization (e.g. \cite{Baity-Jesi2018, Simsekli2019}). 
Alternatively, another good reason to consider the power of mean-field methods lies in the observation rooted in the tradition of theoretical physics that one can learn from models a priori far from the exact neural networks desired, but that retain some key properties, while being amenable to theoretical characterization. For example, \cite{Mannelli2019} studied a high-dimensional non-convex optimization problem inspired by the physics of spin glasses apparently unrelated to neural networks, but gained insights on the dynamics of gradient descent (and Langevin) that is of primal interest. Another example of this surely promising approach is \cite{Wang2018}, who built and analyzed a minimal model of GANs.

Moreover, the possibility to combine well-studied simple settings to obtain a mean-field theory for more complex models, as recently demonstrated in a series of work \cite{Tramel2015, Tramel2016,Tramel2018, Manoel2017b, Fletcher2018a, Gabrie2018, Aubin2019}, constitutes an exciting direction of research that should broaden considerably the limit of applications of mean-field methods.
% \missing{same story with the Lazy training}
% This was our strategy in \citechap~\ref{sec:chap5}, where using a teacher-student scenario and a specific weight constraint we were able to compute entropies in large non-linear neural networks trained with SGD. 

\textbf{Patching the pieces together and going further}
Thus the mean-field approach alone cannot to this day provide complete answers to the still numerous puzzles on the way towards a deep learning theory. Yet, considering different limits and special cases, combining solutions to approach ever more complex models, the approach should help uncover more and more corners of the big black box. Hopefully, intuition gained at the edge will help revealing the broader picture.

\section*{Acknowledgements}  
This paper is based on the introductory chapters of my PhD dissertation, written under the supervision of Florent Krzakala and in collaboration with Lenka Zdeborová, to whom both I am very grateful. I would like to thank also Benjamin Aubin, Cédric Gerbelot, Adrien Laversanne-Finot, and Guilhem Semerjian for their comments on the manuscript. I gratefully acknowledge the support of the `Chaire de recherche sur les modèles et sciences des données' by Fondation CFM pour la Recherche-ENS and of Fondation L'Oréal For Women In Science. I also thank the Kalvi Institute for Theoretical Physics, where part of this work was written.

\appendix
\renewcommand{\thesection}{\Alph{section}}
\section{Index of notations and abbreviations}

\begin{itemize}
\item[] {[N]} - Set of integers from $1$ to $N$
\item[] {$\dirac(\cdot)$} - Dirac distribution
\item[] {$\sigma(x) = (1 + e^{-x})^{-1}$} - Sigmoid
\item[] {$\mathrm{relu}(x)=\max(0,x)$} - Rectified Linear Unit
\item[] {$\mat{X}$} - Matrix
\item[] {$\vect{x}$} - Vector
\item[] {$\mat{I}_N \in \R^{N\times N}$} - Identity matrix
\item[] {$\langle \cdot \rangle$} - Average with respect to the Boltzmann distribution
\item[] {$\mathrm{O}(N) \subset \R^{N\times N}$} - Orthogonal ensemble
% \item[] Neighbors of node $i$}{blu}{$\partial i$}
%
\item[] {1RSB} - 1 Step Replica Symmetry Breaking
\item[] {AMP} - Approximate message passing
\item[] {BP} - Belief Propagation
\item[] {cal-AMP} - Calibration Approximate Message Passing
\item[] {CD} - Contrastive Divergence
\item[] {CS} - Compressed Sensing
\item[] {CSP} - Constrain Satisfaction Problem
\item[] {DAG} - Directed Acyclic Graph
\item[] {DBM} - Deep Boltzmann Machine
\item[] {EC} - Expectation Consistency
\item[] {EP} - Expectation Propagation 
\item[] {GAMP} - Generalized Approximate message passing
\item[] {GAN} - Generative Adversarial Networks 
\item[] {GD} - Gradient Descent
\item[] {GLM} - Generalized Linear Model
\item[] {G-VAMP} - Generalized Vector Approximate Message Passing
\item[] {i.i.d.} - independent identically distributed
\item[] {PCD} - Persistent Contrastive Divergence
\item[] {r-BP} - relaxed Belief Propagation
\item[] {RS} - Replica Symmetric
\item[] {RSB} - Replica Symmetry Breaking
\item[] {RBM} - Restricted Boltzmann Machine
\item[] {SE} - State Evolution
\item[] {SGD} - Stochastic Gradient Descent
\item[] {SK} - Sherrington-Kirkpatrick
\item[] {TAP} - Thouless Anderson Palmer 
\item[] {VAE} - Variational Autoencoder
\item[] {VAMP} - Vector Approximate message passing
\end{itemize}

\section{Statistical models representations}
\label{app:chap2-graphs}
Graphical representations have been developed to represent and efficiently exploit (in)dependencies between random variables encoded in joint probability distributions. They are useful tools to concisely present the model under scrutiny, as well as direct supports for some derivations of inference procedures. Let us briefly present two types of graphical representations. 

\paragraph{Probabilistic graphical models}
Formally, a probabilistic graphical model is a graph $G = (V, E)$ with nodes $V$ representing random variables and edges $E$ representing direct interactions between random variables. In many statistical models of interest, it is not necessary to keep track of all the possible combinations of realizations of the variables as the joint probability distribution can be broken up into factors involving only subsets of the variables. The structure of the connections $E$ reflects this factorization.  

There are two main types of probabilistic graphical models: \emph{directed graphical models} (or Bayesian networks) and \emph{undirected graphical models} (or Markov Random Fields). They allow to represent different independencies and factorizations. In the next paragraphs we provide intuitions and remind some useful properties of graphical models, a good reference to understand all the facets of this powerful tool is \cite{Koller2009}. 
% In the following we abusively use the same notations for nodes and the variables they carry when no ambiguity is possible.

\subparagraph{Undirected graphical models} 
In undirected graphical models
the direct interaction between a subset of variables $C \subset V$ is represented by undirected edges interconnecting each pair in $C$. This fully connected subgraph is called a \emph{clique} and associated with a real positive \emph{potential} function $\psi_C$ over the variable $\x_C=\{x_i\}_{i \in C}$ carried by $C$. The joint distribution over all the variables carried by $V$, $\x_V$  is the normalized product of all potentials
    \begin{gather}
        p(\x_V) = \frac 1 \cZ \prod_{C \in \mathcal{C}} \psi_C(\x_C).
    \end{gather}

% \begin{enumerate}
% \item[
% \begin{minipage}{0.15\textwidth}
% \end{minipage}
\begin{flushright}
\begin{minipage}{0.95\linewidth}
    \hypertarget{item:chap2-rbm}{Example} \hyperlink{item:chap2-rbm}{(i)}: the Restricted Boltzmann Machine,
    \begin{gather}
        \label{eq:chap2-rbm-def}
        p(\x, \hid) = \frac{1}{\cZ} e^{\x\T\W\hidd}p_x(\x)p_t(\hidd)
    \end{gather}
    with factorized $p_x$ and $p_t$ is handily represented using an undirected graphical model depicted in \citefig~\ref{fig:chap2-rbm-bis}. The corresponding set of cliques is the set of all the pairs with one input unit (indexed by $i = 1 \cdots N$) and one hidden unit (indexed by $\alpha = 1 \cdots M$), joined with the set of all single units. The potential functions are immediately recovered from \eqref{eq:chap2-rbm-def},
    \begin{gather}
        \mathcal{C} = \{\{i\}, \{\alpha\}, \{i,\alpha\}\,; \,i=1 \cdots N, \,\alpha = 1 \cdots M \} \, , \qquad
        % _{i=1 \cdots N}^{\alpha = 1 \cdots M} %\\
        \psi_{i\alpha}(x_i, t_\alpha) = e^{x_i W_{i\alpha}\hiddv_\alpha}\, ,\\
        p(\x, \hid) = \frac{1}{\cZ}\prod_{\{i, \alpha\} \in \mathcal{C}}\psi_{i\alpha}(x_i, t_\alpha)\prod_{i=1}^Np_x(x_i)\prod_{\alpha=1}^Mp_t(t_\alpha).
    \end{gather}
    It belongs to the subclass of \emph{pairwise} undirected graphical models for which the size of the cliques is at most two. 
\end{minipage}
\end{flushright}

Undirected graphical models handily encode \emph{conditional independencies}. Let $A, B, S \subset V$ be three disjoint subsets of nodes of $G$. $A$ and $B$ are said to be independent given $S$ if $p(A,B |S) = p(A|S)p(B|S)$. In the graph representation it corresponds to cases where $S$ separates $A$ and $B$: there is no path between any node in $A$ and any node in $B$ that is not going through $S$. 

\begin{flushright}
    \begin{minipage}{0.95\linewidth}
        Example \hyperlink{item:chap2-rbm}{(i)}: In the RBM, hidden units are independent given the inputs, and conversely: 
    \begin{gather}
        p(\hidd | \x) = \prod_{\alpha =1}^M p(\hiddv_\alpha | \x), \qquad
        p(\x|\hidd) = \prod_{i=1}^M p(x_i | \hidd).
    \end{gather}
    This property is easily spotted by noticing that the graphical model (\citefig~\ref{fig:chap2-rbm-bis}) is \emph{bipartite}.
\end{minipage}
\end{flushright}

\begin{figure}[t]
    \centering
    % \captionsetup{width=.3\linewidth}
    % \subfloat[Restricted Boltzmann Machine]{\includegraphics[width=0.4\textwidth, valign=m]{chap2_rbm_GM.pdf}
    % \label{fig:chap2-rbm}}
    % \hspace{1cm}
    \captionsetup{width=.4\linewidth}
    % \subfloat[Feed forward neural network. ]
    {\includegraphics[width=0.3\textwidth, valign=m]{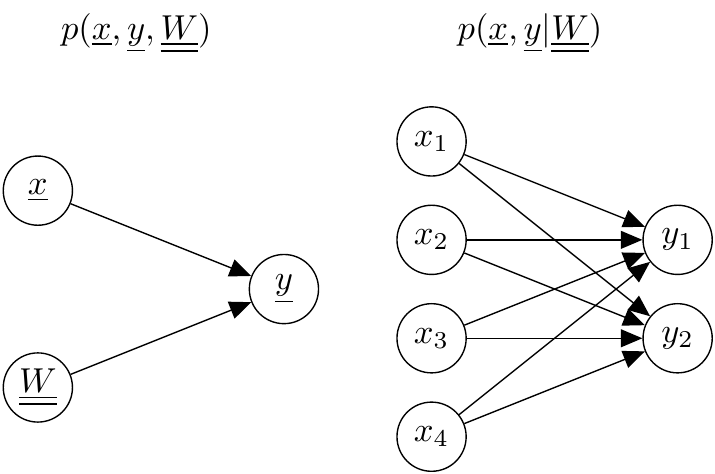}
    }
    \captionsetup{width=.9\linewidth}
    \caption{
        % \textbf{(a)} Undirected probabilistic graphical model (left) and factor graph representation (right). \textbf{(b)} 
        Left: Directed graphical model for $p(\x,\y,\W)$ without assumptions of factorizations for the channel and priors. Right: Directed  graphical model reflecting factorization assumptions for $p(\x,\y|\W)$.\label{fig:chap2-glm}
        %  Right: Corresponding factor graph representation.
        }
\end{figure}
    
\subparagraph{Directed graphical model } A directed graphical model  uses a Directed Acyclic Graph (DAG), specifying directed edges $E$ between the random variables $V$. It induces an ordering of the random variables and in particular the notion of parent nodes  $\pi_i \subset V$  of any given vertex $i \in V$: the set of vertices $j$ such that $j \to i \in E$. The overall joint probability distribution factorizes as
\begin{gather}
    p(\x) = \prod_{i \in V} p(x_i | \x_{\pi_i}).
\end{gather}

\begin{flushright}
    \begin{minipage}{0.95\linewidth}
        \hypertarget{item:chap2-glm}{Example} \hyperlink{item:chap2-glm}{(ii)}: The stochastic single layer feed forward network $\y = g(\W\x ; \eps)$, where $g( \cdot; \eps)$ is a function applied component-wise including a stochastic noise $\eps$ that is equivalent to a conditional distribution $p_{\rm out}(\y | \W\x)$, and where inputs and weights are respectively drawn from distributions $p_x(\x)$ and $p_W(\W)$, has a joint probability distribution
    \begin{gather}
        \label{eq:chap2-glm-def}
        p(\y, \x, \W) = p_{\rm out}(\y | \W\x) p_x(\x) p_W(\W),
    \end{gather}  
     precisely following such a factorization. It can be represented with a three-node DAG as in \citefig~\ref{fig:chap2-glm}. Here we applied the definition at the level of vector/matrix valued random variables. By further assuming that $\pout$, $p_W$ and $p_x$ factorize over their components, we keep a factorization compatible with a DAG representation 
    \begin{gather}
        p(\y, \x, \W) = \prod_{i=1}^{N} p_x(x_i) \prod_{\mu = 1}^M \pout(y_\mu | \sum_{i=1}^N W_{\mu i}x_i) \prod_{\mu, i} p_W(W_{\mu i}).
    \end{gather}
    For the purpose of reasoning it may be sometimes necessary to get to the finest level of decomposition, while sometimes the coarse grained level is sufficient. 
\end{minipage}
\end{flushright}

While a statistical physicist may have never been introduced to the formal  definitions of graphical models, she inevitably already has drawn a few - for instance when considering the Ising model. She also certainly found them useful to guide physical intuitions. The following second form of graphical representation is probably newer to her. 

\paragraph{Factor graph representations}
Alternatively, high-dimensional joint distributions can be represented with factor graphs, that are undirected bipartite graphs  $G = (V, F, E)$ with two subsets of nodes. The variable nodes $V$ representing the random variables as in the previous section (circles in the representation) and the factor nodes $F$ representing the interactions (squares in the representation) associated with potentials. The edge $(i \mu)$ between a variable node $i$ and a factor node $\mu$ exists if the variable $i$ participates in the interaction $\mu$. We note $\partial i$ the set of factor nodes in which variable $i$ is involved, they are the neighbors of $i$ in $G$. Equivalently we note $\partial \mu$ the neighbors of factor $\mu$ in $G$, they carry the arguments $\{x_i\}_{i \in \partial \mu}$, shortened as $\x_{\partial \mu}$, of the potential $\psi_\mu$.  The overall distributions is recovered in the factorized form:
\begin{align}
p(\x) = \frac 1 \cZ \prod_{\mu = 1}^M \psi_\mu(\x_{\partial \mu}).
\end{align}
Compared to an undirected graphical model, the cliques are represented by the introduction of factor nodes.

\begin{flushright}
    \begin{minipage}{0.95\linewidth}
    Examples: The factor-graph representation of the RBM \hyperlink{item:chap2-rbm}{(i)} is not much more informative than the pairwise undirected graph (see \citefig~\ref{fig:chap2-rbm-bis}). 
    For the feed forward neural networks \hyperlink{item:chap2-glm}{(ii)} we draw the factor graph of $p(\y, \x| \W)$ (see \citefig~\ref{fig:chap2-glm-bis}). 
    \end{minipage}
\end{flushright}

\section{Mean-field identity}
We derive the exact identity \eqref{eq:chap3-mf-identity} for fully connected Ising model with binary spins $\x \in \{0,1\}^N$,
\label{app:chap3-mf-identity}
\begin{align}
    \label{eq:app-chap3-1}
    \langle x_i \rangle_p &= \frac{1}{\cZ}
    \sum_{\x\in \{0,1\}^N} \, x_i \, \exp\left(\displaystyle\beta \sum_j b_j x_j + \frac 1 2 \sum_{jk}W_{jk}x_j x_k\right) \\
    &= \frac{1}{\cZ}\sum_{\x_{\setminus i}\in \{0,1\}^{N-1}} \exp\left(\displaystyle\beta \sum_{j\neq i} b_j x_j + \frac 1 2 \sum_{\underset{j\neq i}{k \neq i}}W_{ij}x_i x_j\right) \sum_{x_i \in \{0,1\}} x_i e^{\beta b_i x_i + \sum_j W_{ij}x_i x_j} \notag
    % & = \frac{1}{\cZ}\sum_{\x\in \{0,1\}^N} \, x_i \, e^{\displaystyle\beta \sum_j b_j x_j + \frac 1 2 W_{ij}x_i x_j}  =
\end{align}
where $\x_{\setminus i}$ is the vector of $\x$ without its $i$-th component.
Yet
\begin{gather}
    \sigm(\beta b_i + \sum_{j \in \partial i}\beta W_{ij} x_j) = \frac
    {\sum_{x_i \in \{0,1\}} \, x_i\,  e^{\beta b_i x_i + \sum_j W_{ij}x_i x_j}}
    {\sum_{x_i \in \{0,1\}} e^{\beta b_i x_i + \sum_j W_{ij}x_i x_j}},
\end{gather}
so that multiplying and dividing \eqref{eq:app-chap3-1} by the denominator above we obtain the identity \eqref{eq:chap3-mf-identity} in \citesec~\ref{sec:chap3-nmf}
\begin{align}
    \langle x_i \rangle_p = \langle \sigm(\beta b_i + \sum_{j \in \partial i}\beta W_{ij} x_j) \rangle_p\ .
\end{align}

\section{Georges-Yedidia expansion for generalized Boltzmann machines}
\label{app:chap3-real-GY}

We here present a derivation of the Georges-Yedidia for real-valued degrees of freedom on the example of a Boltzmann machine as in \cite{Tramel2018}.
Formally we consider $\x \in \R^N$ governed by the energy function and parametrized distribution
\begin{gather}
    \label{eq:chap4-real-meas-fully}
    E(\x) = - \sum_{(ij)} W_{ij}x_i x_j - \frac{1}{\beta} \sum_{i=1}^N\log p_x(x_i; \theta_i) \,, \quad
    p(\x) = \frac{1}{\cZ}  e^{\frac{\beta}{2} \x\T\W\x}  \prod_{i=1}^N p_x(x_i ; \theta_i) ,
\end{gather}
where $p_x(x_i;\theta_i)$ is an arbitrary prior distribution with parameter $\theta_i$. For a Bernoulli prior with parameter $\sigm(\beta b_i)$ we recover the measure of binary Boltzmann machines. However we choose here a prior that does not depend on the temperature a priori. We now derive the expansion for this general case following the outline discussed in \ref{sec:chap3-GY}, and highlighting the differences with the binary case.

Note that inference in the generalized fully connected Boltzmann machine is somehow related to the symmetric rank-1 matrix factorization problem, which also features pairwise interactions. Similarly, inference for the bi-partite RBM maps to the asymmetric rank-1 matrix factorization. However, conversely to the Boltzmann inference, these factorizations are reconstruction problems. The mean-field techniques, derived in \cite{Lesieur2016,Lesieur2017}, allow there to compute the MMSE estimator of unknown signals from approximate marginals. Here we focus on the evaluation of the free energy. 

\paragraph{Minimization for fixed marginals}
While fixing the value of the first moment is sufficient for binary variables, more than one constraint is now needed in order to minimize the Gibbs free energy at a given value of the marginals.
In the same spirit of the AMP algorithm we assume a Gaussian parametrization of the marginals. We note $\am$ the first moment of $\x$ and $\cm$ its variance.
We wish to compute the constrained minimum over the distributions $q$ on $\R^N$
\begin{gather}
    G(\am, \cm) = \min_{q} \left[ \langle E(\x) \rangle_{q} - H(q)/\beta \; | \;  \langle \x\rangle_q = \am \,, \langle \x^2\rangle_q = \am^2 + \cm  \right],
\end{gather}
where the notation of squared vectors corresponds here and below to the vectors of squared entries.
It is equivalent to an unconstrained problem with Lagrange multipliers $\lbd(\am,\cm, \beta)$ and $\vect{\xi}(\am,\cm, \beta)$
% \begin{gather}
%     G(\am, \cm) = \min_{q} \left[ \langle E(\x) \rangle_{q} - H(q)/\beta - \sum_{i=1}^Nh_i(\am,\cm, \beta)(\langle x_i\rangle_q - a_i) - \sum_{i=1}^{N}g_i(\am,\cm, \beta)(\langle x_i^2\rangle_q - a_i^2 - c_i)\right].
% \end{gather}
\begin{gather}
    \label{eq:chap4-real-GY-G01}
    G(\am, \cm) = \min_{q} \left[ \langle E(\x) \rangle_{q} - H(q)/\beta - \lbd\T(\langle \x\rangle_q - \am) / \beta - \vect{\xi}(\langle \x^2\rangle_q - \am^2 - \cm)/ \beta \right].
\end{gather}
 The terms depending on the distribution $q$ in the functional to minimize  above can be interpreted as a Gibbs free energy for the effective energy functional
\begin{gather}
    \tilde{E}(\x) = E(\x) -  \lbd\T\x /\beta - \vect{\xi}\T \x^2/\beta .
\end{gather}
The solution of the minimization problem \eqref{eq:chap4-real-GY-G01} is therefore the corresponding Boltzmann distribution 
\begin{gather}
q_{\am, \cm}(\x) = \frac{e^{-\beta\tilde{E}(\x)}}{\tilde{\cZ}} = \frac{1}{\tilde{\cZ}} e^{-\beta E(\x) + \lbd(\am,\cm, \beta)\T\x +  \vect{\xi}(\am,\cm, \beta)\T \x^2 } \, 
\end{gather}
and the minimum $G(\am, \cm)$  is 
\begin{align}
    \label{eq:chap4-real-GY-G1}
    -\beta G(\am, \cm) 
    & = - \lbd\T \am - \vect{\xi}\T(\am^2+\cm) +  \log \int \dd{\x} e^{-\beta E(\x) + \lbd\T\x + \vect{\xi}\T \x^2 } \notag\\
    & =  \log \int \dd{\x} e^{-\beta E(\x) + \lbd\T(\x -\am) + \vect{\xi}\T (\x^2 - \am^2 -\cm)},
\end{align}
where the Lagrange multipliers $\lbd(\am,\cm, \beta)$ and $\vect{\xi}(\am,\cm, \beta)$ enforcing the constraints are still implicit.
Defining a functional $\tilde{G}$ for arbitrary vectors $\tilde{\lbd} \in \R^N$ and $\vect{\tilde \xi} \in \R^N$, 
\begin{gather}
    -\beta \tilde{G}(\am,\cm, \tilde{\lbd}, \tilde{\vect{\xi}}) =  \log \int \dd{\x} e^{-\beta E(\x) + \tilde{\lbd}\T(\x -\am) + \tilde{\vect{\xi}}\T (\x^2 - \am^2 -\cm)}, 
\end{gather}
we have
\begin{align}
    \label{eq:chap4-real-GY-stationary-lbd}
    & a_i = \langle x_i \rangle_{q_{\am,\cm}} \Rightarrow -\beta \left.\frac{\partial\tilde{G}}{\partial \tilde \lambda_i}\right|_{\lbd, \vect{\xi}} = 0, && -\beta \left.\frac{\partial^2\tilde{G}}{\partial \tilde\lambda_i^2}\right|_{\lbd, \vect{\xi}}= \langle x_i^2 \rangle_{q_{\am,\cm}} - a_i^2 > 0 ,\\
    \label{eq:chap4-real-GY-stationary-xi}
    &c_i + a_i^2 = \langle x_i^2 \rangle_{q_{\am,\cm}}  \Rightarrow -\beta \left.  \frac{\partial  \tilde{G}}{\partial \tilde \xi_i}\right|_{\lbd, \vect{\xi}} = 0,  && -\beta \left.\frac{\partial ^2\tilde{G}}{\partial \tilde \xi_i^2} \right|_{\lbd, \vect{\xi}}= \langle (x^2_i)^2 \rangle_{q_{\am,\cm}} - (c_i + a_i^2)^2 > 0 .
\end{align}
Hence the Lagrange multipliers are identified as minimizers of $-\beta\tilde{G}$ and
\begin{gather}
    - \beta G(\am, \cm) = - \beta \tilde{G}(\am,\cm, \lbd(\am,\cm, \beta), \vect{\xi}(\am,\cm, \beta)) = \min_{\tilde{\lbd}, \tilde{\vect{\xi}}} - \beta \tilde{G}(\am,\cm, \tilde{\lbd}, \tilde{\vect{\xi}}).
\end{gather}
The true free energy $F = - \log \cZ / \beta$ would eventually be recovered by minimizing the constrained minimum $G(\am, \cm)$ with respect to its arguments.
% \begin{gather}
%     F = \min_{\am, \cm} G(\am, \cm) = \min_{\am, \cm} \max_{\vect{h}, \vect{g}} \tilde{G}(\am,\cm, \vect{h}, \vect{g}),
% \end{gather}
Nevertheless, the computation of $G$ and $\tilde{G}$ involves an integration over $\x \in \R^N$ and remains intractable. The following step of the Georges-Yedidia derivation consists in approximating these functionals by a Taylor expansion at infinite temperature where interactions are neutralized.

\paragraph{Expansion around $\beta=0$}
To perform the expansion we introduce the notation $A(\beta, \am, \cm) = - \beta G(\am, \cm) $.
% \begin{gather}
%     \lbd(\beta, \am, \cm) = \beta \vect{h}(\am,\cm, \beta) \, , \quad
%     \vect{\xi}(\beta, \am, \cm) = \beta \vect{g}(\am,\cm, \beta) \, , \quad
%     A(\beta, \am, \cm) = - \beta G(\am, \cm) \, . % \quad \tilde{A}(\beta, \am, \cm, \lbd, \vect{\xi}) = - \beta \tilde{G}(\am, \cm, \vect{g}, \vect{h}).
% \end{gather}
We also define the auxiliary operator 
\begin{gather}
    U(\x; \beta) = -\frac{1}{2} \x\T\W\x  + \frac{1}{2}\langle \x\T\W\x \rangle_{q_{\am, \cm}} - \sum_{i=1}^N \frac{\partial \lambda_i}{\partial \beta} (x_i - a_i) - \sum_{i=1}^N \frac{\partial \xi_i}{\partial \beta} (x_i^2 - a_i^2 - c_i), 
\end{gather}
that allows to write concisely for any observable $O$ the derivative of its average with respect to $\beta$, 
\begin{gather}
    \frac{\partial \langle O(\x; \beta) \rangle_{q_{\am,\cm}}}{\partial \beta} = \left\langle \frac{\partial O(\x; \beta)}{\partial \beta} \right\rangle_{q_{\am,\cm}} - \langle U(\x;\beta)O(\x;\beta) \rangle_{q_{\am,\cm}}.
\end{gather}
To compute the derivatives of $\lbd$ and $\vect{\xi}$ with respect to $\beta$ we note that 
\begin{gather}
    \label{eq:chap4-real-GY-derivative-A-a-c}
    \frac{\partial A}{\partial a_i} =  -\beta  \frac{\partial \tilde G}{\partial a_i} =  - \lambda_i(\beta, \am, \cm)- 2 a_i \xi_i(\beta, \am, \cm) \, ,  \\
    \frac{\partial A}{\partial c_i} = -\beta  \frac{\partial \tilde G}{\partial c_i} = - \xi_i(\beta, \am, \cm),
\end{gather}
where we used that $\partial \tilde G / \partial \tilde{\lambda_i} = 0 $ and $\partial \tilde G / \partial \tilde{\xi}_i = 0 $ when evaluated for $\lbd(\am,\cm, \beta)$ and $\vect{\xi}(\am,\cm, \beta)$. Consequently,
\begin{gather}
    \label{eq:chap4-real-GY-deriv-lbd}
    \frac{\partial \xi_i}{\partial \beta} = - \frac{\partial}{\partial c_i} \frac{\partial A }{\partial \beta} \, , \qquad 
    \frac{\partial \lambda_i}{\partial \beta} = - \frac{\partial}{\partial a_i} \frac{\partial A }{\partial \beta} + 2 a_i \frac{\partial \xi_i}{\partial \beta}.
\end{gather}
We can now proceed to compute the first terms of the expansion that will be performed for the functional $A$. 

\subparagraph{Zeroth order}
Substituting $\beta=0$ in the definition of $A$ we have
\begin{gather}
    A(0,\am,\cm) = - \lbd(0,\am,\cm)\T \am - \vect{\xi}(0,\am,\cm)\T (\am^2 + \cm) + \log \tilde{\cZ}^0(\lbd(0,\am,\cm), \vect{\xi}(0,\am,\cm)),
\end{gather}
with 
\begin{align}
    \tilde{\cZ}^0(\lbd(0,\am,\cm), \vect{\xi}(0,\am,\cm)) &= \int \dd{\x} e^{\lbd(0,\am,\cm)\T \x+ \vect{\xi}(0,\am,\cm)\T \x^2}\prod_{i=1}^N p_x(x_i ; \theta_i) \\
        & = \prod_{i=1}^N  \int \dd{x_i} e^{\lambda_i(0,\am,\cm) x_i+ \xi_i(0,\am,\cm) x_i^2}p_x(x_i ; \theta_i).
\end{align}
At infinite temperature the interaction terms of the energy do not contribute so that the integral in $\tilde{\cZ}^0$ factorizes and can be evaluated numerically in the event that it does not have a closed-form.

\subparagraph{First order} We compute the derivative of $A$ with respect to $\beta$. We use again that $\lbd(\am,\cm,\beta)$ and  $\vect{\xi}(\am,\cm,\beta)$ are stationary points of $\tilde{G}$ to write
\begin{align}
    \frac{\partial A}{\partial \beta} & = - \beta \frac{\partial \tilde{G}}{\partial \beta} = \frac{\partial }{\partial \beta} \left[ \log \int \dd{\x} e^{-\beta E(\x) + \lbd(\am,\cm, \beta)\T(\x -\am) + \vect{\xi}(\am,\cm, \beta)\T (\x^2 - \am^2 -\cm)}  \right]  \\
    % & = \frac{1}{\tilde{\cZ}} \int \dd{\x} \left( 
    %     \frac{\partial }{\partial \beta} (-\beta E(\x))
    %     + \frac{\partial \lbd }{\partial \beta}\T (\x -\am) 
    %     + \frac{\partial \vect{\xi}}{\partial \beta}\T (\x^2 - \am^2 -\cm)   \right) e^{-\beta E(\x) + \beta \vect{h}(\am,\cm, \beta)\T\x + \beta \vect{g}(\am,\cm, \beta)\T \x^2 }
     & = \left\langle 
        \frac{\partial }{\partial \beta} (-\beta E(\x))
        + \frac{\partial \lbd }{\partial \beta}\T (\x -\am) 
        + \frac{\partial \vect{\xi}}{\partial \beta}\T (\x^2 - \am^2 -\cm)   \right\rangle_{q_{\am, \cm}} \\
    & =  \frac{1}{2} \langle \x\T\W\x \rangle_{q_{\am, \cm}}.
\end{align}
At infinite temperature the average over the product of variables becomes a product of averages so that we have
\begin{gather}
   \left. \frac{\partial A}{\partial \beta} \right|_{\beta = 0} =  \frac{1}{2}\am\T\W\am =  \sum_{(ij)} W_{ij} a_i a_j .
\end{gather}

\subparagraph{Second order} Using the first order derivative of $A$ we can compute the derivatives of the Lagrange parameters \eqref{eq:chap4-real-GY-deriv-lbd} and the auxillary operator at infinite temperature, 
\begin{gather}
    \left. \frac{\partial \xi_i}{\partial \beta} \right|_{\beta = 0} = 0 \, , \qquad \left. \frac{\partial \lambda_i}{\partial \beta} \right|_{\beta = 0} = - \sum_{j\in\partial i} W_{ij} a_j \,, \qquad  U(\x;0) = - \sum_{(ij)} W_{ij} (x_i - a_i)(x_j - a_j). \notag
\end{gather}
The second order derivative is then easily computed at infinite temperature
\begin{align}
    \left. \frac{\partial^2 A}{\partial \beta^2} \right|_{\beta = 0} & = \frac{1}{2}\left. \frac{\partial}{\partial \beta} \Big(\langle \x\T\W\x \rangle_{q_{\am, \cm}}\Big) \right|_{\beta = 0}= - \frac{1}{2} \langle  U(\x;0) (\x\T\W\x )\rangle^{\beta=0}_{q_{\am, \cm}} \\
    & = \sum_{(ij)} W_{ij}^2 \langle (x_i -a_i)x_i (x_j - a_j)\rangle^{\beta=0}_{q_{\am, \cm}} = \sum_{(ij)} W_{ij}^2 c_i c_j. 
\end{align}

\paragraph{TAP free energy for the generalized Boltzmann machine}
\label{sec:chap4-tap-fe-grbm}
Stopping at the second order of the systematic expansion, and gathering the different terms derived above we have 
\begin{align}
    -\beta G(\am, \cm) = - \lbd(0,\am,\cm)\T \am & - \vect{\xi}(0,\am,\cm)\T (\am^2 + \cm) + \log \tilde{\cZ}^0(\lbd(0,\am,\cm), \vect{\xi}(0,\am,\cm))  \\
    &+ \beta \sum_{(ij)} W_{ij} a_i a_j + \frac{\beta^2}{2}\sum_{(ij)} W_{ij}^2 c_i c_j, \notag
\end{align}
where the values of the parameters $\lbd(0,\am,\cm)$ and $\vect{\xi}(0,\am,\cm)$ are implicitly defined through the stationary conditions \eqref{eq:chap4-real-GY-stationary-lbd}-\eqref{eq:chap4-real-GY-stationary-xi}.
The TAP approximation of the free energy also requires to consider the stationary points of the expanded expression as a function of $\am$ and $\cm$. 

This second condition yields the relations 
\begin{gather}
    \label{eq:chap4-real-GY-A}
    -2 \xi_i (0,\am,\cm) = -\beta^2 \sum_{j\in \partial i } W_{ij}^2 c_j = A_i \\
    \label{eq:chap4-real-GY-B}
    \lambda_i (0,\am,\cm) = A_i a_i + \beta \sum_{j\in \partial i } W_{ij} a_j = B_i \,
\end{gather}
where we define new variables $A_i$ and $B_i$. 
While the extremization with respect to the Lagrange multipliers gives
\begin{gather}
    \label{eq:chap4-real-GY-a}
    a_i = \frac{1}{\cZ^x_i} \int \dd{x_i} x_i p_x(x_i; \theta_i) e^{-\frac{A_i}{2}x_i^2 + B_i x_i} = f_1^x(B_i,A_i;\theta_i), \\
    \label{eq:chap4-real-GY-c}
    c_i = \frac{1}{\cZ^x_i} \int \dd{x_i} x_i^2 p_x(x_i; \theta_i) e^{-\frac{A_i}{2}x_i^2 + B_i x_i} - a_i^2 = f_2^x(B_i,A_i;\theta_i) ,
\end{gather}
where we introduce update functions $f_1^x$ and $f_2^x$ with respect to the partition function 
\begin{gather}
    \cZ^x_i(B_i,A_i;\theta_i) = \int \dd{x_i} p_x(x_i; \theta_i) e^{-\frac{A_i}{2}x_i^2 + B_i x_i} .
\end{gather}
Finally we can rewrite the TAP free energy as
\begin{align}
    \label{eq:chap4-real-GY-final-G}
    -\beta G(\am, \cm) =  - \vect{B}\T \am  + \vect{A}\T (\am^2 + \cm)/2  + \sum_{i=1}^N &\log \tilde{\cZ_i}^x(B_i, A_i; \theta_i)   \\
    &+ \beta \sum_{(ij)} W_{ij} a_i a_j + \frac{\beta^2}{2}\sum_{(ij)} W_{ij}^2 c_i c_j, \notag
\end{align}
with the values of the parameters set by the self-consistency conditions \eqref{eq:chap4-real-GY-A}, \eqref{eq:chap4-real-GY-B}, \eqref{eq:chap4-real-GY-a} and \eqref{eq:chap4-real-GY-c}, which are the TAP equations of the generalized Boltzmann machine at second order. Note that the naive mean-field equations are recovered by ignoring the second order terms in $\beta^2$.

\subparagraph{Relation to message passing}
The TAP equations obtained above must correspond to the fixed points of the Approximate Message Passing (AMP) following the derivation from Belief Propagation (BP) that is presented in \citesec~\ref{sec:chap3-gamp}. In the Appendix B of \cite{Tramel2018} the relaxed-BP equations are derived for the generalized Boltzmann machine: 
\begin{gather}
    \msg{B}{i}{j}^{(t)} = \sum_{k \in \partial i \setminus j}\beta W_{ik} \msg{a^{(t)}}{k}{i} , \quad 
    \msg{A}{i}{j}^{(t)} = - \sum_{k \in \partial i \setminus j} \beta^2 W^2_{ik} \msg{c^{(t)}}{k}{i} ,\\
    \msg{a}{i}{j}^{(t)} = f_1^x(\msg{B}{i}{j}^{(t-1)},\msg{A}{i}{j}^{(t-1)};\theta_i) , \quad
    \msg{c}{i}{j}^{(t)} = f_2^x(\msg{B}{i}{j}^{(t-1)},\msg{A}{i}{j}^{(t-1)};\theta_i).
\end{gather}
To recover the TAP equations for them we define
\begin{gather}
    B_i^{(t)} = \sum_{k \in \partial i}\beta W_{ik} \msg{a^{(t)}}{k}{i} , \quad 
    A_i^{(t)} = - \sum_{k \in \partial i} \beta^2 W^2_{ik} \msg{c^{(t)}}{k}{i} ,\\
    \label{eq:chap4-real-GY-TAP-ac}
    a_i^{(t)} = f_1^x(B_i^{(t-1)},A_i^{(t-1)};\theta_i) , \quad
    c_i^{(t)} = f_2^x(B_i^{(t-1)},A_i^{(t-1)};\theta_i).
\end{gather}
As $B_i^{(t)} = \msg{B}{i}{j}^{(t)} + \beta W_{ij} \msg{a^{(t)}}{j}{i}$ and $A_i^{(t)} = \msg{A}{i}{j}^{(t)} - \beta^2 W^2_{ij} \msg{c^{(t)}}{j}{i}$ we have by developing $f_2^x$ that $c_i^{(t)} = \msg{c}{i}{j}^{(t)} + O(\beta)$ so that
\begin{gather}
    \label{eq:chap4-real-GY-TAP-A}
    A_i^{(t)} = - \beta^2 \sum_{j \in \partial i}  W^2_{ij} c_j^{(t)} + o(\beta^2).
\end{gather}
By developing $f_1^x$ we also have
\begin{align}
    a_k^{(t)} &= f_1^x(\msg{B}{k}{j}^{(t-1)} + \beta W_{kj} \msg{a^{(t-1)}}{j}{i}\, , \; \msg{A}{k}{j}^{(t-1)}- \beta^2 W^2_{kj} \msg{c^{(t-1)}}{j}{k};\theta_i) \\
    & = \msg{a^{(t)}}{k}{j} + \frac{\partial f_1^x}{\partial B_k} \beta W_{kj} \msg{a^{(t-1)}}{j}{k} + O(\beta^2),
\end{align}
with $\displaystyle \frac{\partial f_1^x}{\partial B_k}(B_k^{(t-1)},A_k^{(t-1)};\theta_k) = c_k^{(t)}$. Finally, by replacing in the definition of $B_i$ the messages we obtain
\begin{gather}
    B_i^{(t)}  = \sum_{k \in \partial i}\beta W_{ik} \msg{a^{(t)}}{k}{i}  = \sum_{k \in \partial i}\beta W_{ik} a_k^{(t)} - \beta W_{ki}c_k^{(t)} \msg{a^{(t-1)}}{i}{k}  .
\end{gather}
As $\msg{a^{(t-1)}}{i}{k} = {a^{(t-1)}_i + O(\beta)}$ and using the definition of $A_i^{(t)}$, we finally recover
\begin{gather}
    \label{eq:chap4-real-GY-TAP-B}
    B_i^{(t)}  =  \sum_{k \in \partial i}\beta W_{ik} a_k^{(t)} + A_i^{(t)}a^{(t-1)}_i.
\end{gather}

Hence we indeed recover the TAP equations as the AMP fixed points in \eqref{eq:chap4-real-GY-TAP-ac}, \eqref{eq:chap4-real-GY-TAP-A} and \eqref{eq:chap4-real-GY-TAP-B}. Beyond the possibility to cross-check our results, the message passing derivation also specifies a scheme of updates to solve the self-consistency equations obtained by the Georges-Yedidia expansion. In the applications we consider below we should resort to this time indexing with good convergence properties \cite{bolthausen2014iterative}.

\subparagraph{Solutions of the TAP equations}
As already discussed in \citesec~\ref{sec:chap3-tap}, the TAP equations do not necessarily admit a single solution. In practice, different fixed points are reached when considering different initializations of the iteration of the self-consistent equations. 
%In the applications of the TAP formalism discussed in the remaining of this \citechap~we will discuss the strategy we adopt to take into account this multiplicity.

\section{Vector Approximate Message Passing for the GLM}
\label{sec:app-chap3-vamp}
\begin{figure}
    \centering
    \includegraphics[width=0.5\textwidth]{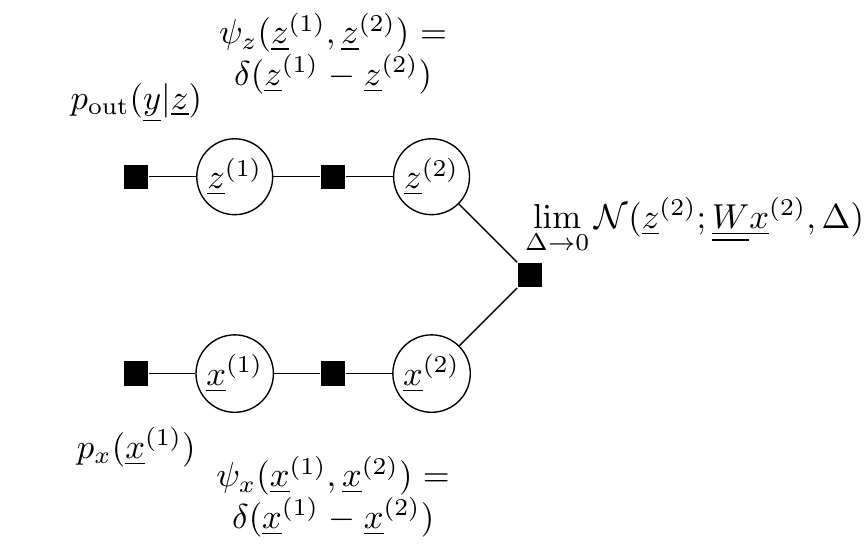}
    \caption{Factor graph representation of the GLM for the derivation of VAMP (reproduction of \citefig~\ref{fig:chap3-vamp-glm} \label{fig:app-chap3-vamp-glm}to help following the derivation described here in the \citeapp).}
\end{figure}
 
We recall here a possible derivation of G-VAMP discussed in \citechap~\ref
{sec:chap3} (\citealg~\ref{alg:chap3-vamp}). We consider a projection of the BP equations for the factor graph \citefig~\ref{fig:app-chap3-vamp-glm}.

\subparagraph{Gaussian assumptions} We start by parametrizing marginals as well as messages coming out of the Dirac factors. For $a = 1, 2$: 
\begin{gather}
    m_{x,{(a)}}(\x^{(a)}) = \cN(\x^{(a)}, \xh^{(a)}, \Cx^{(a)}) \, ,\qquad
    m_{z,{(a)}}(\z^{(a)})  = \cN(\z^{(a)}, \hat{\z}^{(a)}, \Cz^{(a)}) \, ,
\end{gather}
and
\begin{gather}
    \msgt{m}{\psi_x}{\x^{(a)}}(\x^{(a)})  \propto e^{-\frac 1 2 {\x^{(a)}}\T\A_x^{(a)} \x^{(a)} + {\B_x^{(a)}}\T {\x^{(a)}}  } \, ,\\
    \msgt{m}{\psi_z}{\z^{(a)}}(\z^{(a)})  \propto e^{-\frac 1 2 {\z^{(a)}}\T\A_z^{(a)} \z^{(a)} + {\B_z^{(a)}}\T {\z^{(a)}}  } \, .
\end{gather}

\subparagraph{Self consistency of the parametrizations at Dirac factor nodes} 
Around $\psi_x$ the message passing equations are simply
\begin{gather}
    \label{eq:chap3-vamp-trick1}
    \msgt{m}{\psi_x}{\x^{(2)}}(\x^{(2)}) = \msg{m}{\x^{(1)}}{\psi_x}(\x^{(2)}), \qquad \msgt{m}{\psi_x}{\x^{(1)}}(\x^{(1)}) = \msg{m}{\x^{(2)}}{\psi_x}(\x^{(1)})
\end{gather}
and similarly around $\psi_z$. Moreover, considering that messages are marginals to which the contribution of the opposite message is retrieved we have
\begin{gather}
    \msg{m}{\x^{(1)}}{\psi_x}(\x^{(1)}) \propto m_{x,{(1)}}(\x^{(1)}) / \msgt{m}{\psi_x}{\x^{(1)}}(\x^{(1)}), \\
     \msg{m}{\x^{(2)}}{\psi_x}(\x^{(2)}) \propto m_{x,{(2)}}(\x^{(2)}) / \msgt{m}{\psi_x}{\x^{(2)}}(\x^{(2)}) \, .
\end{gather}
Combining this observation along with \eqref{eq:chap3-vamp-trick1} leads to updates \eqref{alg:chap3-vamp-Ax1}
%-\eqref{alg:chap3-vamp-Bx1} 
and \eqref{alg:chap3-vamp-Ax2}.
%-\eqref{alg:chap3-vamp-Bx2}. 
The same reasoning can be followed for the messages around $\psi_z$ leading to updates \eqref{alg:chap3-vamp-Az1}
%-\eqref{alg:chap3-vamp-Bz1} 
and \eqref{alg:chap3-vamp-Az2}.
%-\eqref{alg:chap3-vamp-Bz2}.

\subparagraph{Input and output update functions}
The update functions of means and variances of the marginals are deduced from the parametrized message passing. For the variable $\x^{(1)}$ taking into account the prior $p_x$, the updates are very similar to GAMP input functions: 
\begin{align}
    \xh^{(1)} 
    % & \propto \int \dd{\x^{(1)}} \x^{(1)} m_{x,{(1)}}(\x^{(1)}) \\
     & \propto \int \dd{\x^{(1)}} \x^{(1)} p_x(\x^{(1)}) \msgt{m}{\psi_x}{\x^{(1)}}(\x^{(1)}) \\
     & = \frac{1}{\cZ_x^{(1)}} \int \dd{\x^{(1)}} \x^{(1)} p_x(\x^{(1)}) e^{-\frac 1 2 {\x^{(1)}}\T\A_x^{(1)} \x^{(1)} + {\B_x^{(1)}}\T {\x^{(1)}}} = f_1^x( {\B_x^{(1)}}, \A_x^{(1)}) \, , \\
     {\Cx}^{(1)} &= \frac{1}{\cZ_x^{(1)}} \int \dd{\x^{(1)}} \x^{(1)}{\x^{(1)}}\T 
     p_x(\x^{(1)}) e^{-\frac 1 2 {\x^{(1)}}\T\A_x^{(1)} \x^{(1)} + {\B_x^{(1)}}\T {\x^{(1)}}} \\
     &\qquad \qquad \qquad \qquad \qquad \qquad \qquad \qquad \qquad  - f_1^x( {\B_x^{(1)}}, \A_x^{(1)}) {f_1^x( {\B_x^{(1)}}, \A_x^{(1)}) }\T \notag \\
     &= f_2^x( {\B_x^{(1)}}, \A_x^{(1)})  \, ,
\end{align}
where $\cZ_x^{(1)}$ is as usual the partition ensuring the normalization.

Similarly for the variable $\z^{(1)}$, the update functions are very similar to the GAMP output functions including the information coming from the observations:
\begin{align}
    \hat{\z}^{(1)} & \propto \int \dd{\z^{(1)}} \pout(\y|\z^{(1)}) \msgt{m}{\psi_z}{\z^{(1)}}(\z^{(1)}) \\
    & = \frac{1}{\cZ_z^{(1)}} \int \dd{\z^{(1)}} \pout(\y|\z^{(1)}) e^{-\frac 1 2 {\z^{(1)}}\T\A_z^{(1)} \z^{(1)} + {\B_z^{(1)}}\T {\z^{(1)}}} = f_1^z( {\B_z^{(1)}}, \A_z^{(1)}) \, , \\
    {\Cz}^{(1)} &= \frac{1}{\cZ_z^{(1)}} \int \dd{\z^{(1)}} \z^{(1)}{\z^{(1)}}\T 
     \pout(\y|\z^{(1)}) e^{-\frac 1 2 {\z^{(1)}}\T\A_z^{(1)} \z^{(1)} + {\B_z^{(1)}}\T {\z^{(1)}}} \\
     &\qquad \qquad \qquad \qquad \qquad \qquad \qquad \qquad \qquad  - f_1^z( {\B_z^{(1)}}, \A_z^{(1)}) {f_1^z( {\B_z^{(1)}}, \A_z^{(1)}) }\T \notag \\
     &= f_2^z( {\B_z^{(1)}}, \A_z^{(1)}) \, .
\end{align}

\subparagraph{Linear transformation}
For the middle factor node we consider the vector variable concatenating $\bar{\x}=[\x^{(2)} \z^{(2)}] \in \R^{N + M}$. The computation of the corresponding marginal with the message passing then yields
\begin{gather}
    m_{\bar{\x}}(\bar{\x}) \propto 
    % \int \dd{\x} \int \dd{\z} 
    \lim\limits_{\Delta \to 0}\cN(\z^{(2)}; \W\x^{(2)} , \Delta \mat{I}_M) e^{-\frac 1 2 {\x}\T\A_x^{(2)} \x + {\B_x^{(2)}}\T {\x}  } e^{-\frac 1 2 {\z}\T\A_z^{(2)} \z + {\B_z^{(2)}}\T {\z}}.
\end{gather}
The means of $\x^{(2)}$ and $\z^{(2)}$ are then updated through
\begin{gather}
    \xh^{(2)}, \hat{\z}^{(2)} = \argminn{\x, \z}\left[ 
        \Vert\W\x - \z \Vert^2 /\Delta + {\x}\T\A_x^{(2)} \x - 2 {\B_x^{(2)}}\T {\x} + {\z}\T\A_z^{(2)} \z - 2{\B_z^{(2)}}\T {\z}
         \right],
\end{gather}
at $\Delta \to 0$.
At this point it is advantageous in terms of speed to consider the singular value decomposition $\W=\mat{U}\mat{S}\mat{V}\T$ and to simplify the form of the variance matrices by taking them proportional to the identify, i.e. $\A_z^{(2)} = A_z^{(2)} \mat{I}_{M}$ etc. Under this assumption the solution of the minimization problem is
\begin{gather}
    \xh^{(2)} = g^x_1({\B_x^{(2)}}, A_x^{(2)}, {\B_z^{(2)}}, A_z^{(2)}) = \mat{V} \, \mat{D} \left( {A^{(2)}_z}^{-2} \mat{S} \mat{U}\T \B_z^{(2)} + {A^{(2)}_x}^{-2} \mat{V}\T \B_x^{(2)}   \right) \, , \\
    \hat{z}^{(2)} = g^z_1({\B_x^{(2)}}, A_x^{(2)}, {\B_z^{(2)}}, A_z^{(2)}) = \W g^x_1({\B_x^{(2)}}, A_x^{(2)}, {\B_z^{(2)}}, A_z^{(2)}) \, ,
\end{gather}
with $\mat{D}$ a diagonal matrix with entries $D_{ii} = ({A_z^{(2)}}^{-1} S_{ii}^2+ {A_x^{(2)}}^{-1})^{-1} $.
The scalar variances are then updated using the traces of the Jacobians with respect to the $\B^{(2)}$-s
\begin{align}
    \Cx^{(2)} &= \frac{A_x^{(2)}}{N} \mathrm{tr}\left(\partial g^x_2 / \partial{\B_x^{(2)}}\right) \mat{I}_N = \frac{1}{N} \sum_{i=1}^N ({A_z^{(2)}}^{-1} S_{ii}^2+ {A_x^{(2)}}^{-1})^{-1} \mat{I}_N \\ &= g^x_2({\B_x^{(2)}}, A_x^{(2)}, {\B_z^{(2)}}, A_z^{(2)})\\
    \Cz^{(2)} &= \frac{A_z^{(2)}}{M} \mathrm{tr}\left(\partial g^z_2 / \partial{\B_z^{(2)}}\right) \mat{I}_M = \frac{1}{M} \sum_{i=1}^N S_{ii} ({A_z^{(2)}}^{-1} S_{ii}^2+ {A_z^{(2)}}^{-1})^{-1} \mat{I}_M \\ &= g^z_2({\B_x^{(2)}}, A_x^{(2)}, {\B_z^{(2)}}, A_z^{(2)}).
\end{align}

\section{Multi-value AMP derivation for the GLM}
\label{app:chap6-vect-amp}
% Before specializing to the calibration problem we are interested in, we present the AMP algorithm on vector variables.
% We assume that a set of $P$ observations $\y\kk \in \R^M$ are generated from a set of unknown signals $\x_{0,(k)}$ by a noisy channel following
% \begin{gather}
%     \y\kk = g_0(\W\x_{0,(k)} / \sqrt{N};\eps) \quad \eps \sim p_\epsilon(\eps)  \quad \forall k  = 1 \cdots  P,
% \end{gather}
% , the unknown signals have also i.i.d entries following a prior distribution $p_x$ and $\eps$ plays the role of the noise and is therefore also unobserved. We further assume that the channel is factorized and use the following factorization of the posterior
We here present the derivation of the multi-value AMP and its SE motivated in \citesec~\ref{sec:chap3-multivalue}, focusing on the multi-value GLM. These derivations also appear in \cite{Gabrie2019}.

\subsection{Approximate Message Passing}
The systematic procedure to write AMP for a given joint probability distribution consists in first writing BP on the factor graph, second project the messages on a parametrized family of functions to obtain the corresponding relaxed-BP and third close the equations on a reduced set of parameters by keeping only leading terms in the thermodynamic limit.

\begin{figure}[t]
    \centering
    {\includegraphics[width=0.35\textwidth]{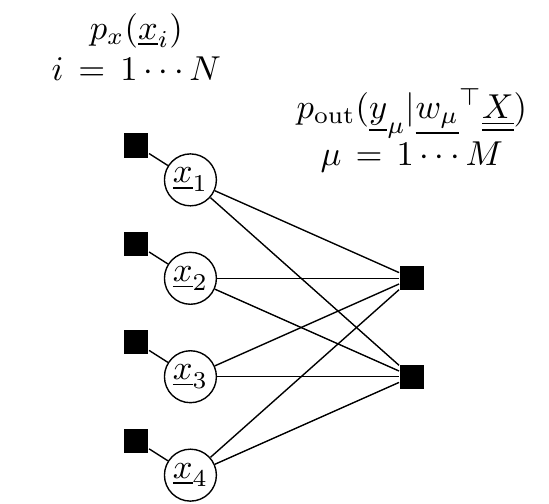}
    }
    % \subfloat[.]{\includegraphics[width=0.4\textwidth]{chap6_glm_vec_cal.pdf}
    % \label{fig:chap6-vect-amp-cal}}
    % \subfloat[blu]{\includegraphics[width=0.35\textwidth]{}
    \caption{Factor graph of the Generalized Linear Model (GLM) on vector variables corresponding to the joint distribution \eqref{eq:chap6-glm-vec-meas}. \label{fig:chap6-vect-amp}}
\end{figure}

For the generic multi-value GLM the posterior measure we are interested in is
\begin{gather}
    \label{eq:chap6-glm-vec-meas}
    p(\X | \Y, \W) = \frac{1}{\cZ(\Y, \W)} \prod_{i=1}^N p(\x_i)\prod_{\mu=1}^M \pout(\y_\mu | \vect{w}_\mu\T\X / \sqrt{N}), \quad \x_i \in \R^P, \quad \y_\mu \in \R^P. 
\end{gather}
where the known entries of matrix $\W$ are drawn i.i.d from a standard normal distribution (the scaling in $1/\sqrt{N}$ is here made explicit). 
The corresponding factor graph is given on \citefig~\ref{fig:chap6-vect-amp}. 
% Conversely to the original derivation of the cal-AMP algorithm \cite{Schulke2013, Schulke2015} we do not further factorize the measure on the $P$ different examples to write the message passing and deduce the corresponding AMP. 
% This setting was actually recently treated for a model of the committee machine in \cite{Aubin2018}. 
We are considering the simultaneous reconstruction of $P$ signals $\x_{0,(k)} \in \R^N$ and therefore write the message passing on the variables $\x_i \in \R^P$.
The major difference with the scalar version (P=1) of AMP is that we will consider covariance matrices between variables coming from the $P$ observations instead of scalar variances. 

\paragraph{Belief propagation (BP)} 
We start with BP on the factor graph of \citefig~\ref{fig:chap6-vect-amp}. For all pairs of index $i-\mu$, we define the update equations of messages function
\begin{gather}
    \label{eq:chap6-bp-calamp-1}
    \msg{\tilde{m}^{(t)}}{\mu}{i} (\x_i) = \frac{1}{\msg{\cZ}{\mu}{i}}
    \int \prod_{i'\neq i} \dd{\x_{i'}}\pout \left(\y_\mu |  \sum_j \frac{W_{\mu j}}{\sqrt{N}}\x_j\right) \prod_{i'\neq i} \msg{m^{(t)}}{i'}{\mu}(\x_{i'})\\
    \label{eq:chap6-bp-calamp-2}
    \msg{m^{(t+1)}}{i}{\mu} (\x_i) = \frac{1}{\msg{\cZ}{i}{\mu}}
    p_x(\x_i)\prod_{\mu' \neq \mu} \msg{\tilde{m}^{(t)}}{\mu'}{i}(\x_i),
\end{gather}
where $\msg{\cZ}{\mu}{i}$ and $\msg{\cZ}{i}{\mu}$ are normalization function that allow to interpret messages as probability distributions.
To improve readability, we drop the time indices in the following derivation, and only specify them in the final algorithm.

\paragraph{Relaxed BP (r-BP)} The second step of the derivation is to develop messages keeping only terms up to order $O(1/N)$ as we take the thermodynamic limit $N \to + \infty$ (at fixed $\alpha = M/N$). At this order, we will find that it is consistent to consider the messages to be approximately Gaussian, i.e. characterized by their means and co-variances. Thus we define
\begin{gather}
\label{eq:chap6-vect-amp-rbp-def-xh}
\msg{\xh}{i}{\mu} = \int \dd{\x} \x \; \msg{m}{i}{\mu}(\x) \\
\label{eq:chap6-vect-amp-rbp-def-Cx}
\msg{\Cx}{i}{\mu} = \int \dd{\x}  \x \x^T \; \msg{m}{i}{\mu}(\x) - \msg{\xh}{i}{\mu}\msg{\xh\T}{i}{\mu}
\end{gather}
and
\begin{gather}
\label{eq:chap6-vect-amp-rbp-def-w}
\displaystyle
\msg{\w}{\mu}{i} = \sum_{i' \neq i} \frac{\Wuip}{\sqrt{N}}\msg{\xh}{i'}{\mu} \\
\label{eq:chap6-vect-amp-rbp-def-V}
\displaystyle
\msg{\V}{\mu}{i} = \sum_{i' \neq i}\frac{\Wuip^2}{N} \msg{\Cx}{i'}{\mu}, 
\end{gather}
where $\msg{\w}{\mu}{i}$ and $\msg{\V}{\mu}{i}$ are related to the intermediate variable $\z_\mu = \vect{w}_\mu\T \X$.

\subparagraph{Expansion of $\msgt{m}{\mu}{i}$ -} 
We defined the Fourier transform $\hat{p}_{\rm out}$ of $\pout(\y_\mu|\z_\mu)$ with respect to its argument $\z_\mu = \vect{w}_\mu\T\X$,
\begin{gather}
    \hat{p}_{\rm out}(\y_\mu|\vect{\xi}_\mu) = \int \dd{\z_\mu} \hat{p}_{\rm out}(\y_\mu | \z_\mu) \, e^{- i \vect{\xi}_\mu\T \z_\mu}.
\end{gather}
Using reciprocally the Fourier representation of $\pout(\y_\mu|\z_\mu)$,
\begin{gather}
    \pout(\y_\mu|\z_\mu) = \frac{1}{(2\pi)^M} \int \dd{\vect{\xi}_\mu} \hat{p}_{\rm out}(\y_\mu | \vect{\xi}_\mu) \, e^{i \vect{\xi}_\mu\T \z_\mu},
\end{gather}
we decouple the integrals over the different $\x_{i'}$ in \eqref{eq:chap6-bp-calamp-1},
\begin{align}
    \label{eq:chap6-deric-calamp-1}
    \msgt{m}{\mu}{i} (\x_i) &
    \propto  \int \dd{\vect{\xi}_\mu} \hat{p}_{\rm out} \left(\y_\mu | \vect{\xi}_\mu\right) e^{i \frac{\Wui}{\sqrt{N}}\vect{\xi}_\mu\T\x_i} \prod_{i'\neq i} \int  \dd{\x_{i'}} \msg{m}{i'}{\mu}(\x_{i'})e^{i \frac{\Wuip}{\sqrt{N}}\x_i \vect{\xi}_\mu\T\x_{i'}} \\
    \label{eq:chap6-deric-calamp-2}
    & \propto \int \dd{\vect{\xi}_\mu} \hat{p}_{\rm out} \left(\y_\mu | \vect{\xi}_\mu\right) e^{i \underline{\xi}\T \left(\frac{\Wui}{\sqrt{N}}\x_i + \msg{\w}{\mu}{i}\right) - \frac 1 2 \underline{\xi}\T \msg{{\V}^{-1}}{\mu}{i} \underline{\xi}}
\end{align}
where developing the exponentials of the product in \eqref{eq:chap6-bp-calamp-1}  allows to express the integrals over the $\x_{i'}$ as a function of the definitions \eqref{eq:chap6-vect-amp-rbp-def-w}-\eqref{eq:chap6-vect-amp-rbp-def-V}, before re-exponentiating to obtain the final result \eqref{eq:chap6-deric-calamp-2}.
Now reversing the Fourier transform and performing the integral over $\vect{\xi}$ we can further rewrite 
\begin{align}
    \msgt{m}{\mu}{i} (\x_i) &
    \propto  \int \dd{\z_\mu} \pout \left(\y_\mu | \z_\mu\right) e^{- \frac{1}{2} 
    \left( \z_\mu - \frac{\Wui}{\sqrt{N}}\x_i - \msg{\w}{\mu}{i}\right)\T
    \msg{{\V}^{-1}}{\mu}{i}
    \left( \z_\mu - \frac{\Wui}{\sqrt{N}}\x_i - \msg{\w}{\mu}{i}\right)
    } \\
    % &
    % \propto  \int \dd{\z_\mu} \pout \left(\y_\mu | \z_\mu\right) e^{- \frac{1}{2} 
    % \left( \z_\mu - \msg{\w}{\mu}{i}\right)\T
    % \msg{{\V}^{-1}}{\mu}{i}
    % \left( \z_\mu - \msg{\w}{\mu}{i}\right) \times
    % } \\
    % & \qquad \qquad \qquad \qquad \qquad \qquad e^{ \left( \z_\mu - \msg{\w}{\mu}{i} \right)\T
    % \msg{{\V}^{-1}}{\mu}{i}
    % \frac{\Wui}{\sqrt{N}}\x_i 
    % - \frac{\Wui^2}{2N}\x_i\T \msg{{\V}^{-1}}{\mu}{i} \x_i
    % }
    \label{eq:chap6-vect-amp-rbp-1}
    &
    \propto  \int \dd{\z_\mu} \mathbb{P}_{\rm out}(\z_\mu; \msg{\w}{\mu}{i}, \msg{\V}{\mu}{i} ) e^{ \left( \z_\mu - \msg{\w}{\mu}{i} \right)\T
    \msg{{\V}^{-1}}{\mu}{i}
    \frac{\Wui}{\sqrt{N}}\x_i 
    - \frac{\Wui^2}{2N}\x_i\T \msg{{\V}^{-1}}{\mu}{i} \x_i
    },
\end{align}
where we are led to introduce the \emph{output update functions},
\begin{gather}
    \label{eq:chap6-vect-amp-Pout}
    \mathbb{P}_{\rm out}(\z_\mu; \msg{\w}{\mu}{i}, \msg{\V}{\mu}{i} ) =  \pout \left(\y_\mu | \z_\mu\right) \cN(\z_\mu; \msg{\w}{\mu}{i}, \msg{\V}{\mu}{i} ) \, ,\\
    \label{eq:chap6-vect-amp-Zout}
    \Zout(\y_\mu , \msg{\w}{\mu}{i}, \msg{\V}{\mu}{i} ) = \int \dd{z_\mu} \pout \left(\y_\mu | \z_\mu\right) \cN(\z_\mu; \msg{\w}{\mu}{i}, \msg{\V}{\mu}{i} ) \, ,\\
    \label{eq:chap6-vect-amp-gout-dgout}
    \gout(\y_\mu , \msg{\w}{\mu}{i}, \msg{\V}{\mu}{i} ) = \frac{1}{\Zout} \frac{\partial \Zout}{\partial \w} \quad \text{ and } \quad
    \dgout = \frac{\partial \gout}{\partial \w},
\end{gather}
where $\cN(\z;\w,\V)$ is the multivariate Gaussian distribution of mean $\w$ and covariance $\V$.
Further expanding the exponential in \eqref{eq:chap6-vect-amp-rbp-1} up to order $O(1/N)$ leads to the Gaussian parametrization 
\begin{align}
    \msgt{m}{\mu}{i} (\x_i) & \propto 1 + \frac{{\Wui}}{\sqrt{N}} \gout \x_i + \frac{{{\Wui}}^2}{2 N} {\x_i}^T (\gout\gout^T + \dgout^1) \x_i \\
    & \propto e^{{\msg{\B}{\mu}{i}}^T\x_i - \frac 1 2 {\x_i}^T\msg{\A}{\mu}{i}\x_i}, 
\end{align}
with
\begin{gather}
    \label{eq:chap6-vect-amp-rb-B}
    \msg{\B}{\mu}{i}  = \frac{{\Wui}}{\sqrt{N}} \gout (\y_\mu , \msg{\w}{\mu}{i}, \msg{\V}{\mu}{i} ) \\
    \label{eq:chap6-vect-amp-rb-A}
    \msg{\A}{\mu}{i}  = - \frac{{{\Wui}}^2}{ N} \dgout(\y_\mu , \msg{\w}{\mu}{i}, \msg{\V}{\mu}{i} ) .
\end{gather}

\subparagraph{Consistency with $\msg{m}{i}{\mu}$ -}
Inserting the Gaussian approximation of $\msgt{m}{\mu}{i}$ in the definition of $\msg{m}{i}{\mu}$, we get the parametrization
\begin{align}
    \msg{m}{i}{\mu}(\x_i) & \propto p_x(\x_i) \prod_{\mu' \neq \mu} e^{{\msg{\B}{\mu'}{i}}^T\x_i - \frac 1 2 {\x_i}^T\msg{\A}{\mu'}{i}\x_i}  \propto p_x(\x_i) e^{-\frac{1}{2}(\x_i - \msg{\lbd}{i}{\mu})^T \msg{\sig}{i}{\mu}^{-1} (\x_i - \msg{\lbd}{i}{\mu})}
\end{align}
with
\begin{gather}
    \label{eq:chap6-vect-amp-rbp-lbd}
\msg{\lbd}{i}{\mu}  = \msg{\sig}{i}{\mu}\left( \sum_{\mu' \neq \mu} \msg{\B}{\mu'}{i} \right)  \\
\label{eq:chap6-vect-amp-rbp-sig}
\msg{\sig}{i}{\mu}  = \left( \sum_{\mu' \neq \mu} \msg{\A}{\mu'}{i} \right)^{-1} .
\end{gather} 

\subparagraph{Closing the equations -}
Ensuring the consistency with the definitions \eqref{eq:chap6-vect-amp-rbp-def-xh}-\eqref{eq:chap6-vect-amp-rbp-def-Cx} of mean and covariance of $\msg{m}{i}{\mu}$ we finally close our set of equations by defining the \emph{input update functions},
\begin{gather}
    \label{eq:chap6-vect-amp-Zx}
    \cZ^x = \int \dd{x} p_x(\x)e^{-\frac 1 2 (\x-\lbd)\T\sigma^{-1}(\x-\lbd)} \\
    \label{eq:chap6-vect-amp-f1x}
    \vect{f}^x_1(\lbd, \sig) = \frac{1}{\cZ^x}\int \dd{\x} \x \, p_x(\x)e^{-\frac 1 2 (\x-\lbd)\T\sig^{-1}(\x-\lbd)} \\
    \label{eq:chap6-vect-amp-f2x}
    \mat{f}^x_2(\lbd, \sig) =  \frac{1}{\cZ^x} \int \dd{x} \x\x\T \, p_x(\x)e^{-\frac 1 2 (\x-\lbd)\T\sig^{-1}(\x-\lbd)} - \vect{f}^x_1(\lbd, \sig)\vect{f}^x_1(\lbd, \sig)\T,
\end{gather}
so that
\begin{gather}
    \label{eq:chap6-vect-amp-rb-xh}
    \msg{\xh}{i}{\mu}  =  \vect{f}^x_1(\msg{\lbd}{i}{\mu} , \msg{\sig}{i}{\mu}) \\
    \label{eq:chap6-vect-amp-rb-Cx}
    \msg{\Cx}{i}{\mu}  = \mat{f}^x_2(\msg{\lbd}{i}{\mu} , \msg{\sig}{i}{\mu}) .
\end{gather}

The closed set of equations \eqref{eq:chap6-vect-amp-rbp-def-w}, \eqref{eq:chap6-vect-amp-rbp-def-V},  \eqref{eq:chap6-vect-amp-rb-B} \eqref{eq:chap6-vect-amp-rb-A}, \eqref{eq:chap6-vect-amp-rbp-lbd}, \eqref{eq:chap6-vect-amp-rbp-sig}, \eqref{eq:chap6-vect-amp-rb-xh} and \eqref{eq:chap6-vect-amp-rb-Cx}, with restored time indices, defines the r-BP algorithm. At convergence of the iterations, we obtain the approximated marginals
\begin{align}
    \label{eq:chap6-vect-amp-marginal-def}
    m_i(\x_i) = \frac 1 {\cZ_i} p_x(\x_i) e^{-\frac 1 2 (\x-\lbd_i)\T\sig_i^{-1}(\x-\lbd_i)} 
\end{align}
with 
\begin{gather}
    % \begin{array}{l}
    \lbd_i  = \sig_i\left( \sum\limits_{\mu=1}^M \msg{\B}{\mu}{i} \right)  \\
    \sig_i  = \left( \sum\limits_{\mu}^M \msg{\A}{\mu}{i} \right)^{-1} .
    % \end{array}
\end{gather}.

As usual, while BP requires to follow iterations over $M \times N$ message distributions over vectors in $\R^P$, r-BP only requires to track $O(M \times N \times P)$ variables, which is a great simplification. Nonetheless, r-BP can be further reduced to the more practical GAMP algorithm, given the scaling of the weights in $O(1/\sqrt{N})$.

\paragraph{Approximate message passing}
% Given the scaling of the weights in $O(1/\sqrt{N})$ it is possible to further simplify the algorithm in the thermodynamic limit.
We define parameters $\w_\mu$, $\V_\mu$ and $\xh_i$, $\Cx_i$, likewise $\lbd_i$ and $\sig_i$ defined above and consider their relations to the original $\msg{\lbd}{i}{\mu}$, $\msg{\sig}{i}{\mu}$, $\msg{\w}{\mu}{i}$, $\msg{\V}{\mu}{i}$, $\msg{\xh}{i}{\mu}$ and $\msg{\Cx}{i}{\mu}$. As a result we obtain the vectorized AMP for the GLM presented in \citealg~\ref{alg:chap6-vect-amp}. 
Note that, similarly to GAMP, relaxing the Gaussian assumption on the weight matrix entries to any distribution with finite second moment yields the same algorithm using the Central Limit Theorem.

%Furthermore, the algorithmic procedure should also generalize to a wider class of random matrices allowing for correlations between entries: the ensemble of orthogonally invariant matrices. In the singular value decomposition of such weight matrices $\W=\mat{U}\,\mat{S}\,\mat{V}\T \in \R^{M\times N}$ the orthogonal basis matrices $\mat{U}$ and $\mat{V}$ are drawn uniformly at random from respectively $\mathrm{O}(M)$ and $\mathrm{O}(N)$, while the diagonal matrix of singular values $\mat{S}$ has an arbitrary spectrum. For the GLM, without calibration variables, the signal is recovered in such cases by the (Generalized) Vector-Approximate Message Passing (G-VAMP) algorithm \cite{Rangan2016,Schniter2016}, inspired from prior works in statistical physics \cite{Opper2001,Opper2001prl,Kabashima2008, Shinzato2009, Kabashima2014} and statistical inference \cite{Minka2001, Opper2005}.

\subsection{State Evolution}
We consider the limit  $N \to + \infty$ at fixed $\alpha = M/N$ and a quenched average over the disorder (here the realizations of $\X_0$, $s_0$, $\Y$ and $\W$), to derive a State Evolution analysis of the previously derived AMP.
To this end, our starting point will be the r-BP equations.

% In \citechap~\ref{sec:chap3} we have seen that there are two ways of deriving the State Evolution, either from the direct averaging of the AMP steps or from 

\subsubsection{State Evolution derivation in mismatched prior and channel setting}
\label{sec:chap6-se-cal-amp}
\paragraph{Definition of the overlaps}
The important quantities to follow the dynamic of iterations and fixed points of AMP are the overlaps. Here, they are the $P \times P$ matrices
\begin{gather}
    \q = \frac{1}{N} \sum_{i=1}^N \xh_i \xh_i^T, \quad \mm = \frac{1}{N} \sum_{i=1}^N \xh_i {\x_{0,i}}^T, \quad \q_0 = \frac{1}{N} \sum_{i=1}^N \x_{0,i} {\x_{0,i}}^T.
\end{gather} 

\paragraph{Output parameters}
Under independent statistics of the entries of $\W$ and under the assumption of independent incoming messages, the variable $\msg{\w}{\mu}{i}$ defined in \eqref{eq:chap6-vect-amp-rbp-def-w} is a sum of independent variables and follows a Gaussian distribution by the Central Limit Theorem. Its first and second moments are
\begin{align}
    \EE{\W}{\msg{\w}{\mu}{i}} & = \frac{1}{\sqrt{N}} \sum_{i'\neq i} \EE{\W}{W_{\mu i'}} \msg{\xh}{i'}{\mu} = 0 \, ,
\end{align}
\begin{align}
\EE{\W}{\msg{\w}{\mu}{i} \msg{\w}{\mu}{i} ^T} 
    & =  \frac{1}{N} \sum_{i'\neq i} \sum_{i''\neq i} \EE{\W}{W_{\mu i''}W_{\mu i'}}\msg{\xh}{i''}{\mu}\msg{\xh}{i'}{\mu}^T \\
	& =  \frac{1}{N} \sum_{i'\neq i} \EE{\W}{W^2_{\mu i'}} \msg{\xh}{i'}{\mu}\msg{\xh}{i'}{\mu}^T =  \frac{1}{N} \sum_{i'=1}^N \msg{\xh}{i'}{\mu}\msg{\xh}{i'}{\mu}^T + O\left({1}/{N}\right)  \notag \\
	& =  \frac{1}{N} \sum_{i=1}^N \xh_{i'}\xh_{i'}^T - \partial_{\lbd} \vect{f}^x_1\sig_i\msg{B}{\mu}{i} \xh_i^T - \left(\partial_{\lbd} \vect{f}_1^x\sig_i\msg{B}{\mu}{i} \xh_i^T \right)^T +O\left({1}/{N}\right) \notag\\
	& = \frac{1}{N} \sum_{i'=1}^N \xh_{i'}\xh_{'i}^T + O\left({1}/{\sqrt{N}}\right)\,
\end{align}
where we used the facts that the $W_{\mu i}$-s are independent with zero mean, and that $\msg{B}{\mu}{i}$, defined in \eqref{eq:chap6-vect-amp-rb-B}, is of order $O(1/\sqrt{N})$.
Similarly, the variable $\msg{\z}{\mu}{i} = \sum_{i'\neq i} \frac{W_{\mu i'}}{\sqrt{N}} \x_{i'}$ is Gaussian with first and second moments
\begin{gather}
    \EE{\W}{\msg{\z}{\mu}{i}} 
    % & 
    = \frac{1}{\sqrt{N}} \sum_{i'\neq i} \EE{\W}{W_{\mu i'}} \x_{0,i'} = 0 \, ,
    \\
    % \quad
% \end{gather}
% \begin{gather}
    \EE{\W}{\msg{\z}{\mu}{i} \msg{\z}{\mu}{i} ^T} 
        % & 
        = \frac{1}{N} \sum_{i'=1}^N \x_{0,i'}{\x_{0,i'}}^T + O\left({1}/{\sqrt{N}}\right). 
    \end{gather}
Furthermore, their covariance is 
\begin{align}
    \EE{\W}{\msg{\z}{\mu}{i} \msg{\w}{\mu}{i}^T} 
        & =  \frac{1}{N} \sum_{i'\neq i} \EE{\W}{W^2_{\mu i'}} \x_{0,i'}\msg{\xh}{i'}{\mu}^T =  \frac{1}{N} \sum_{i'=1}^N \x_{0,i'}\msg{\xh}{i'}{\mu}^T + O\left({1}/{N}\right)  \\
        & =  \frac{1}{N} \sum_{i'=1}^N \x_{0,i'}\xh_{i'}^T -  \x_{0,i}\partial_{\lbd} \vect{f}^x_1\sig_i\msg{\B}{\mu}{i}^T +O\left({1}/{N}\right) \\
        & = \frac{1}{N} \sum_{i'=1}^N \x_{0,i'}\xh_{i'}^T + O\left({1}/{\sqrt{N}}\right). 
    \end{align} 
Hence we find that for all $\mu$-s and all $i$-s, $  \msg{\w}{\mu}{i}$ and $ \msg{\z}{\mu}{i}$ are approximately jointly Gaussian in the thermodynamic limit following a unique distribution $\cN\left( \msg{\z}{\mu}{i}, \msg{\w}{\mu}{i}; \; \vect{0}, \, \Q \right) $ with the block covariance matrix
\begin{gather}
        \Q = 
        \begin{bmatrix}
        \q_0 & \mm \\
        \\
        {\mm}\T & \q \\ 
        \end{bmatrix}.
\end{gather}
For the variance message $\msg{\V}{\mu}{i}$, defined in \eqref{eq:chap6-vect-amp-rbp-def-V}, we have 
\begin{align}
    \EE{\W}{\msg{\V}{\mu}{i}} &= \sum_{i'\neq i} \EE{\W}{\frac{W_{\mu i'}}{N}^2} \msg{\Cx}{i'}{\mu} = \sum_{i'=1}^N \frac{1}{N} \msg{\Cx}{i'}{\mu} + O\left({1}/{N}\right) \\
    &= \sum_{i'=1}^N \frac{1}{N} \Cx_{i'} + O\left({1}/{\sqrt{N}}\right) ,
\end{align}
where using the developments of $\msg{\lbd}{i}{\mu}$ and $\msg{\sig}{i}{\mu}$ \eqref{eq:chap6-vect-amp-rbp-lbd}-\eqref{eq:chap6-vect-amp-rbp-sig}, along with the scaling of $\msg{\B}{\mu}{i} $ in $O({1}/{\sqrt{N}})$ we replaced 
\begin{align}
    \msg{\Cx}{i}{\mu} = \mat{f}_2^x(\msg{\lbd}{i}{\mu}, \msg{\sig}{i}{\mu}) = \mat{f}_2^x(\lbd_i, \sig_i) - \partial_{\lbd}\mat{f}^x_2 \sig_i \msg{\B}{\mu}{i}^T = \mat{f}_2^x(\lbd_i, \sig_i) + O\left({1}/{\sqrt{N}}\right).
\end{align}
Futhermore, we can check that 
\begin{gather}
    \lim_{N\to + \infty} \EE{\W}{\msg{\V}{\mu}{i}^2 - \EE{\W}{\msg{\V}{\mu}{i}}^2} = 0 ,
\end{gather}
meaning that all $\msg{\V}{\mu}{i}$ concentrate on their identical mean in the thermodynamic limit, which we note
\begin{gather}
    \V = \sum_{i=1}^N \frac{1}{N} \Cx_i .
\end{gather}

\paragraph{Input parameters} Here we use the re-parametrization trick to express $\y_\mu$ as a function $g_0(\cdot)$ taking
%  the calibration variable $s_\mu$ and 
 a noise $\eps_\mu \sim p_\epsilon(\eps_\mu)$ as inputs:
% \begin{gather}
    $\y_\mu = g_0(\vect{w}_\mu\T\X_0, \eps_\mu)$.
% \end{gather}
Following \eqref{eq:chap6-vect-amp-rb-A}-\eqref{eq:chap6-vect-amp-rb-B} and \eqref{eq:chap6-vect-amp-marginal-def}, 
\begin{align}
    \sig_i^{-1}\lbd_i 
    & = \sum_{\mu=1}^M \frac{W_{\mu i}}{\sqrt{N}} \gout\left(\y_\mu, \msg{\w}{\mu}{i},  \msg{\V}{\mu}{i}\right) \\
    & =  \sum_{\mu=1}^M \frac{W_{\mu i}}{\sqrt{N}} \gout\left(g_0\left( \sum_{i'\neq i} \frac{W_{\mu i'}}{\sqrt{N}} \x_{0,i'} + \frac{W_{\mu i}}{\sqrt{N}} \x_{0,i}, \eps_\mu \right), \msg{\w}{\mu}{i},  \msg{\V}{\mu}{i}\right) \\
    & =  \sum_{\mu=1}^M \frac{W_{\mu i}}{\sqrt{N}} \gout\left(g_0\left( \sum_{i'\neq i} \frac{W_{\mu i'}}{\sqrt{N}} \x_{0,i'}, \eps_\mu \right), \msg{\w}{\mu}{i},  \msg{\V}{\mu}{i}\right)  \notag\\
    &\qquad \qquad + \sum_{\mu=1}^M \frac{W^2_{\mu i}}{N}  \partial_{z} \mat{\gouts}\left(g_0\left( \msg{\z}{\mu}{i}, \eps_\mu\right), \msg{\w}{\mu}{i}, \msg{\V}{\mu}{i}\right) \x_{0,i}.
\end{align}
The first term is again a sum of independent random variables, given the $W_{\mu i}$ are i.i.d. with zero mean, of which the messages of type $\mu \to i$ are assumed independent. The second term has non-zero mean and can be shown to concentrate. Finally recalling that all $\msg{\V}{\mu}{i}$ also concentrate on $\V$ we obtain the distribution
\begin{gather}
    \sig_i^{-1}\lbd_i \sim \cN\left(\sig_i^{-1}\lbd_i;\; \alpha \mh \x_{0,i}, \sqrt{\alpha \qh} I_P\right) 
\end{gather}
with
\begin{gather}
    \label{eq:chap6-se-non-nishi-qh}
    \qh  = \int \dd{\eps} p_{\epsilon}(\eps) \dd{s_0}p_{s_0}(s_0) \int \dd{\w} \dd{\z} \cN(\z, \w ; \underline{0}, \Q) 
    \gout(g_0\left( \z  ,  \eps\right) ,\w,  \V) \times \\
    \qquad \qquad \qquad \qquad \qquad \qquad \qquad \qquad \qquad \qquad \qquad \qquad \qquad \gout(g_0\left( \z  , \eps\right), \w,  \V)^T \, ,\notag \\
	\label{eq:chap6-se-non-nishi-mh}
    \mh  = \int \dd{\eps} p_{\epsilon}(\eps) \dd{s_0}p_{s_0}(s_0) \int \dd{\w} \dd{\z} \cN(\z, \w ; \underline{0}, \Q)
	\partial_{\z} \mat{\gouts}(g_0\left( \z  ,  \eps\right), \w , \V) .
\end{gather}
For the inverse variance $\sig_i^{-1}$ one can check again that it concentrates on its mean 
\begin{gather}
    \sig_i^{-1}  = \sum\limits_{\mu=1}^M \frac{{\Wui}^2}{ N} \dgout(\y_\mu , \msg{\w}{\mu}{i}, \msg{\V}{\mu}{i} ) \simeq \alpha \chih \, , \\
    \label{eq:chap6-se-non-nishi-chih}
    \chih  = - \int \dd{\eps} p_\epsilon(\eps) \dd{s_0} p_{s_0}(s_0)\int \dd{\eps} \dd\z \cN(\z, \w ; \underline{0}, \Q) 
	 \partial_{\omega} \mat{\gouts}( g_0\left( \z , \eps\right), \w , \V) \, .
\end{gather}

\paragraph{Closing the equations} These statistics of the input parameters must ensure that consistently
\begin{gather}
    \V = \frac{1}{N}\sum\limits_{i=1}^N \Cx_i = \EE{\lbd, \sig}{\mat{f}^x_2(\lbd, \sig)} ,\\
    \q = \frac{1}{N} \sum\limits_{i=1}^N \xh_i \xh_i\T = \EE{\lbd, \sig}{\mat{f}^x_1(\lbd, \sig)\mat{f}^x_1(\lbd, \sig)\T},\\
    \mm = \frac{1}{N} \sum\limits_{i=1}^N \xh_i {\x_{0,i}}\T = \EE{\lbd, \sig}{ \mat{f}^x_1(\lbd, \sig){\x_{0,i}}\T } ,
\end{gather}
which gives upon expressing the computation of the expectations
\begin{gather}
    \label{eq:chap6-se-non-nishi-V}
    \V = \int \dd{\x_0}p_{x_0}(\x_0) \int \D{\vect{\xi}} \mat{f}^x_2 \left( (\alpha \chih)^{-1}\left({\sqrt{\alpha \qh} \underline{\xi} + \alpha \mh\x_0}\right); (\alpha \chih)^{-1}  \right) \, , \\
    \label{eq:chap6-se-non-nishi-m}
    \mm = \int \dd{\x_0}p_{x_0}(\x_0) \int \D{\vect{\xi}} \vect{f}^x_1 \left( (\alpha \chih)^{-1}\left({\sqrt{\alpha \qh} \underline{\xi} + \alpha \mh\x_0}\right); (\alpha \chih)^{-1}  \right){\x_0}\T \, , \\
    \q =  \int \dd{\x_0}p_{x_0}(\x_0) \int \D{\vect{\xi}} \vect{f}^x_1 \left( (\alpha \chih)^{-1}\left({\sqrt{\alpha \qh} \underline{\xi} + \alpha \mh\x_0}\right); (\alpha \chih)^{-1}  \right) \times \notag\\
    \label{eq:chap6-se-non-nishi-q}
    \qquad \qquad \qquad \qquad \qquad \qquad \qquad \qquad \vect{f}^x_1 \left( (\alpha \chih)^{-1}\left({\sqrt{\alpha \qh} \underline{\xi} + \alpha \mh\x_0}\right); (\alpha \chih)^{-1}  \right)\T .
\end{gather}
The State Evolution analysis of the GLM on the vector variables finally consists in iterating alternatively the equations \eqref{eq:chap6-se-non-nishi-qh}, \eqref{eq:chap6-se-non-nishi-mh}, \eqref{eq:chap6-se-non-nishi-chih}, and the equations \eqref{eq:chap6-se-non-nishi-V}, \eqref{eq:chap6-se-non-nishi-m} \eqref{eq:chap6-se-non-nishi-q} until convergence.

% \paragraph{Reconstruction of the calibration variable} 
% In parallel, one can follow the reconstruction of $\vect{s}$ by introducing the scalar overlaps
% \begin{gather}
%     r = \frac{1}{M} \sum_{\mu=1}^M \sh_\mu ^2 , \quad \nu = \frac{1}{M} \sum_{\mu=1}^M \sh_\mu s_{0,\mu},  \quad r_0 = \frac{1}{M} \sum_{\mu=1}^M s^2_{0,\mu} .
% \end{gather}
% Recalling the definition of the estimator $\vect{\sh}$ \eqref{eq:chap6-vect-amp-sh}, and after following the steps of the above derivation, one can see that the calibration overlaps can be computed from the previously introduced SE variables,
% \begin{gather}
%     \label{eq:chap6-se-non-nishi-r}
%     r =  \int \dd{\eps} p_{\epsilon}(\eps) \dd{s_0}p_{s_0}(s_0) \int \dd{\w} \dd{\z} \cN(\z, \w ; \underline{0}, \Q) \,
%     \sh\left(g_0\left( \z  ,  \eps\right), \w , \V\right)^2 , \\
%     \label{eq:chap6-se-non-nishi-rho}
%     \nu =  \int \dd{\eps} p_{\epsilon}(\eps) \dd{s_0}p_{s_0}(s_0) \int \dd{\w} \dd{\z} \cN(\z, \w ; \underline{0}, \Q) \,
%     \sh\left(g_0\left( \z  ,  \eps\right), \w , \V\right)  s_0 .
% \end{gather}

\paragraph{Performance analysis}
The mean squared error (MSE) on the reconstruction of $\X$ by the AMP algorithm is then predicted by 
\begin{gather}
\MSE(\X) = q - 2 m + q_0,
\end{gather}
where the scalar values used here correspond to the (unique) value of the diagonal elements of the corresponding overlap matrices. This MSE can be computed throughout the iterations of State Evolution. 
% Similarly, the MSE in the reconstruction of the calibration variable can be computed as
% \begin{gather}
%     \MSE(\vect{s}) = r - 2 \nu + r_0,
% \end{gather}
% throughout the iterations.
Remarkably, the State Evolution MSEs follow precisely the MSE of the cal-AMP predictors along the iterations of the algorithm provided the procedures are initialized consistently. A random initialization of $\xh_i$ in cal-AMP corresponds to an initialization of zero overlap $m = 0$, $\nu = 0$, with variance of the priors $q = q_0$ in the State Evolution.
% The fixed points of the iterated equations can subsequently be plugged into \eqref{eq:chap6-se-non-nishi-r} and \eqref{eq:chap6-se-non-nishi-rho} to 

\subsubsection{Bayes optimal State Evolution}
The SE equations can be greatly simplified in the Bayes optimal setting where the statistical model used by the student (priors $p_x$ and $p_s$, and channel $\pout$) is known to match the teacher. 
In this case, the true unknown signal $\X_0$ is in some sense statistically equivalent to the estimate $\mat{\hat{X}}$ coming from the posterior. More precisely one can prove the Nishimori identities \cite{Opper1991, Iba1999, Nishimori2001} (or \cite{Kabashima2016} for a concise demonstration and discussion) implying that
\begin{gather}
    \q =  \mm, \quad  \V = \q_0  - \mm, \quad \qh  = \mh =\chih  \quad  \text{ and } \quad r = \nu.
\end{gather}
As a result the State Evolution reduces to a set of two equations
\begin{gather}
    % \label{eq:chap6-se-vect-glm-bayes-opt-r}
    % r =  \int \dd{\eps} p_{\epsilon}(\eps) \dd{s_0}p_{s_0}(s_0) \int \dd{\w} \dd{\z} \cN(\z, \w ; \underline{0}, \Q) \,
    % \sh\left(g_0\left( \z  ,  \eps\right), \w , \q_0 - \m\right)^2 , \\
    \label{eq:chap6-se-vect-glm-bayes-opt-m}
    \mm = \int \dd{\x_0}p_{x_0}(\x_0) \int \D{\vect{\xi}} \vect{f}^x_1 \left( (\alpha \mh)^{-1}\left({\sqrt{\alpha \mh} \underline{\xi} + \alpha \mh\x_0}\right); (\alpha \mh)^{-1}  \right){\x_0}\T \, \\
    \label{eq:chap6-se-vect-glm-bayes-opt-mh}
    \mh  = \int \dd{\eps} p_{\epsilon}(\eps) \dd{s_0}p_{s_0}(s_0) \int \dd{\w} \dd{\z} \cN(\z, \w ; \underline{0}, \Q)
    \gout\left(g_0\left( \z  ,  \eps\right), \w , \q_0 - \m)\right) \times \\
    \qquad \qquad \qquad \qquad \qquad \qquad \qquad \qquad  \qquad \qquad \qquad \gout\left(g_0\left( \z  ,  \eps\right), \w , \q_0 - \m)\right) \notag,
\end{gather}
with the block covariance matrix
\begin{gather}
    \label{eq:chap6-Q-bayes-opt}
    \Q = 
    \begin{bmatrix}
    \q_0 & \mm \\
    \\
    {\mm}\T & \mm \\ 
    \end{bmatrix}.
\end{gather}

\bibliographystyle{alpha}
\bibliography{mini-review}
\end{document}